\documentclass[sigplan]{acmart}

\acmConference[PPoPP'20]{Principles and Practice of Parallel Programming 2020}{February 22--26, 2019}{San Diego, California, USA}
\acmYear{2020}
\acmISBN{} 
\acmDOI{} 
\startPage{1}

\setcopyright{none}

\usepackage{booktabs}   
\usepackage{subcaption} 

\usepackage{microtype}

\usepackage{xcolor}

\usepackage{algorithm2e}
\SetKwProg{Struct}{Struct}{ is}{end}
\SetKwInOut{Input}{input}
\SetKwInOut{Output}{output}
\SetKwProg{Proc}{Procedure}{ is}{end}
\SetKwRepeat{Do}{do}{while}
\DontPrintSemicolon
\LinesNumbered
\SetKwFor{Case}{case}{}{end case}
\SetArgSty{textnormal}
\SetKw{Volatile}{volatile}
\SetKw{New}{new}
\SetKw{Until}{until}
\SetKw{Break}{break}
\SetAlgoVlined

\definecolor{algoColorLines}{named}{lightgray}

\makeatletter

\def\algocf@Hlne{%
  \hbox{
    \leaders \hbox{\vrule width 3pt height 0.6pt depth 0pt \hskip 2pt } \hskip 10pt
  }
}

\def\mydashbox#1#2{%
  \setbox0\hbox{#2}%
  \dimen0\ht0
  \advance\dimen0\dp0
  \setbox2\vbox to \dimen0{{
    \color{#1}\leaders
    \vbox{
      \vskip0pt\hrule height 3pt width .6pt
    }
    \vfill
  }}%
  \ht2=\ht0
  \dp2=\dp0
  \box2
  \unhbox0
}

\renewcommand{\algocf@Vline}[1]{%
  \strut\par\nointerlineskip
  \algocf@push{\skiprule}%
  \hbox{%
    \mydashbox{algoColorLines}{%
      \vtop{%
        \algocf@push{\skiptext}%
        \vtop{\algocf@addskiptotal\advance\hsize by -\skiplength #1}%
        {\color{algoColorLines}\Hlne}
     }%
    }
  }
  \vskip\skiphlne%
  \algocf@pop{\skiprule}%
  \nointerlineskip%
}

\renewcommand{\algocf@Vsline}[1]{%
  \strut\par\nointerlineskip%
  \algocf@push{\skiprule}%
  \hbox{%
    \mydashbox{algoColorLines}{%
      \vtop{
        \algocf@push{\skiptext}%
        \vtop{\algocf@addskiptotal\advance\hsize by -1pt #1}
      }%
    }%
  }
  \algocf@pop{\skiprule}%
}

\makeatother

\usepackage{amsthm}

\newtheorem{invariant}{Invariant}

\usepackage{wrapfig}

\usepackage[utf8]{inputenc}

\usepackage{newunicodechar}
\newunicodechar{§}{\monospaced}
\def\monospaced#1§{$\texttt{#1}$}

\usepackage{mathtools}

\DeclarePairedDelimiter\ceil{\lceil}{\rceil}
\DeclarePairedDelimiter\floor{\lfloor}{\rfloor}
\DeclareMathOperator{\insertion}{insert}
\DeclareMathOperator{\deletion}{delete}
\DeclareMathOperator{\lookup}{lookup}
\DeclareMathOperator{\haskey}{hasKey}

\usepackage{tikz}

\usetikzlibrary{patterns}
\usetikzlibrary{calc}

\usepackage{pgfplots}

\usepackage{listings}

\usepackage[scaled]{beramono}

\definecolor{pdfbgcolor}{RGB}{180,180,180}
\IfFileExists{trbotpc.txt}{\pagecolor{pdfbgcolor}}{}

\newcommand{\paratitle}[1]{\vspace{1mm}\noindent\textbf{#1.}}

\makeatletter
\newcommand{\removelatexerror}{\let\@latex@error\@gobble}
\makeatother

\SetInd{0.35em}{0.35em}

\usepackage[hidecomments]{collab}

\collabAuthor{ap}{red}{Aleksandar Prokopec}
\collabAuthor{tb}{brown}{Trevor Brown}
\collabAuthor{da}{green}{Dan Alistarh}

\let\origthelstnumber\thelstnumber
\makeatletter
\newcommand*\Suppressnumber{%
  \lst@AddToHook{OnNewLine}{%
    \let\thelstnumber\relax%
     \advance\c@lstnumber-\@ne\relax%
    }%
}

\newcommand*\Reactivatenumber[1]{%
  \setcounter{lstnumber}{\numexpr#1-1\relax}
  \lst@AddToHook{OnNewLine}{%
   \let\thelstnumber\origthelstnumber%
   \refstepcounter{lstnumber}
  }%
}

\hypersetup{
  colorlinks=true
}

\usepackage{lipsum}

\usepackage{xcolor}

\definecolor{Cherry}{rgb}{0.7, 0.1, 0.1}
\definecolor{Cerulean}{rgb}{0.0, 0.5, 0.7}
\definecolor{Forest}{rgb}{0.3, 0.7, 0.3}
\definecolor{SeaBlue}{rgb}{0.8, 0.3, 0.7}
\definecolor{Gray}{rgb}{0.5, 0.5, 0.5}
\definecolor{Bluish}{rgb}{0.4, 0.6, 0.9}
\definecolor{DarkBluish}{rgb}{0.2, 0.4, 0.7}
\definecolor{BloodOrange}{rgb}{0.9, 0.4, 0.2}
\definecolor{SunsetYellow}{rgb}{0.95, 0.65, 0.2}
\definecolor{Greenish}{rgb}{0.35, 0.70, 0.35}

\newcommand\crule[3][black]{\textcolor{#1}{\rule{#2}{#3}}}

\usepackage{flushend}

\begin{document}


\title[Analysis and Evaluation of Non-Blocking IST]{Analysis and Evaluation of Non-Blocking Interpolation Search Trees}         


\author{Aleksandar Prokopec}
\orcid{0000-0003-0260-2729}             
\affiliation{
  \institution{Oracle Labs}           
  \country{Switzerland}                   
}
\email{aleksandar.prokopec@gmail.com}         

\author{Trevor Brown}
\affiliation{
  \institution{University of Waterloo}            
  \country{Canada}                    
}
\email{me@tbrown.pro}          

\author{Dan Alistarh}
\orcid{0000-0003-3650-940X}             
\affiliation{
  \institution{Institute of Science and Technology}            
  \country{Austria}                    
}
\email{dan.alistarh@ist.ac.at}          

\ccsdesc[500]{Theory of computation~Concurrent algorithms}
\ccsdesc[300]{Theory of computation~Shared memory algorithms}
\ccsdesc[500]{Computing methodologies~Concurrent algorithms}


\keywords{concurrent data structures, search trees, interpolation, non-blocking algorithms}


\begin{abstract}
  We start by summarizing the recently proposed implementation of
  the first non-blocking concurrent interpolation search tree (C-IST) data structure.
  We then analyze the individual operations of the C-IST,
  and show that they are correct and linearizable.
  We furthermore show that lookup (and several other non-destructive operations) are wait-free,
  and that the insert and delete operations are lock-free.

  We continue by showing that the C-IST has the following properties.
  For arbitrary key distributions,
  this data structure ensures worst-case $O(\log n + p)$ amortized time
  for search, insertion and deletion traversals.
  When the input key distributions are \emph{smooth},
  lookups run in expected $O(\log \log n + p)$ time,
  and insertion and deletion run in expected amortized $O(\log \log n + p)$ time, 
  where $p$ is a bound on the number of threads.

  Finally, we present an extended experimental evaluation
  of the non-blocking IST performance.
\end{abstract}

\maketitle


\section{Introduction}
\label{sec:intro}

The recently proposed non-blocking interpolation search tree (C-IST)
improves the asymptotic complexity of the basic tree operations,
such as insertion, lookup and predecessor queries,
from logarithmic to doubly-logarithmic for a specific class of workloads
whose input key distributions are \emph{smooth}~\footnote{
Intuitively speaking, this property bounds the height of the "spikes"
in the probability density function of the input key set.
The work on sequential interpolation search trees
defines smoothness more precisely~\cite{IST}.
}.
For arbitrary key distributions, C-IST operations
retain standard logarithmic running time bounds.

While C-IST was proposed and described in detail in related work~\cite{c-ist-ppopp2020},
this paper analyzes the C-IST data structure
from the correctness and an asymptotic complexity standpoint.
This work furthermore extends the experimental analysis
of the runtime behavior and performance of a concrete implementation of C-IST,
and compares C-IST to a wider range of data structures.

After summarizing the C-IST data structure and its operations in Section~\ref{sec:algorithm},
we put forth the following contributions:

\begin{itemize}
\item
We prove that the basic C-IST operations
are correct, linearizable and non-blocking
(wait-free lookup, and lock-free modifications)
in Section~\ref{sec:correctness}.
\item
We analyze the asymptotic complexity of the C-IST algorithm
in Section~\ref{sec:complexity-analysis}.
In particular,
we show that the worst-case cost of a lookup operation is $O(p + \log n)$,
where $p$ is the bound on the number of concurrent threads
and $n$ is the number of keys in the C-IST;
and we show that the worst-case amortized running time
of insert and delete operations is $O(\gamma (p + \log n))$,
where $\gamma$ is a bound on average interval contention.
Furthermore, we show that, when the input key distribution is smooth,
the expected worst-case running time of a lookup operation is $O(p + \log \log n)$;
and the expected amortized running time
of insertion and deletion is $O(\gamma (p + \log \log n))$.
\item
We present an extended experimental evaluation of our C-IST implementation
in Section~\ref{sec:evaluation}.
The evaluation includes a performance comparison against seven
state-of-the-art concurrent tree data structures,
a measurement of last-level cache misses between C-IST and best-performing competitors,
a breakdown of execution time across different key-set sizes,
parallelism levels and update rates,
the impact of using a multi-counter in the root node,
and a performance evaluation on a No-SQL database workload,
and a Zipfian distribution workload, at different skew levels.

\end{itemize}

The paper ends with an overview of related work in Section \ref{sec:related},
and a short conclusion in Section \ref{sec:conclusion}.


\newpage
\section{Non-Blocking Interpolation Search Tree}
\label{sec:algorithm}

\subsection{Data Types}
\label{sec:data-types}

\begin{figure*}[]
\begin{minipage}[t]{0.22\textwidth}
\begin{lstlisting}[numbers=none]
struct IST is
  root: Node

struct Single: Node is
  key: KeyType
  val: ValType
\end{lstlisting}
\end{minipage}
\begin{minipage}[t]{0.22\textwidth}
\begin{lstlisting}[numbers=none]
struct Inner: Node is
  initSize: int
  degree: int
  keys: KeyType[]
  children: Node[]
  status: [int, bool, bool]
  count: int
\end{lstlisting}
\end{minipage}
\begin{minipage}[t]{0.22\textwidth}
\begin{lstlisting}[numbers=none]
struct Empty: Node is

struct Rebuild: Node is
  target: Inner
  parent: Inner
  index: int
\end{lstlisting}
\end{minipage}
\begin{minipage}[t]{0.24\textwidth}
\vspace{0.1cm}
\includegraphics[scale=0.21]{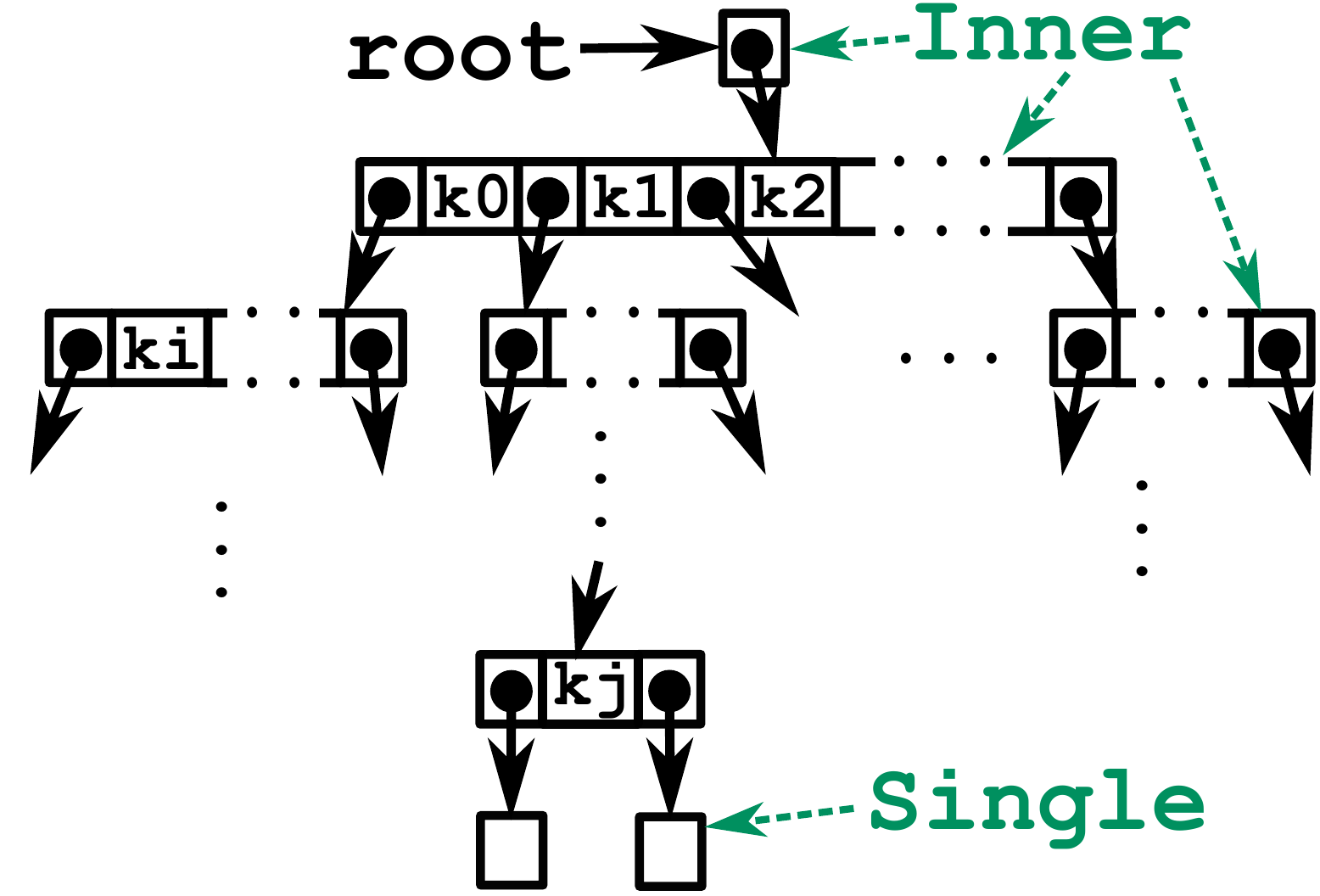}
\end{minipage}
\vspace{-7mm}
\caption{Data Types}
\label{algo:data-types}
\end{figure*}

Non-blocking IST consists of the data types shown in Figure~\ref{algo:data-types}.
The main data type, named §IST§, represents the interpolation search tree.
It has only one field called §root§, which is a pointer to the root node of the data structure.
When the data structure is created,
the §root§ field points to an empty leaf node of type §Empty§.
The data type §Single§ is used as a leaf node
that has a single key and an associated value.
The data type §Inner§ represents inner nodes in the tree.
The relationship between these node types is shown
on the right side of Figure~\ref{algo:data-types}.

The §Inner§ data type contains the search keys, and the pointers to the children nodes.
In addition to that, the §Inner§ data type has the field §degree§,
which holds the number of children of that inner node,
and a field called §initSize§,
which holds the number of keys
the corresponding subtree had at the moment when this inner node was created.
Apart from the pointers in the §children§ array,
all of these fields are written once when the inner node is created,
and are not modified after that.

Finally, §Inner§ contains two more volatile fields, called §count§ and §status§.
These two fields are used to coordinate the rebuilding process.
The §count§ field contains the number of updates that occurred
in the subtree rooted at this inner node since this inner node was created.
The §status§ field contains an integer and two booleans.
When the inner node is created, it is $(0, \bot, \bot)$.
It changes to a non-zero value when this inner node
must be replaced with its copy during a rebuilding operation
(more on that later).

The §Rebuild§ data type is used when a subtree needs to be rebuilt.
It has a field called §target§, which contains a pointer to the root
of the subtree that needs to be rebuilt.
It also has the field §parent§, which contains a pointer to the §Rebuild§ node's parent,
and the §index§ of the §target§ in the §parent§ node's array of pointers to children.
These two fields are sufficient to replace the §Rebuild§ in the parent
once the rebuild operation completes.
The §status§ field and the §Rebuild§ type are explained in more detail later.

To work correctly, the operations of the C-IST
must maintain the following invariants.

\begin{invariant}[Key presence]
\label{inv:presence}
  For any key §k§ reachable in the IST §I§,
  there exists exactly one path of the form §I§
  $\overset{\texttt{root}}{\rightarrow}$ §n§$_0$
  $\overset{\texttt{children[i$_0$]}}{\rightarrow}$
  $(§n§_1 \enskip | \enskip §r§_1 \overset{\texttt{target}}{\rightarrow} §n§_1)$
  $\overset{\texttt{children[i$_1$]}}{\rightarrow}$
  $\ldots$
  $\overset{\texttt{children[i$_{m-1}$]}}{\rightarrow}$
  $(§n§_m \enskip | \enskip §r§_m \overset{\texttt{target}}{\rightarrow} §n§_m)$,
  where §n§$_m$ holds the key §k§, §r§$_m$ is a §Rebuild§ node,
  and $|$ is a choice between two patterns.
\end{invariant}

\begin{definition}[Cover]
\label{def:cover}
A root node §n§
\emph{covers} the interval $\langle -\infty, \infty \rangle$.
Given an inner node §n§ of degree $d$ that covers the interval $[ a, b \rangle$,
and holds the keys $k_0, k_1, \ldots, k_{d-2}$ in its §keys§ array,
its child §n.children[i]§ covers the interval $[ k_{i-1}, k_i \rangle$,
where we define $k_{-1} = a$ and $k_{d-1} = b$.
\end{definition}

\begin{definition}[Content]
A node §n§ \emph{contains} a key §k§ if and only if
the path from the root of the IST §I§ to the leaf with the key §k§
contains the node §n§.
An IST §I§ \emph{contains} a key §k§ if and only if the §root§ contains the key §k§.
\end{definition}

\begin{invariant}[Search tree]
\label{inv:search}
If a node §n§ covers $[ a, b \rangle$ and contains a key §k§,
then §k§ $\in [ a, b \rangle$.
\end{invariant}

\begin{invariant}[Acyclicity]
\label{inv:acyclicity}
There are no cycles in the interpolation search tree.
\end{invariant}

\begin{definition}[Has-key]
\label{def:has-key}
Relation $\haskey(§I§, k)$ holds if and only if §I§ satisfies the invariants,
and contains the key $k$.
\end{definition}

In the C-IST data structure (and in the traditional, sequential IST as well),
the degree $d$ of a node with cardinality $n$ is $\Theta(\sqrt{n})$.
An example of an C-IST is shown in the figure below.
The root has a degree $\Theta(\sqrt{n})$,
its children have the degree $\Theta(\sqrt[4]{n})$,
its grandchildren have the degree $\Theta(\sqrt[8]{n})$ and so on.

\begin{center}
\includegraphics[scale=0.24]{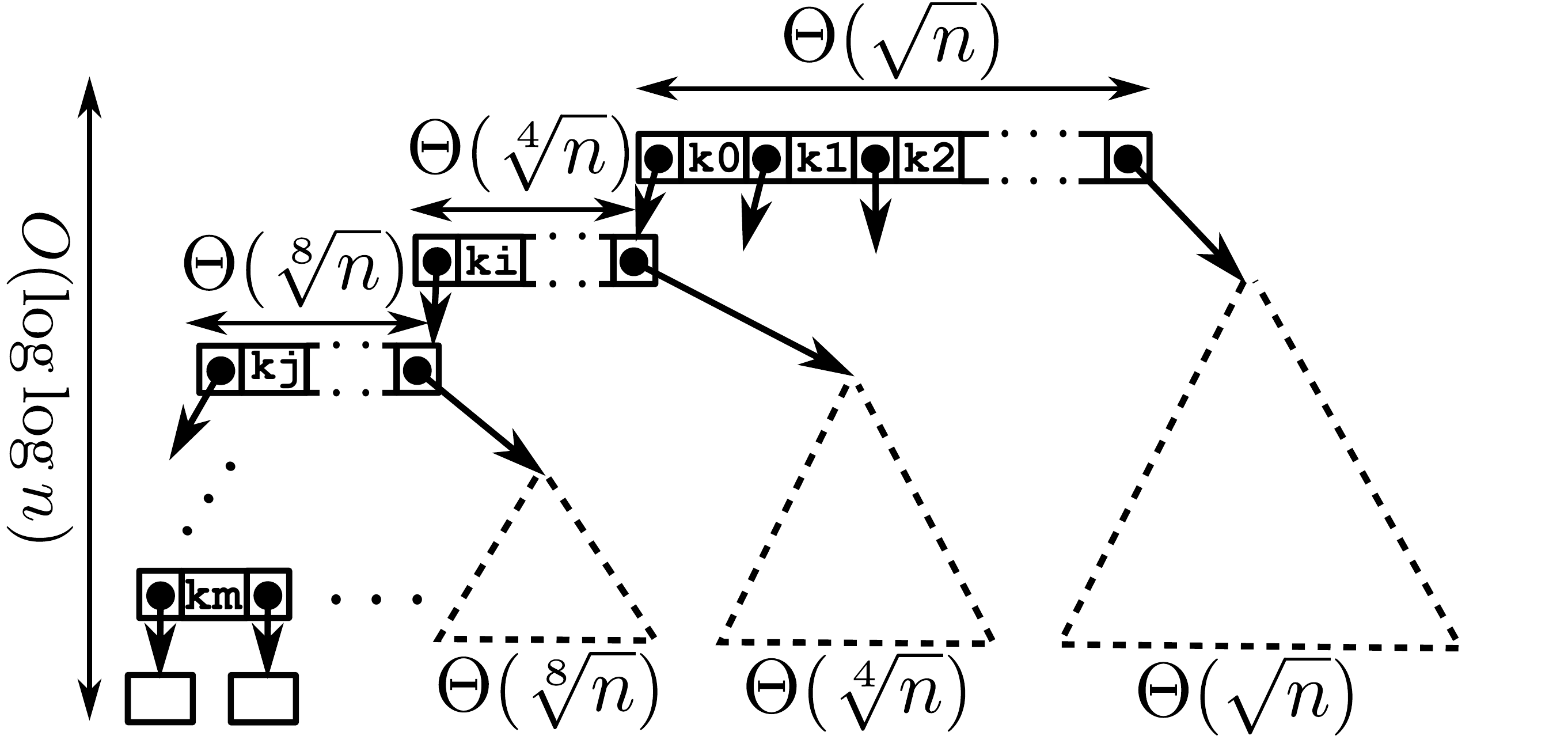}
\end{center}

An \emph{ideal C-IST} is a C-IST that is perfectly balanced.
In such a C-IST, the degree of a node with cardinality $n$
is either $\floor{\sqrt{n}}$ or $\ceil{\sqrt{n}}$,
and the number of keys in each of the node's child subtrees is
either $\floor{\sqrt{n}}$ or $\ceil{\sqrt{n}}$.
This property ensures that depth is bound by $O(\log \log n)$.

The interpolation search tree will generally not be ideal
after a sequence of insertion and deletion operations,
but its subtrees are ideal ISTs immediately after they get rebuilt.

\subsection{Insertion and Deletion}
\label{sec:insertion-deletion}

An insertion operation first searches the tree
to find a §Single§ or §Empty§ node that is in a location corresponding to the input key.
Then, the resulting node is replaced with either one or two new nodes.
An §Empty§ node gets replaced with a new §Single§ node that contains the new key.
A §Single§ node gets replaced either with a new §Single§ node
in case of finding an exact key match,
or otherwise with a new inner node that contains two single child nodes,
which contain the existing key and the newly added key.

To be able to decide when to rebalance a subtree,
C-IST must track the amount of imbalance in each subtree.
To achieve this,
whenever a key is inserted or deleted at some leaf,
the sequential IST increments the value of the §count§ field,
for all the inner nodes on the path from the root to the leaf that was affected~\cite{IST}.
Once the §count§ field reaches a threshold at some inner node,
the subtree below that inner node gets rebuilt.
The concurrent variant, C-IST, uses a similar technique to track the amount of imbalance,
but it avoids the contention at the root, where the contention is most relevant,
with a scalable quiescently-consistent multicounter~\cite{ABKLN}.

After rebalancing gets triggered in some subtree,
the subsequent modification operations in the corresponding subtree must fail,
and help complete that rebalancing operation before retrying themselves.
To ensure that this happens,
the rebalancing operation sets the §status§ field of all the inner nodes
in the subtree that is being rebuilt.
When replacing a child of the inner node,
a modification operation, such as insertion,
atomically checks the §status§ field of that inner node.
To achieve this, C-IST relies on an atomic double-compare-single-swap (§DCSS§) primitive,
which takes two addresses, two corresponding expected values, and one new value as arguments,
and behaves like a §CAS§, but succeeds
only if the second address also matches the expected value for that second address.
§DCSS§ is bundled with a wait-free §DCSS\_READ§ primitive,
which reads the fields that can be concurrently modified by a §DCSS§ operation.
This §DCSS\_READ§ operation is used by the C-IST operation whenever the value
of an inner node's children array needs to be read.

\begin{figure}[!t]
\begin{lstlisting}[firstnumber=1]
procedure insert(ist, key, val)
  path = [] °\hfill° // Stack that saves the path.
  n = ist.root
  while true
    index = interpolationSearch(key, n) °\label{lst:insert-interpolation-search}°
    child = DCSS_READ(n.children[index])
    if child is Inner
      n = child
      path.push( [child, index] )
    else if child is Empty | Single
      r = createFrom(child, key, val)
      result =
        DCSS(n.children[index], child, r, n.status, [0,°$\bot$°,°$\bot$°])°\label{lst:insert-dcss}°
      if result == FAILED_MAIN_ADDRESS
        continue °\hfill° // Retry from the same n.
      else if result == FAILED_AUX_ADDRESS
        return insert(ist, key, val) °\hfill°// Retry from the root.
      else
        for each [n, index] in path
          FETCH_AND_ADD(n.count, 1) °\label{lst:insert-fetch-and-add}°
        parent = ist.root
        for each [n, index] in path
          count = READ(n.count)
          if count >= REBUILD_THRESHOLD * n.initSize °\label{lst:insert-threshold}°
            rebuild(n, parent, index)
            break °\hfill°// exit for °\ap{Should we break as soon as you find the highest node for rebuild?}° °\ap{Can these 2 loops be fused into one? Seems like we can break as soon as we find a rebuild-point.}°
          parent = n
        return true
    else if child is Rebuild
      helpRebuild(child)
      return insert(ist, key, val) °\hfill° // Retry from the root.
\end{lstlisting}
\caption{Insert Operation}
\label{algo:insert}
\end{figure}

Figure~\ref{algo:insert} shows the pseudocode of the §insert§ operation.
The §delete§ operation is similar to §insert§.
The main difference is that it either replaces a §Single§ node with a new §Empty§ node,
or does not change the data structure if the key is not present.
The deletion does not shrink the §Inner§ nodes.
This allows for some §Empty§ nodes to be accumulated in the C-IST,
but the rebuilding operations eventually occur, and remove such empty nodes.

\subsection{Partial Rebuilding}
\label{sec:partial-rebuild}

When an insertion or deletion operation
observes that a subtree rooted at some inner node $target$
(henceforth, we call that subtree the ``target subtree'')
has become sufficiently imbalanced,
it triggers the rebuild operation on that subtree,
which is shown in Figure \ref{algo:rebuild}.
The rebuilding operation occurs in four steps.
First, a thread announces its intention to rebuild the target subtree
by creating a descriptor object of the type §Rebuild§,
and it inserts this descriptor between the $target$ and its $parent$ with a §DCSS§ operation.
After the first step, other threads can cooperate on the rebuild operation.
Second, this thread must prevent further modifications in the target subtree,
so it does a preorder traversal to set a bit in the §status§ field
of each node of that target subtree.
After the second step, no key can be added or removed from the target subtree.
Third, the thread creates an ideal IST (rooted at a new inner node called $ideal$)
using the old subtree's keys (rooted at the target subtree, i.e. $target$).
Finally, the thread uses a §DCSS§ operation to replace the target subtree
with the new ideal subtree in the parent.
All other threads can compete to complete second, third and the fourth step.

\begin{figure}[!ht]
\begin{lstlisting}
procedure rebuild(node, p, i)
  op = new Rebuild(node, p, i)
  result = DCSS(p.children[i], node, op, p.status, [0,°$\bot$°,°$\bot$°]) °\label{lst:rebuild-dcss}°
  if result == SUCCESS then helpRebuild(op)

procedure helpRebuild(op)
  keyCount = markAndCount(op.target) °\label{lst:help-rebuild-mark}°
  ideal = createIdeal(op.target, keyCount) °\label{lst:help-rebuild-create-ideal}°
  p = op.parent
  DCSS(p.children[op.index], op, ideal, p.status, [0,°$\bot$°,°$\bot$°])°\label{lst:help-rebuild-dcss}°

procedure markAndCount(node) °\label{lst:mark-declaration}°
  if node is Empty then return 0
  if node is Single then return 1
  if node is Rebuild then return markAndCount(op.target)
  // node is Inner
  CAS(node.status, [0,°$\bot$°,°$\bot$°], [0,°$\bot$°,°$\top$°]) °\label{lst:mark-cas-0-1}° °\label{lst:mark-end-of-preamble}°
  keyCount = 0 °\label{lst:mark-start-of-postamble}°
  for index in 0 until length(node.children) °\label{lst:mark-loop}°
    child = READ(node.children[index])
    if child is Inner then
      [count, finished, started] = READ(child.status)
      if finished then keyCount += count
      else keyCount += markAndCount(child) °\label{lst:mark-end-loop}°
    else if child is Single then keyCount += 1
  CAS(node.status, [0,°$\bot$°,°$\top$°], [keyCount,°$\top$°,°$\top$°]) °\label{lst:mark-cas-1-keycount}°
  return keyCount °\label{lst:mark-end}°
\end{lstlisting}
\caption{Rebuild Operation}
\label{algo:rebuild}
\end{figure}

\subsection{Collaborative Rebuilding}
\label{sec:concurrent-collaborative-rebuild}

When a lot of threads concurrently attempt to modify the C-IST,
the rebuilding procedure described in Section~\ref{sec:partial-rebuild}
suffers from a scalability bottleneck.
For example, since many threads compete to mark
the old subtree in the §markAndCount§ procedure,
multiple threads will typically try to set the §status§ bits of the same nodes.
Similarly, multiple threads attempt to create the ideal version of the same target subtree
in the §createIdeal§ procedure from Figure~\ref{algo:rebuild},
so part of the overall work gets duplicated.
To reduce contention, a collaborative rebuilding algorithm is used,
which allows threads to mark and rebuild different parts of the subtree in parallel.

\begin{figure}[t]

\begin{minipage}[t]{0.22\textwidth}
\begin{lstlisting}[numbers=none]
struct Rebuild: Node is
  target: Inner
  newTarget: Inner
  parent: Inner
  index: int
\end{lstlisting}
\end{minipage}
\begin{minipage}[t]{0.25\textwidth}
\begin{lstlisting}[numbers=none]
struct Inner: Node is
  initSize: int
  degree: int
  keys: KeyType[]
  children: Node[]
  status: [int, bool, bool]
  count: int
  nextMark: int
\end{lstlisting}
\end{minipage}
\caption{Modified Data Types for Collaborative Rebuilding}
\label{algo:data-types-cr}
\end{figure}

The previous rebuilding algorithm from Section~\ref{sec:partial-rebuild}
is changed in several ways to allow threads to perform rebuilding \textit{collaboratively}.
First of all, the §markAndCount§ procedure,
responsible for setting the §status§ bits in a subtree,
is replaced with a new §markAndCountCollaborative§ procedure.
In this new procedure, helper threads attempt to process different parts
of the target subtree in parallel.
They carefully avoid duplicating the work by picking a different initial index
for traversing the children of the current node: each thread atomically increments
a special field called §nextMark§ to decide on its index.
This is shown in Figure~\ref{algo:mark-and-count-collaborative},
in which we note the changes to the old §markAndCount§ procedure.

Second, the call to §createIdeal§,
inside the procedure §helpRebuild§
in Figure~\ref{algo:rebuild},
is replaced with a call to a new procedure §createIdealCollaborative§,
in which a new root of the subtree is first created
(whose §children§ array initially contains only §null§-pointers)
and then announced to the other threads, as before.
For this purpose, the §newTarget§ field is added to the §Rebuild§ data type,
as shown in Figure~\ref{algo:data-types-cr},
to store the root node of the new subtree.
Each §null§-pointer in the new root of the subtree represents a ``subtask''
that a thread can perform by building the corresponding subtree
(and then changing the corresponding §null§-pointer to point to this new subtree).
Note that many of these subtasks can be performed in parallel.
Until the new ideal IST is complete,
the §newTarget§ node effectively
acts as a lock-free work pool of tasks that the threads compete for.
Note that the afore-mentioned §nextMark§ field in the §Inner§ nodes is set to zero,
so that it can be used for collaborative marking during the next rebuild operation.

The subtlety of these changes is to distribute work
between the threads while at the same time retaining the lock-freedom property,
which mandates that all work is done after a finite number of steps,
even if some threads block.
The collaborative rebuilding algorithm is shown in Figure~\ref{algo:create-ideal-cr}.

\edef\lstMarkDeclaration{\getrefnumber{lst:mark-declaration}}
\edef\lstMarkEndOfPreamble{\getrefnumber{lst:mark-end-of-preamble}}
\edef\lstMarkStartOfPostamble{\getrefnumber{lst:mark-start-of-postamble}}

\begin{figure}[!t]
\begin{lstlisting}[
  firstnumber=\lstMarkDeclaration
]
procedure markAndCountCollaborative(node) °\Suppressnumber°
  // ... same as markAndCount until line °\textit{\ref{lst:mark-end-of-preamble}}°, but with
  // recursive calls to markAndCountCollaborative ... °\Reactivatenumber{\lstMarkEndOfPreamble + 1}°
  if node.degree > COLLABORATION_THRESHOLD °\label{lst:mark-cr-threshold}°
    while true °\label{lst:mark-cr-loop}°
      index = FETCH_AND_ADD(node.nextMark, 1)
      if index >= node.degree then break
      markAndCountCollaborative(node.children[index]) °\Suppressnumber° °\label{lst:mark-cr-end-loop}°
  // ... same as markAndCount from line °\textit{\ref{lst:mark-start-of-postamble}}°, but with
  // recursive calls to markAndCountCollaborative ...
\end{lstlisting}
\caption{The \texttt{markAndCountCollaborative} Procedure}
\label{algo:mark-and-count-collaborative}
\end{figure}

\edef\lstMarkEnd{\getrefnumber{lst:mark-end}}
\Reactivatenumber{\lstMarkEnd + 2}

\begin{figure}[!ht]
\begin{lstlisting}[
]
procedure createIdealCollaborative(op, keyCount)
  if keyCount < COLLABORATION_THRESHOLD then
    newTarget = createIdeal(op.target, keyCount) °\label{lst:create-ideal-cr-sequential}°
  else
    newTarget = new Inner( °\label{lst:create-ideal-cr-new-inner}°
      initSize = keyCount,
      degree = 0, // Will be set to final value in line °\textit{\ref{lst:create-ideal-cr-cas-degree}}°.
      keys = new KeyType[°$\lfloor\sqrt{\texttt{keyCount}}\rfloor$° - 1],
      children = new Node[°$\lfloor\sqrt{\texttt{keyCount}}\rfloor$°],
      status = [0, °$\bot$°, °$\bot$°], count = 0, nextMark = 0)
  if not CAS(op.newTarget, null, newTarget) then °\label{lst:create-ideal-cr-cas-new-target}°
    // Subtree root was inserted by another thread.
    newTarget = READ(op.newTarget) °\label{lst:create-ideal-cr-read-new-target}°
  if keyCount < COLLABORATION_THRESHOLD then
    while true °\label{lst:create-ideal-cr-loop}°
      index = READ(newTarget.degree)
      if index == length(newTarget.children) then break
      if CAS(newTarget.degree, index, index + 1) then °\label{lst:create-ideal-cr-cas-degree}°
        if not rebuildAndSetChild(op, keyCount, index)
          return newTarget °\label{lst:create-ideal-cr-end-loop}°
    for index in 0 until length(newTarget.children) °\label{lst:create-ideal-cr-for-loop}°
      child = READ(newTarget.children[index])
      if child == null then
        if not rebuildAndSetChild(op, keyCount, index)
          return newTarget °\label{lst:create-ideal-cr-end-for-loop}°
  return newTarget

procedure rebuildAndSetChild(op, keyCount, index)
  // Calculate the key interval for this child, and rebuild.
  totalChildren = °$\lfloor\sqrt{\texttt{keyCount}}\rfloor$°
  childSize = °$\lfloor$°keyCount / totalChildren°$\rfloor$°
  remainder = keyCount % totalChildren
  fromKey = childSize * index + min(index, remainder)
  childKeyCount = childSize + (index < remainder ? 1 : 0)
  child = createIdeal(op.target, fromKey, childKeyCount) °\label{lst:rebuild-and-set-create-ideal}°
  if index < length(op.newTarget.keys)
    key = findKeyAtIndex(op.target, fromKey)
    WRITE(op.newTarget.keys[index], key) °\label{lst:rebuild-and-set-write-key}°
  // Set new child, check if failed due to status change.
  result = DCSS(op.newTarget.children[index], °\label{lst:rebuild-and-set-dcss}°
    null, child, op.newTarget.status, [0, °$\bot$°, °$\bot$°])
  return result != FAILED_AUX_ADDRESS
\end{lstlisting}
\caption{The \texttt{createIdealCollaborative} Procedure}
\label{algo:create-ideal-cr}
\end{figure}

\subsection{Lookup}
\label{sec:lookups-and-queries}

The §lookup§ subroutine is shown in Figure~\ref{algo:lookup}.
Similar to §insert§, it starts with an interpolation search,
and repeats it at each inner node until reaching an §Empty§ or a §Single§ node.
If it reaches a §Single§ node that contains the specified key, it returns §true§.
Otherwise, if §lookup§ encounters an §Empty§ node or
a §Single§ node that does \emph{not} contain the specified key, it returns §false§.

If §lookup§ encounters a §Rebuild§ object,
it can safely ignore the rebuild operation --
lookup simply follows the §target§ pointer
to move to the next node, and continues traversal.
Unlike the §insert§ operation,
§lookup§ does not need to help concurrent subtree-rebuilding operations.
Lookups do not need to help to ensure progress,
and so they avoid the unnecessary slowdown.
Apart from the use of the §DCSS\_READ§ operation,
and checking for the §Rebuild§ objects,
\textit{lookup} effectively behaves in the same way
as the sequential interpolation tree search.

\begin{figure}[!t]
\begin{lstlisting}
procedure lookup(ist, key)
  n = ist.root
  while true
    if n is Inner then
      index = interpolationSearch(key, n)
      n = DCSS_READ(n.children[index]) °\label{lst:lookup-dcss-read}°
    else if n is Single then return n.k == key ? n.v : null
    else if n is Empty then return null
    else if n is Rebuild then n = n.target
\end{lstlisting}
\caption{Lookup Operation}
\label{algo:lookup}
\end{figure}

\subsection{Summary}

We presented the C-IST data types and operations precisely,
but without lengthy explanations of the pseudocode.
For a more descriptive discussion and concrete example scenarios,
we refer the readers to related work~\cite{c-ist-ppopp2020}.


\newpage
\section{Correctness}
\label{sec:correctness}

In this section, we formalize the concurrent IST,
and prove that the operations are safe,
linearizable and non-blocking.
We begin by introducing the concept of an abstract set,
and then define the correspondence
between the abstract set and the concurrent IST.

\begin{definition}[Abstract set]
  An \emph{abstract set} $\mathbb{A}$
  is a mapping $\mathbb{A} : \mathbb{K} \rightarrow \{ \top, \bot \}$,
  where $\mathbb{K}$ is the set of all keys,
  and which is true ($\top$)
  for keys that are present in the set.
  The abstract set operations are:
  $\lookup(\mathbb{A}, k) = \top \Leftrightarrow k \in \mathbb{A}$,
  $\insertion(\mathbb{A}, k) = \{ k' : k' \in \mathbb{A} \vee k' = k \}$,
  and
  $\deletion(\mathbb{A}, k) = \{ k' : k' \in \mathbb{A} \wedge k' \neq k \}$.
\end{definition}

\begin{definition}[Validity]
  An IST §I§ is \emph{valid} if and only if the Invariants
  \ref{inv:presence},
  \ref{inv:search} and \ref{inv:acyclicity} hold.
\end{definition}

\begin{definition}[Consistency]
  An IST §I§ is \emph{consistent} with the abstract set $\mathbb{A}$
  if and only if $\forall k \in \mathbb{A} \Leftrightarrow \haskey(§I§, k)$,
  (as per Definition \ref{def:has-key}).
  The IST §lookup§ operation is consistent with an abstract set $\lookup$
  if and only if $\forall k, §lookup(I,k)§ = \lookup(\mathbb{A}, k)$.
  The IST §insert§ and §delete§ operations are consistent
  with the abstract set $\insertion$ and $\deletion$
  if and only if for all keys $k$ and initial states §I§
  consistent with some abstract set $\mathbb{A}$,
  they leave the IST in a state §I§',
  consistent with some abstract set $\mathbb{A}'$,
  such that $\insertion(\mathbb{A}, k) = \mathbb{A}'$
  or $\deletion(\mathbb{A}, k) = \mathbb{A}'$, respectively.
\end{definition}

\begin{theorem}[Safety]
  \label{thm:safety}
  An IST §I§ that uses non-collaborative rebuilding from Figure~\ref{algo:rebuild}
  is always valid and consistent with some abstract set $\mathbb{A}$.
  IST operations are consistent with the abstract set semantics.
\end{theorem}

We prove safety inductively -- it is easy to see that the newly created IST
is consistent with the empty abstract set.
For the inductive case, we must show that after each mutation,
the IST remains valid, and that it either remains consistent
with the same abstract set $\mathbb{A}$,
or becomes consistent with some other set $\mathbb{A}'$,
according to the abstract set semantics.
After that, we extend the proof to the IST variant
that uses collaborative rebuilding.

\begin{lemma}[Freezing]
  \label{lem:freeze}
  After §markAndCount§
  in line \ref{lst:help-rebuild-mark} of Figure \ref{algo:rebuild} completes,
  all the nodes in the corresponding subtree have §status§
  different than $[0, \bot, \bot]$.
\end{lemma}

\begin{proof}
  Consider the §CAS§ in line \ref{lst:mark-cas-0-1} of Figure \ref{algo:rebuild}.
  Exactly one thread will successfully execute this §CAS§.
  After this happens, none of the entries in the §children§ array will be modified,
  since all the §DCSS§ invocations are conditional on §status§ being zero.
  Assuming that the statement holds inductively for the children,
  when a thread reaches the §CAS§ in line \ref{lst:mark-cas-1-keycount},
  all the inner nodes in the corresponding subtree have a non-zero §status§.
\end{proof}

\begin{lemma}[End of Life]
  \label{lem:end-of-life}
  If the field §status§ of an inner node §n§ is $[0, \bot, \bot]$ at $t_0$,
  then the node §n§ is reachable from the root at some time $t_0$.
\end{lemma}

\begin{proof}
  Assume the converse -- node §n§ is unreachable at $t_0$
  and it has §status§ set to $[0, \bot, \bot]$.
  The only instruction that makes inner nodes unreachable
  is the §DCSS§ in line \ref{lst:help-rebuild-dcss} of Figure \ref{algo:rebuild},
  which is executed on §n§'s parent.
  This instruction is executed after
  §markAndCount§ in line \ref{lst:help-rebuild-mark} completes,
  so the inner nodes in the corresponding subtree
  must have a §status§ value different than $[0, \bot, \bot]$, by Lemma \ref{lem:freeze}.
\end{proof}

\begin{lemma}[Cover stability]
  \label{lem:cover-stability}
  If a node §n§ covers the interval $[ a, b \rangle$ at some time $t_0$
  (as per Definition \ref{def:cover}),
  then §n§ covers the same interval $\forall t > t_0$,
  as long as §n§ is reachable at the time $t$.
\end{lemma}

\begin{proof}
  We prove this inductively.
  For the base case, the statement trivially holds for the root.
  Next, assume that it holds for the parent of some node §n§.
  If §n§ is reachable at some time $t$,
  then so is its parent, whose cover did not change.
  Since the cover of §n§ is defined based on the cover of its parent
  and the contents of the immutable §keys§ array
  (as specified in Definition \ref{def:cover}),
  the cover of §n§ also does not change after its creation.
\end{proof}

\begin{proof}[Proof of Theorem \ref{thm:safety}]
  We consider all execution steps that mutate the data structure,
  and reason about the respective change in the corresponding abstract set.

  Consider the §DCSS§ in line \ref{lst:insert-dcss} of Figure \ref{algo:insert}.
  If this §DCSS§ succeeds at $t_0$,
  then the §status§ field of the corresponding §node§ must be $[0, \bot, \bot]$,
  and the node is reachable at $t_0$, by Lemma \ref{lem:end-of-life}.
  Before $t_0$, the tree is inductively valid and consistent with some set $\mathbb{A}$.
  After $t_0$, the new node (created in the §createFrom§ subroutine)
  becomes reachable from the root.
  The new node contains the new key,
  and (if §child§ was non-§Empty§, and its key different than the new key)
  the old key from §child§.
  Therefore, this change is consistent with the abstract set semantics.
  Furthermore, Invariants \ref{inv:presence},
  and \ref{inv:acyclicity}
  are trivially preserved.
  To see that Invariant \ref{inv:search} holds,
  note that all the nodes on the path from the root to §node§ are reachable,
  since §node§ is reachable by Lemma \ref{lem:end-of-life}.
  Each of those nodes covers exactly the same interval
  that it covered when §interpolationSearch§ ran
  in line~\ref{lst:insert-interpolation-search} of Figure~\ref{algo:insert},
  by Lemma~\ref{lem:cover-stability}.
  The node that was created by the §createFrom§ subroutine
  also respects Invariant~\ref{inv:search}.
  Therefore, the new key is inside the intervals covered by those nodes,
  and the search tree property is preserved.

  Next, consider the §DCSS§ in line \ref{lst:help-rebuild-dcss} of Figure \ref{algo:rebuild}.
  After the §markAndCount§ call in line \ref{lst:help-rebuild-mark} completes,
  that §DCSS§ cannot succeed, by Lemma~\ref{lem:freeze}.
  Thus, there will be no further updates of the §children§ arrays in that subtree,
  so the corresponding subtree is effectively immutable.
  All §createIdeal§ calls occur after the subtree becomes immutable,
  and each such call will create a new subtree
  that corresponds to the same abstract set as the old subtree,
  so the §DCSS§ in line \ref{lst:help-rebuild-dcss}
  does not change the consistency or the validity of the IST.

  It is easy to see that
  the §DCSS§ in line \ref{lst:rebuild-dcss} of Figure \ref{algo:rebuild}
  changes neither the validity nor the consistency of an IST,
  since it only inserts up to one §Rebuild§ node between some two §Inner§ nodes.

  Finally, neither the §CAS§ instructions
  in lines \ref{lst:mark-cas-0-1} and \ref{lst:mark-cas-1-keycount}
  of Figure \ref{algo:rebuild},
  nor the §FETCH\_AND\_ADD§ in line \ref{lst:insert-fetch-and-add}
  of Figure \ref{algo:insert},
  affect validity or consistency of the IST,
  since these properties are independent
  of the §status§ and the §count§ fields.
\end{proof}

Next, we need to show safety for the collaborative IST variant.
The crux of the proof is to show that the procedure §markAndCountCollaborative§
and the procedure §createIdealCollaborative§ have the same semantics
as their non-collaborative variants.

\begin{lemma}[Collaborative Freezing]
  \label{lem:collaborative-marking}
  After the procedure §markAndCountCollaborative§
  from Figure~\ref{algo:mark-and-count-collaborative} completes,
  all the nodes in the corresponding subtree
  have §status§ different than $[0, \bot, \bot]$.
\end{lemma}

\begin{proof}
  After lines~\ref{lst:mark-cr-threshold}-\ref{lst:mark-cr-end-loop}
  of §markAndCountCollaborative§ complete,
  some subset of the nodes below the target §node§
  will have their §status§ different than $[0, \bot, \bot]$,
  due the CASes in lines~\ref{lst:mark-cas-0-1} and \ref{lst:mark-cas-1-keycount}.
  This is not necessarily for all of the nodes below the target §node§,
  since some threads might still be processing their subtrees
  at the time when the current thread exits the loop
  in lines~\ref{lst:mark-cr-threshold}-\ref{lst:mark-cr-end-loop}.
  Nevertheless, the loop from Figure~\ref{algo:rebuild}
  in lines~\ref{lst:mark-loop}-\ref{lst:mark-end-loop},
  which also appears in §markAndCountCollaborative§
  (after the loop in lines~\ref{lst:mark-cr-threshold}-\ref{lst:mark-cr-end-loop}),
  ensures that each descendant node has their §status§
  set to a non-$[0, \bot, \bot]$ value before the procedure ends,
  so the proof is similar to that of Lemma~\ref{lem:freeze}.
\end{proof}

\begin{corollary}[Collaborative End of Life]
  \label{cor:collaborative-end-of-life}
  Lemma~\ref{lem:end-of-life} holds when
  the §markAndCount§ procedure is replaced with
  the call to §markAndCountCollaborative§.
\end{corollary}

\begin{lemma}[Collaborative Ideal Tree Creation]
  \label{lem:collaborative-tree-creation}
  Let §T§ be a subtree below the §target§ node,
  which corresponds to some abstract set $\mathbb{S}$,
  and on which §markAndCountCollaborative§ completed.
  Once the §createIdealCollaborative§ completes,
  either the tree §T§ gets replaced with a new ideal tree §T'§
  that corresponds to the same abstract set $\mathbb{S}$,
  or the status of the §target§ node changed to a non-$[0, \bot, \bot]$ value.
\end{lemma}

\begin{proof}
  First, not that the tree §T§ is effectively-immutable
  by Lemma~\ref{lem:collaborative-marking} and
  Corollary~\ref{cor:collaborative-end-of-life},
  so the entries in its §children§ arrays never change.

  The case in which the §keyCount§ is below the threshold creates 
  an already-complete ideal tree §newTarget§.
  In the other case, local variable §newTarget§ is set to an empty root node
  for the new ideal tree.
  If the CAS in line~\ref{lst:create-ideal-cr-cas-new-target} succeeds,
  then §op.newTarget§ points to this new root.
  Conversely, if that CAS fails, then §op.newTarget§
  points to a new root created by another thread,
  which, due to the effectively-immutable §T§ based on which it was created,
  must be structurally identical to the node in the local variable §newTarget§
  (but is a different memory object in the heap).
  Thus, all threads that move beyond the line~\ref{lst:create-ideal-cr-read-new-target},
  have the pointer to the same root node in their §newTarget§ variable,
  which corresponds to the pointer in the §op.newTarget§ field.

  Next, consider the case in which the §status§ field of the §op.target§ node
  never changed to a non-$[0, \bot, \bot]$ value.
  In this case, the §DCSS§ in line~\ref{lst:rebuild-and-set-dcss}
  of the §rebuildAndSetChild§ procedure can never fail
  due to a change in the auxiliary address.
  Therefore, in this case, §rebuildAndSetChild§ always returns §true§,
  so the §createIdealCollaborative§ cannot return early
  in lines~\ref{lst:create-ideal-cr-end-loop} or \ref{lst:create-ideal-cr-end-for-loop},
  and both loops must finish before §createIdealCollaborative§ returns.

  Consider the loop in lines~\ref{lst:create-ideal-cr-loop}-\ref{lst:create-ideal-cr-end-loop}.
  After any thread exits this loop,
  an entry $i$ of the §newTarget.children§ array
  contains either §null§,
  or the $i$-th child of the ideal tree corresponding to $\mathbb{S}$.
  Moreover, after any thread exits this loop, it must be true that
  §newTarget.degree§ is equal to the length of the §newTarget.children§ array.

  Next, consider the loop
  in lines~\ref{lst:create-ideal-cr-for-loop}-\ref{lst:create-ideal-cr-end-for-loop}.
  After any thread exits this loop,
  it must be true that every entry $i$ of the §newTarget.children§ array
  contains the $i$-th child of the ideal tree corresponding to $\mathbb{S}$.
  Moreover, $i$-th entry of the §newTarget.keys§ array
  contains the $i$-th key of the ideal tree corresponding to $\mathbb{S}$
  (due to the write in line~\ref{lst:rebuild-and-set-write-key}).

  Therefore, if the §status§ field of the §op.target§ node
  did not change to a non-$[0, \bot, \bot]$ value,
  then the procedure §createIdealCollaborative§ returns
  an ideal tree corresponding to $\mathbb{S}$.

  Finally, consider the case in which the §status§ field of the §op.target§ node
  did change to a non-$[0, \bot, \bot]$ value.
  Here, the claim trivially holds.
\end{proof}

\begin{theorem}[Collaborative Safety]
  \label{thm:collaborative-safety}
  An IST §I§ that uses collaborative rebuilding from
  Figures~\ref{algo:mark-and-count-collaborative} and \ref{algo:create-ideal-cr}
  is always valid and consistent with some abstract set $\mathbb{A}$.
  IST operations are consistent with the abstract set semantics.
\end{theorem}

\begin{proof}
  Similar to the proof for Theorem~\ref{thm:collaborative-safety},
  using Lemma~\ref{lem:collaborative-marking}
  and Lemma~\ref{lem:collaborative-tree-creation}
  to show that the §DCSS§ in line~\ref{lst:help-rebuild-dcss}
  does not cause a change in the abstract set.

  In particular, if the §createIdealCollaborative§ returns prematurely
  due to a non-$[0, \bot, \bot]$ value in the §status§ field,
  then the §DCSS§ in line~\ref{lst:rebuild-dcss} in Figure~\ref{algo:rebuild}
  will fail and cause a restart from the root,
  due to a rebuild operation in one of the ancestors.
\end{proof}

Linearizability follows naturally from the safety proofs,
since all the state-changing execution steps occur at the atomic operations.

\begin{corollary}[Linearizability]
  Lookup, insertion and deletion operations are linearizable,
  in both the non-collaborative and collaborative variant
  of the concurrent IST data structure.
\end{corollary}

\begin{proof}
  For insertion and deletion,
  this follows immediately from the proofs of Theorems~\ref{thm:safety}
  and \ref{thm:collaborative-safety},
  since the §DCSS§ in line \ref{lst:insert-dcss} is the linearization point.
  The linearization point of a lookup is
  the last §DCSS\_READ§ in line \ref{lst:lookup-dcss-read}
  of Figure \ref{algo:lookup}
  that was done on a node whose §status§ field was $[0, \bot, \bot]$.
\end{proof}

We first state the non-blocking properties of the non-collaborative IST variant.
After that, we extend the proof to the collaborative variant.

\begin{theorem}[Non-Blocking]
  \label{thm:non-blocking}
  In the non-collaborative IST variant,
  insertion and deletion are lock-free,
  and lookup is wait-free.
\end{theorem}

\begin{proof}
  By Invariant \ref{inv:acyclicity}, every path in the IST is finite.
  The §lookup§ subroutine in Figure \ref{algo:lookup}
  executes a finite number of steps in each loop iteration
  before advancing to the next node on the path.
  At any point, there can be up to $p$ insertion threads
  that are racing to extend this path.
  The §lookup§ operation can potentially race
  with these inserting threads
  (for example, if they are all repetitively extending the same leaf node).
  However, after a finite number of steps,
  each of these $p$ threads will increment the §count§ field beyond a threshold
  and trigger the rebuild of the enclosing subtree.
  At that point, the path can no longer be extended,
  and the §lookup§ observes a §Rebuild§ node,
  and terminates in a finite number of steps.
  Therefore, §lookup§ is wait-free.

  For the insertion and deletion,
  identify all the instructions that do not change the abstract state $\mathbb{A}$,
  but can fail an instruction that is a linearization point.
  We call these instructions \emph{housekeeping}.
  The §DCSS§ in line \ref{lst:rebuild-dcss} of Figure \ref{algo:rebuild}
  and the §CAS§ in line \ref{lst:mark-cas-0-1} of Figure \ref{algo:rebuild},
  are the only such housekeeping instructions.
  These instructions can be executed only finitely many times
  before some instruction changes $\mathbb{A}$.

  To see this, define the \emph{pending count} of an inner node §n§
  as the value of its field §count§ increased by the number of threads $p$
  that are currently preparing to increment §n.count§
  (note: each such thread can increment §n.count§ at most once
  without changing $\mathbb{A}$).
  Next, identify all nodes whose pending count crossed the threshold for rebuilding,
  and call them \emph{triggered}.
  Define an abstract variable $\mathbb{V}_0$
  as the triggered nodes that are directly
  pointed to by the §children§ array.
  Define $\mathbb{V}_1$
  as the total number of inner nodes in the subtrees of all triggered nodes
  whose §status§ field is $0$.
  Note that $\mathbb{V} = \mathbb{V}_0 + \mathbb{V}_1$ is finite.
  Furthermore, as long as $\mathbb{A}$ stays the same,
  housekeeping instructions can only strictly decrease $\mathbb{V}$.
  When $\mathbb{V}$ is $0$,
  neither the §DCSS§ in line \ref{lst:rebuild-dcss} of Figure \ref{algo:rebuild}
  nor the §CAS§ in line \ref{lst:mark-cas-0-1} of Figure \ref{algo:rebuild},
  can succeed, so they cannot fail an instruction
  that would cause a linearization point.
  To complete the proof, note that two invocations of the
  §DCSS§ in line \ref{lst:insert-dcss} of Figure \ref{algo:insert}
  are distanced by a finite number of steps,
  by the same argument as for the §READ§s in the §lookup§ subroutine.
\end{proof}

In the collaborative variant, the §lookup§ operation is wait-free by the same arguments.
To prove that insertion and deletion are also non-blocking,
we need to show that the §markAndCountCollaborative§ procedure,
as well as the §createIdealCollaborative§ procedure,
both complete in a finite number of steps.

\begin{lemma}
  \label{lem:wait-free-collaborative-rebuild}
  The §markAndCountCollaborative§ procedure
  and the §createIdealCollaborative§ procedure are wait-free.
\end{lemma}

\begin{proof}
  First, it is easy to see that helper procedures, such as §createIdeal§,
  finish in a finite number of steps (because the tree is finite).

  Next, note that the loops
  in lines~\ref{lst:mark-loop}-\ref{lst:mark-end-loop} of Figure~\ref{algo:rebuild},
  and the lines~\ref{lst:create-ideal-cr-for-loop}-\ref{lst:create-ideal-cr-end-for-loop}
  of Figure~\ref{algo:create-ideal-cr},
  both make progress in each iteration, and finish within a finite number of execution steps.

  The same is true of the loop in lines~\ref{lst:mark-cr-loop}-\ref{lst:mark-cr-end-loop}
  of Figure~\ref{algo:mark-and-count-collaborative} --
  after every iteration of the loop, the §nextMark§ field is strictly increased.
  Thus, §markAndCountCollaborative§ is wait-free.

  The loop in lines~\ref{lst:create-ideal-cr-loop}-\ref{lst:mark-cr-end-loop}
  of Figure~\ref{algo:create-ideal-cr}
  contains a §CAS§ on the §degree§ field that makes progress when successful.
  However, that §CAS§ fails only if another thread successfully executes the same §CAS§
  on the §degree§ field, incrementing it by $1$.
  Therefore, an iteration of that loop makes progress in both cases,
  and completes in a finite number of iterations.
  This proves that the §createIdealCollaborative§ procedure is wait-free.
\end{proof}

\begin{corollary}
  \label{cor:non-blocking-collaborative}
  In the collaborative IST variant,
  insertion and deletion are lock-free,
  and lookup is wait-free.
\end{corollary}

\begin{proof}
  Follows from Theorem~\ref{thm:non-blocking}
  and Lemma~\ref{lem:wait-free-collaborative-rebuild}.
\end{proof}


\section{Complexity Analysis}
\label{sec:complexity-analysis}

The complexity proof for C-IST follows the argument for sequential ISTs~\cite{IST},
with a few modifications -- most notably, it accounts for the fact that
at any time there can be up to $p$ threads that are concurrently modifying the C-IST.

\begin{lemma}
\label{lem:worst-case-depth}
Let $p$ be the number of concurrent threads that are modifying a C-IST.
Worst-case depth of a C-IST that contains $n$ keys is $O(p + \log n)$.
\end{lemma}

\begin{proof}
Let §u§ be some node and §v§ be its parent.
In the worst case, all keys that were updated below the node §v§,
were also below its child §u§.
In our case, the rebuild threshold $R = 1 / 4$,
so there were at most $v.initSize / 4 + p_v$ such updates,
where $v.initSize$ is the initial size of the node §v§,
and $p_v$ is the number of concurrent threads
that are currently modifying the subtree below §v§.
We add $p_v$ because those threads possibly inserted the new keys,
but did not yet update the counts on the path to the root.

Since §u§ was an ideal tree at the time when its parent §v§ was created,
we have that:

\begin{equation}
\label{eq:child-size-bound}
|u| \leq \sqrt{v.initSize} + v.initSize / 4 + p_v \leq v.initSize / 2 + p_v
\end{equation}

Next, consider the longest path $L$ from the root §r§ to some leaf.
On this path, the total number of concurrent updates
that have not yet updated the §count§ fields is $p_L \leq \sum_v p_v = p$.
The new nodes inserted by these $p_L \leq p$ updates, which we denote by $w_1, w_2, \ldots, w_{p_L}$ must be at the suffix of the path,
that is:

\begin{equation}
\label{eq:longest-path}
L =
r \rightarrow v_1 \rightarrow \ldots \rightarrow v_{d_0} \rightarrow
w_{1} \rightarrow \ldots \rightarrow w_{p_L}
\end{equation}

Otherwise, if the nodes represented by those updates were moved elsewhere,
then a rebuild must have been triggered,
meaning that the §count§ fields \emph{had already been updated},
which is a contradiction with how we defined $w_{1}, \ldots, w_{p_L}$.

Therefore, these $p_L \leq p$ updates can only increase the depth by $p_L$.
Let us ignore them in the size consideration momentarily.
For the prefix $r \rightarrow v_1 \rightarrow \ldots \rightarrow v_{d_0}$ of the path $L$,
we have:

\begin{equation}
\label{eq:root-size-bound}
r.initSize \geq 2 \cdot |v_1| \geq \ldots \geq 2^{d_0} \cdot 1
\end{equation}

Furthermore, $r.initSize \leq |r| \cdot (1 - R)^{-1}$
so the depth $d$ of the path $L$ is:

\begin{equation}
\label{eq:init-size-depth}
d = p_L + d_0 \leq p + \log_2(r.initSize) \leq O(p + \log |r|)
\end{equation}

which proves the claim of the lemma.
\end{proof}

\begin{lemma}
\label{lem:amortized-cost}
The worst-case amortized cost of insert and delete operations,
without including the cost of searching for the node in the C-IST,
is $O(\gamma (p + \log n))$, where $\gamma$ is a bound on average interval contention.
\end{lemma}

\begin{proof}
Let us consider the execution at a node $v$ between two rebuilding phases. 
We can split the work performed at the node itself and by operations which traverse it into three disjoint categories:
\begin{itemize}
	\item Work performed to rebuild the node.
	\item Work by operations which traverse the node $v$ but \emph{fail} in line 17 and restart from the root. 
	\item Work performed by \emph{successful} operations, which do not restart from the root. 
\end{itemize}

First, note that the amortized cost of a single successful operation is no worse than the worst-case length of a root-to-leaf path, which is $O(p + log n)$, 
by Lemma~\ref{lem:worst-case-depth}. 
Second, notice that, by standard non-blocking amortization argument, a successful operation can be on average responsible for at most $\gamma$ distinct \emph{failed} operations, each of which has worst-case cost $O( p + \log n )$. 
Third, notice that the work necessary to rebuild the subtree rooted at node $v$ is in the order of $nodes(v) \leq (1 + R) \cdot v.initSize + p$. 
However, this rebuild process is triggered only when at least $R \cdot v.initSize$ operations have \emph{succeeded} in the subtree, 
as only successful operations increment the counter and therefore trigger a rebuild. 

We therefore obtain that the total amortized cost of a successful operation is upper bounded by 
\begin{align*}
O( \gamma ( p + \log n ) + & ((1 + R) \cdot v.initSize + p) / R \cdot v.initSize)  \\
= & O (\gamma ( p + \log n ) ).
\end{align*}

\end{proof}

To analyze the expected amortized cost of operations, 
we introduce a concept of a $\mu$-random C-IST, identically to the sequential variant. 

\begin{definition}
\label{def:mu-random}
Let $\mu$ be a probability density function on reals.
A $\mu$-random C-IST of size $n$ is generated as follows:

\begin{enumerate}
\item
Independently draw $n_s$ real keys according to the probability density $\mu$,
and use them to create an ideal IST.
\item
Do a sequence of updates $U_1, \ldots, U_m$,
with $i$ insertions and $d$ deletions,
so that $m = i + d$ and $n = n_s + i - d$.
An insertion picks a random key according to the probability density $\mu$.
A deletion picks a random key that is present in the C-IST,
according to a uniform probability density.
\end{enumerate}
\end{definition}

We will now establish some bounds on the expected size of subtrees in C-IST.
We will track two components of the C-IST size --
the size contribution due to the updates that are already completed,
and the size contribution due to slow updates due to straggler threads
that have inserted a node, but have not yet incremented the §count§ fields,
nor checked whether a rebuild is required. 
We will assume a standard non-malicious \emph{oblivious} shared-memory scheduler, 
whose actions are independent of the input distribution and of the threads random coin flips. 

\begin{definition}
Let the \emph{initial size} of a node denote the node count of its corresponding subtree at the point where it was last rebuilt. Let the \emph{stable size} denote the number of keys that were inserted or deleted,
and for which the §count§ fields of the nodes on the respective paths were updated, up to the current node, at the current moment in time. 
\end{definition}

We show that,
in a $\mu$-random C-IST, as long as nodes have size $\Omega(p)$, the stable size distributes evenly over the subtrees.

\begin{lemma}
\label{lem:square-root-child}
Let $\mu$ be a density function with finite support $[a, b]$,
let §v§ be the root of a $\mu$-random C-IST of initial size $n_0$, such that $n_0 R > p$, of stable size $n$ at the current point. 
Let §u§ be a child of the node §v§.
Then, the stable size of the subtree rooted at §u§ will be $O(n^{1/2})$, at any future point in time until the next rebuild of node §u§
with probability  $\geq 1 - O(n^{-1/2})$.
\end{lemma}

\begin{proof}
We start by noticing that, by the rebuild conditions, and since there can be at most $p - 1$ pending operations at any given point before the next rebuild, we have that $n \leq (1 + R) n_0 + p \leq (1 + 2R)n_0$. 
Also, we know that there could have been at most $i \leq R n_0$ inserts into $v$ since it was last rebuilt. 

At the time when §v§ was rebuilt, its child §u§ had the size $\sqrt{n_0} \pm 1$.
Let $X$ be a random variable that denotes the number of elements that were stored in §u§ since that time.
Using the result from Theorem A from sequential ISTs~\cite{IST},
the expectation and the variance of $X$ can be bounded as follows:

\begin{equation}
\label{eq:expectation-inserts}
E(X) = \sqrt{n_0} \cdot \frac{i}{n_0 + 1}
\leq R \cdot \sqrt{n_0}.
\end{equation}
\begin{equation}
\label{eq:variance-inserts}
Var(X) \leq E(X) \cdot (1 + \frac{i}{n_0 + 1}) \leq 2 \cdot R \cdot \sqrt{n_0}.
\end{equation}

We can therefore apply Chebyshev's inequality $P(|X - E(X)| \geq t) \leq \frac{Var(x)}{t^2}$  
for $t = (1 + R) n_0^{1/2}$,
to bound the probability that $X$ exceeds the square root size
by a factor of $1 + R$:

\begin{equation}
\label{eq:probability-larger-than-square-root}
P(X \geq (1 + R) \cdot n_0^{1/2}) \leq \frac{2R}{(1 + R)^2} \cdot n_0^{-1/2}.
\end{equation}

Thus, the claim about the stable size follows.

\end{proof}

\begin{lemma}
\label{lem:amortized-expected-log-log}
Let $\mu$ be a probability density with a finite support $[a, b]$.
The expected total cost of processing
a sequence of $n$ $\mu$-random insertions and uniformly random deletions
into an initially empty C-IST is $O(n(\log \log n + p) \gamma)$, 
where $\gamma$ is a bound on average interval contention. 
\end{lemma}

\begin{proof}
We proceed by defining the following token generation scheme for \emph{successful} operations:
we say that a traversal corresponding to an insert or remove operation generates a token at a node $v$ if it passes the node on its walk from the root to a leaf. 
(Notice that this also counts the number of fetch-and-increment operations the walk will generate on its backward path.) 
We can split such generated tokens into \emph{internal tokens}, which are generated at internal nodes, and \emph{leaf tokens}, which are generated at leaves.  
Notice that the only way in which a walk can generate multiple leaf tokens is if it encounters additional concurrent operations. 

We proceed to first bound the expected number of internal tokens generated by a successful root-to-leaf walk. If we denote this number of additional internal tokens generated by the successful update when starting from a node of stable size $m$ by $T(m)$, we have that, by Lemma~\ref{lem:square-root-child} and Lemma~\ref{lem:worst-case-depth}, 

\begin{equation}
T(m) \leq 1 + T(O(m^{1/2})) + O(m^{-1/2}) \cdot O(\log m),
\end{equation}

\noindent where the recursion has to stop at nodes of maximum initial size $p$. However, the number of tokens generated  nodes below this size is at most $p$, and therefore, by the above recursion, we obtain that the expected number of generated internal tokens by a traversal can be at most $O( p + \log \log n )$. 
We are therefore left with the task of bounding the number of \emph{leaf tokens} generated by the successful walk. 
Since the operations are non-blocking, the average number of such tokens per walk is $O( p )$, which yields a total average token bound of $O( p + \log \log n)$. 
Similarly to Lemma~\ref{lem:amortized-cost}, we notice that each token corresponds to a constant amount of work by some operation. 
Therefore, the amortized cost of successful operations is $O( p + \log \log n )$. 

Our argument so far has only taken into account successful traversals, which do not restart from the root. If we simply amortize each failed traversal against the successful one which caused it to fail, we obtain that each successful traversal $i$ can be charged for $O( \gamma_i (p + \log \log n))$ tokens, where $\gamma_i$ is its interval contention. 
Therefore, we get that the amortized expected cost of an operation is $O( \gamma (p + \log \log n))$, as claimed. 
 
We note that stronger bounds on amortized expected cost could potentially be obtained by attempting to leverage the structure of the distribution to bound the probability of a collision. However, this is not immediate, since the smoothness property does not imply a small upper bound on the probability of a key.
\end{proof}

\begin{lemma}
\label{lem:constant-time-interpolation-search}
Let $\mu$ be a smooth probability density,
as defined by Mehlhorn and Tsakalidis~\cite{IST},
for a parameter $\alpha$, such that $\frac{1}{2} \leq \alpha < 1$,
and let §v§ be the root of a $\mu$-random C-IST with parameter $\alpha$.
Then, the expected search time in the §v.children§ is $O(1)$.
\end{lemma}

\begin{proof}
Since the §children§ array is immutable,
the proof is precisely identical to the one in the related work
on sequential ISTs~\cite{IST}.
\end{proof}

\begin{lemma}
Let $\mu$ be a smooth probability density
for a parameter $\alpha$, such that $\frac{1}{2} \leq \alpha < 1$.
The expected search time in a $\mu$-random IST of size $n$ is
$O(\log \log n + p)$.
\end{lemma}

\begin{proof}
The proof follows from Lemma~\ref{lem:constant-time-interpolation-search} and the argument in Lemma~\ref{lem:amortized-expected-log-log}, and the bound of $\leq p$ on the size of a root extension before a rebuild is triggered from Theorem~\ref{thm:non-blocking}.
\end{proof}


\section{Evaluation}
\label{sec:evaluation}

\subsection{Basic Experimental Evaluation}
\label{sec:original-evaluation}

This section summarizes the experimental setup,
and the results reported in the related work
that proposes and describes C-IST in detail~\cite{c-ist-ppopp2020}.

We implemented the concurrent IST in C++,
and we ran the benchmarks on a NUMA system with four Intel Xeon Platinum 8160 3.7GHz CPUs,
each of which has 24 cores and 48 hardware threads.
Within each CPU, cores share a 33MB LLC,
and cores on different CPUs do not share any caches.
The system has 384GB of RAM, and runs Ubuntu Linux 18.04.1 LTS.
Our code was compiled with GCC 7.4.0-1, with the highest optimization level (\texttt{-O3}).
Threads were \textit{pinned} to cores such that
thread counts up to 48 ran on only one CPU,
thread counts up to 96 run on only two CPUs, and so on.
We used the fast scalable allocator jemalloc 5.0.1-25.
When a memory page is allocated on our 4-CPU Xeon system,
it has an \textit{affinity} for a single CPU,
and other CPUs pay a penalty to access it.
We used the \texttt{numactl --interleave=all} option to ensure
that pages are evenly distributed across CPUs.

The goal of this section is to show
that the amortized $O(\log \log n)$ running time
induces performance improvements on datasets that are reasonably large.
We therefore evaluate the C-IST operations
against other comparable data structures in Section~\ref{sec:evaluation:basic-operations},
where we show, for 1 billion keys, improvements ranging from $15$-$50\%$
compared to the $(a,b)$-tree~\cite{BrownPhD} (the next best alternative),
depending on the ratio of updates and lookups.
To further characterize the performance,
we compare the average key depth and the impact on cache behavior
in Section \ref{sec:evaluation:average-depth-and-cache}.
We conclude with a comparison of memory footprints
in Section~\ref{sec:evaluation:memory-footprint}.

\subsubsection{Comparison of the Basic Operations}
\label{sec:evaluation:basic-operations}

Figure~\ref{fig:appendix:experiments:basic-operations}
shows the throughput of concurrent IST operations,
compared against other sorted set data structures,
for dataset sizes of $k = 2 \cdot 10^8$ and $k = 2 \cdot 10^9$ keys,
and for $u = 0\%$, $u = 1\%$, $u = 10\%$ and $u = 40\%$,
where $u$ is the ratio of update operations among all operations.

In all cases, C-IST operations have much higher throughput than
Natarajan and Mittal's non-blocking binary search tree (NM),
and concurrent AVL trees due to Bronson (BCCO).
For update ratios $u = 0\%$ and $u = 1\%$, concurrent IST also has a higher throughput
compared to Brown's non-blocking $(a,b)$-tree.
The underlying cause for better throughput is a lower rate of LLC misses
due to IST's doubly-logarithmic depth.
For higher update ratios $u = 10\%$ and $u = 40\%$,
the cost of concurrent rebuilds starts to dominate the gains of doubly-logarithmic searches,
and ABTree has a better throughput for $k = 2 \cdot 10^8$ keys.
Above $k = 2 \cdot 10^9$ keys,
ISTree outperforms ABTree even for the update ratio of $u = 40\%$.

\subsubsection{Average Depth and Cache Behavior}
\label{sec:evaluation:average-depth-and-cache}

The main benefit of C-IST's expected-$O(\log \log n)$ depth
is that the key-search results in less cache misses
compared to other tree data structures.
The plot shown below compares the average number of pointer hops required to reach a key
(error bars show min/max values over all trials),
for dataset sizes from $2 \cdot 10^6$ to $2 \cdot 10^9$ keys.
While the average depth is $20$-$40$ for NM and BCCO,
the average ABTree depth is between $6$ and $10$,
and the average C-IST depth is below $5$.

\begin{wrapfigure}{l}{0.22\textwidth}
\vspace{-5mm}
\noindent\includegraphics[scale=0.17]{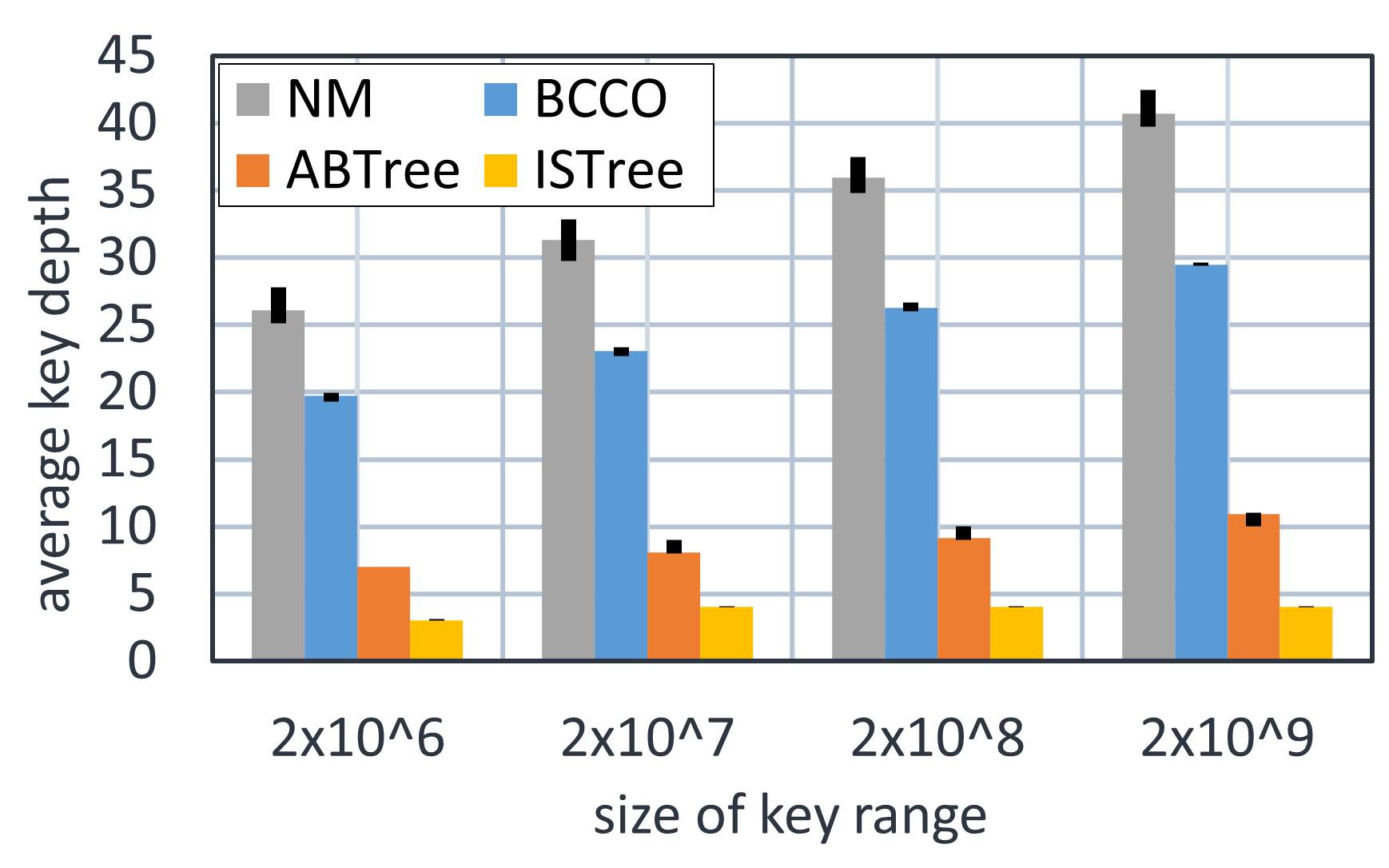}
\vspace{-10mm}
\end{wrapfigure}
The differences in average depths between these data structures
correlate with the average number of cache misses.
Figure~\ref{fig:appendix:experiments:llc} compares the average number of last-level cache-misses
between the different data structures, for different update ratios $u$.
For the dataset size of $2 \cdot 10^9$ keys,
ISTree operations undergo $2\times$ less cache misses,
and slightly fewer cache misses than ABTree.

\begin{figure}[t]
\includegraphics[scale=0.20]{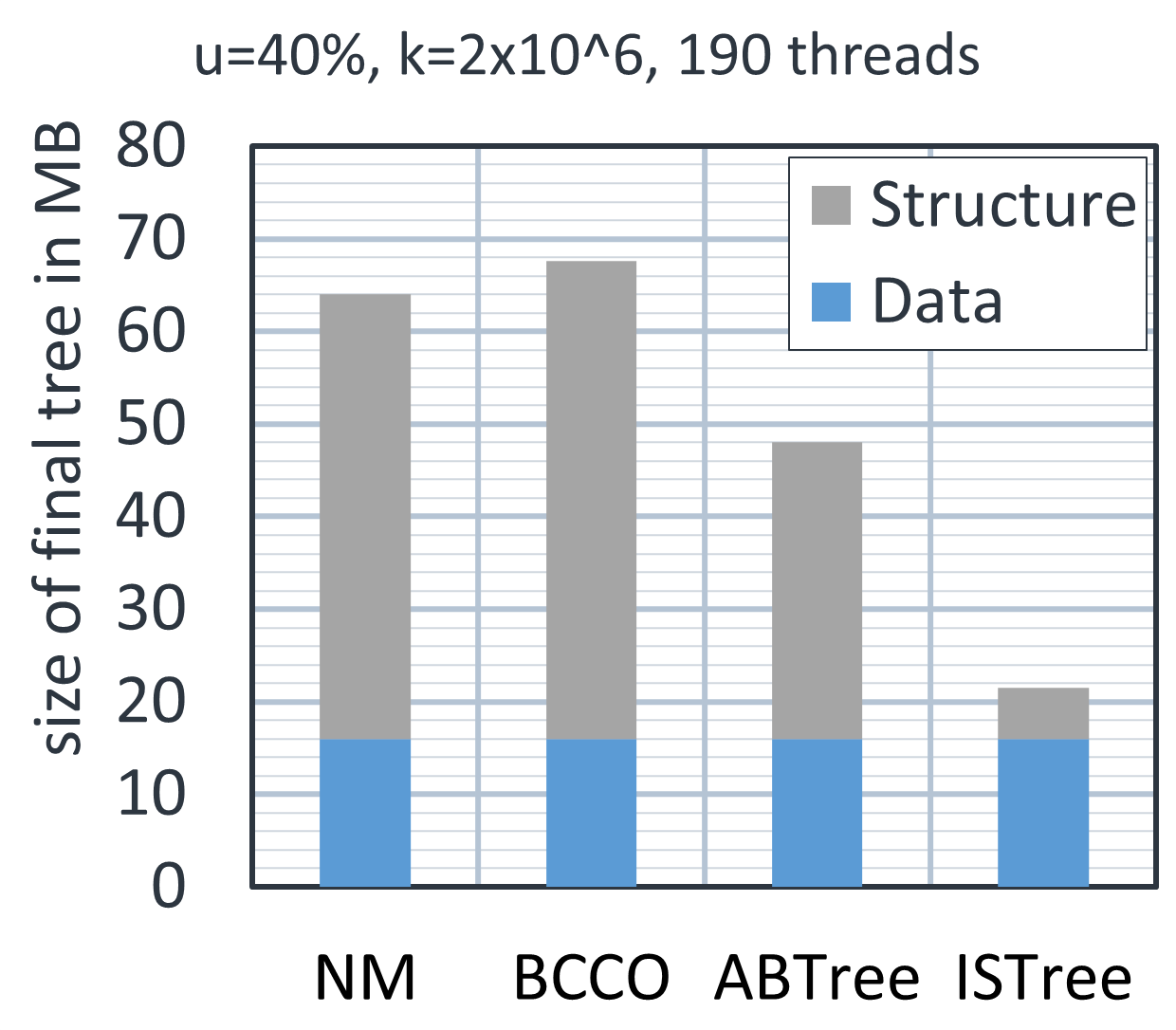}
\includegraphics[scale=0.20]{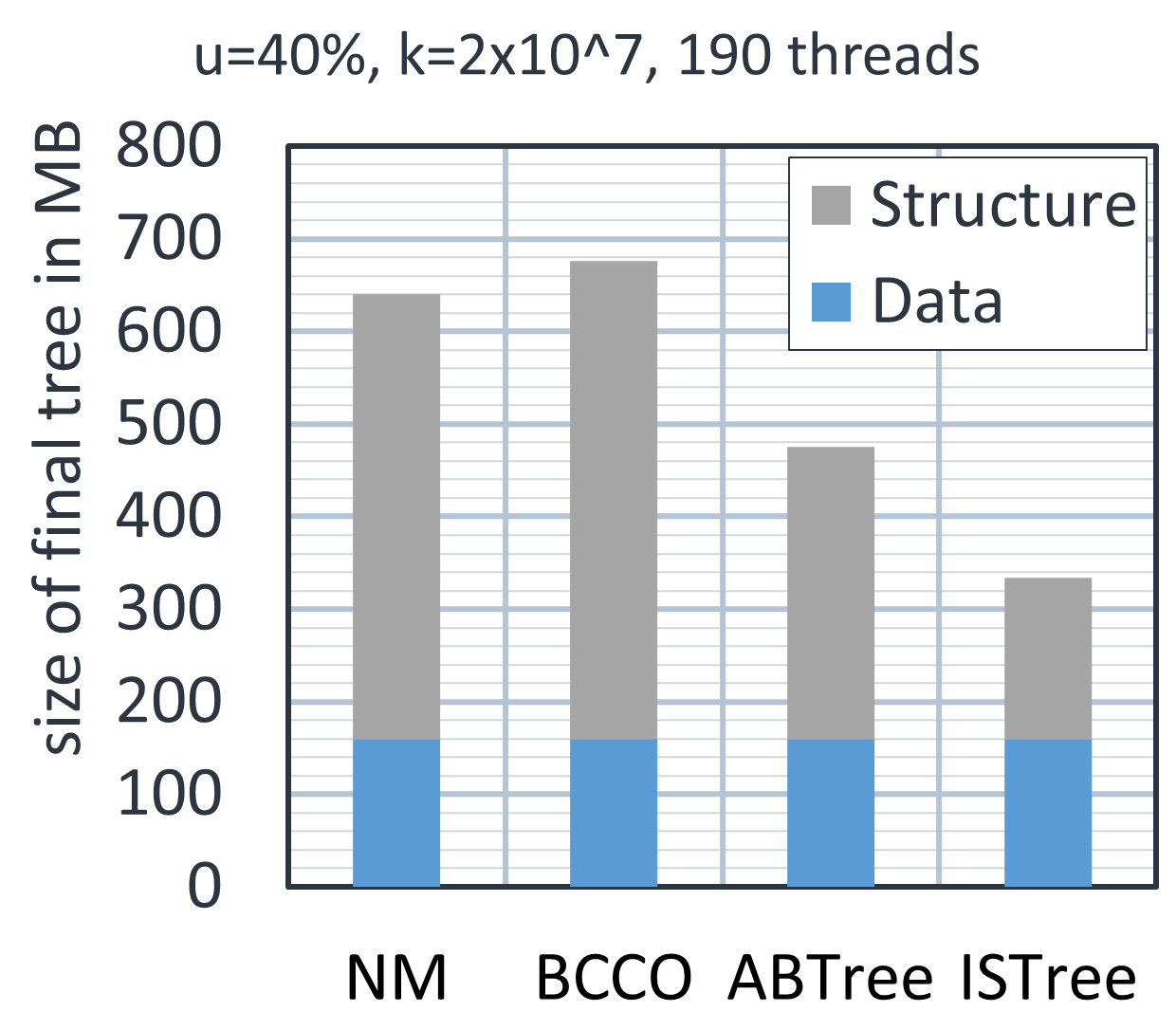}
\includegraphics[scale=0.20]{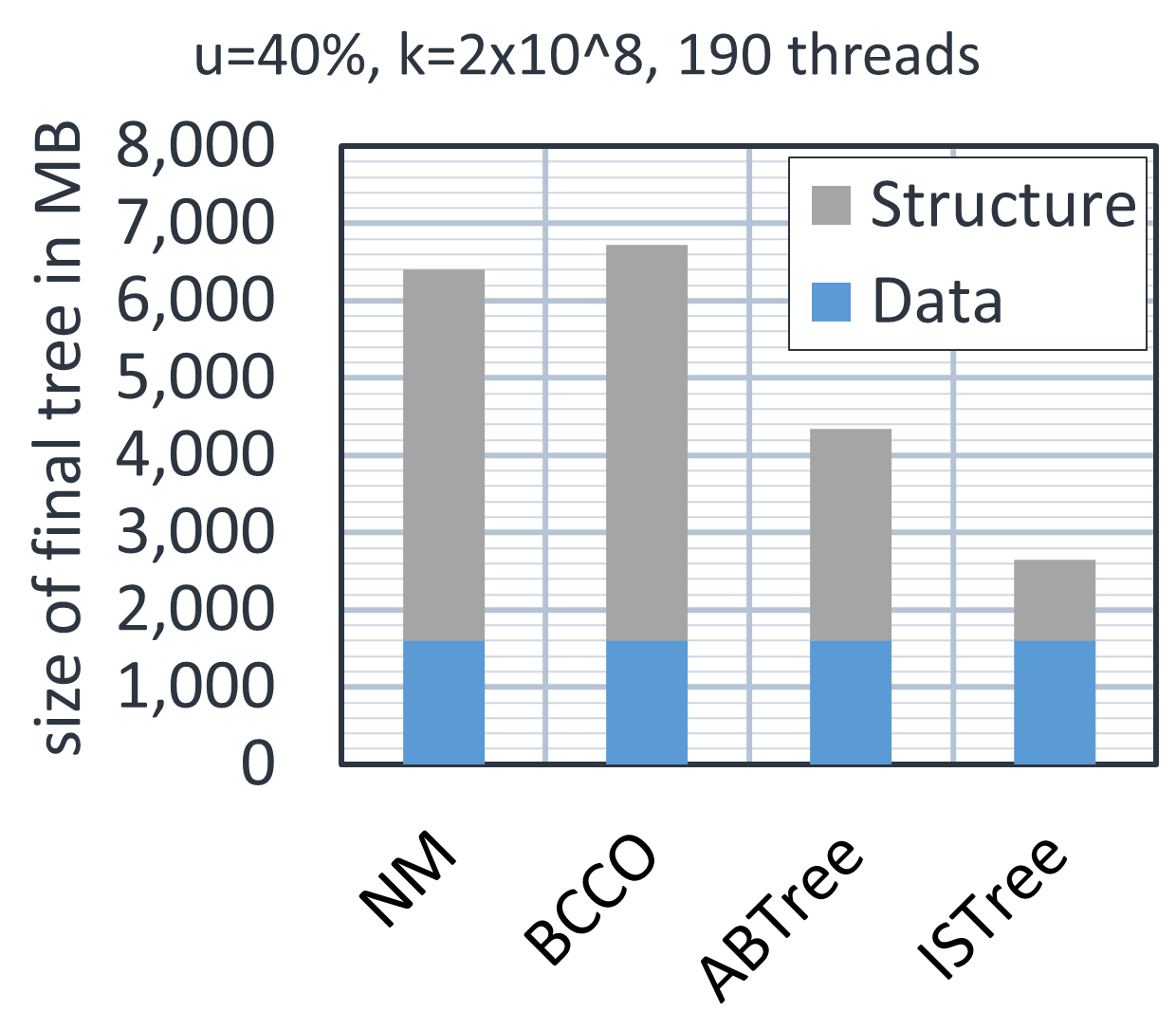}
\includegraphics[scale=0.20]{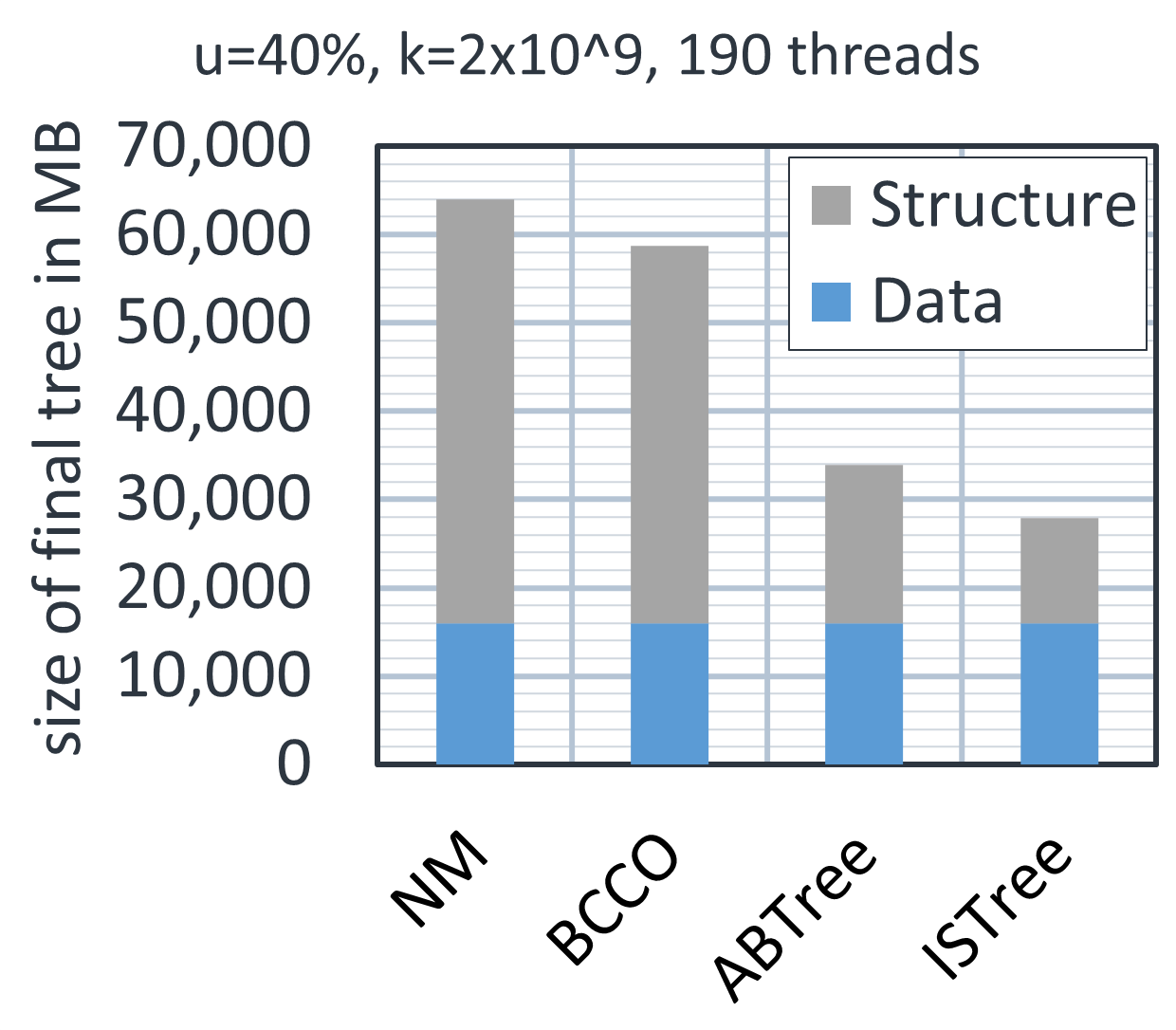}
\caption{
Memory Footprint Comparison
}
\label{fig:evaluation:memory-footprint}
\end{figure}

\subsubsection{Memory Footprint}
\label{sec:evaluation:memory-footprint}

Due to using a lower number of nodes for the same dataset,
the average space overhead is lower for the C-IST than the other data structures.
Figure~\ref{fig:evaluation:memory-footprint} shows
the different memory footprints for four different dataset sizes.
C-IST has a relative space overhead of $\approx$$30$-$100\%$,
whereas the overhead of the other data structures is between $\approx$$120$-$400\%$.

\subsection{Extended Experimental Evaluation}
\label{sec:appendix:experiments}

In this section,
we show additional experimental data that did not fit
into the main body of the paper.

\paratitle{Excluding data structures from graphs}
As we mentioned in Section~\ref{sec:original-evaluation}, we also compared ISTree with many
other concurrent search trees, including the external lock-free BST
of Ellen et~al.~\cite{Ellen:2010:NBS:1835698.1835736} (EFRB),
the internal lock-free BST of Howley and Jones~\cite{Howley:2012} (HJ),
the partially external lock-free BST of Drachsler et~al.~\cite{Drachsler} (DVY),
the internal lock-free BST of Ramachandran and Mittal~\cite{intlf} (RM),
the external lock-based BST of David et~al.~\cite{David:2015} (DGT),
and the external lock-free BST (BER), lock-free relaxed AVL tree (RAVL),
and lock-free Chromatic tree (Chromatic) of Brown et~al.~\cite{BrownPhD}.

\begin{figure}[t]
\includegraphics[width=0.42\textwidth]{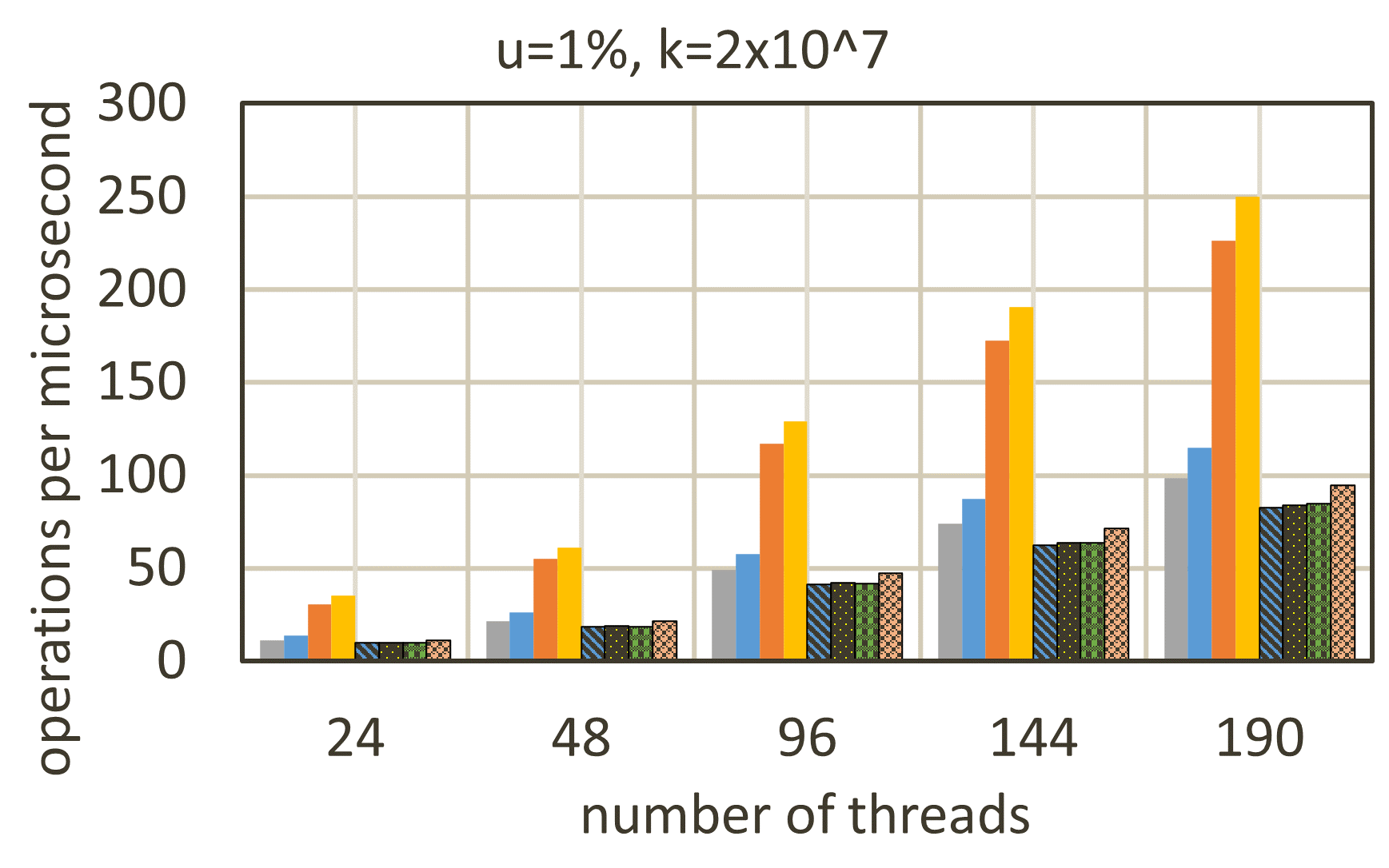}
\includegraphics[width=0.42\textwidth]{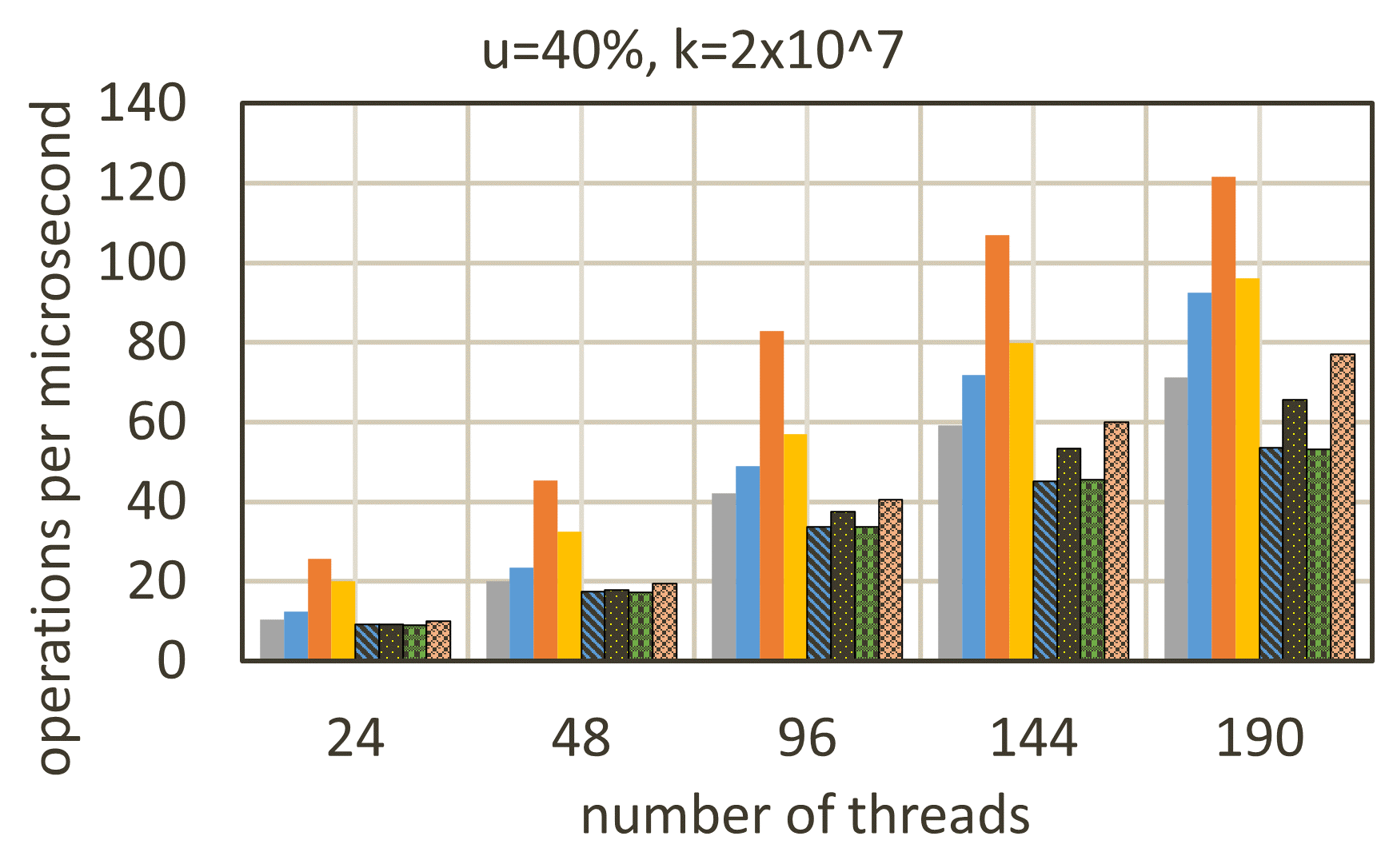}

\includegraphics[scale=0.25]{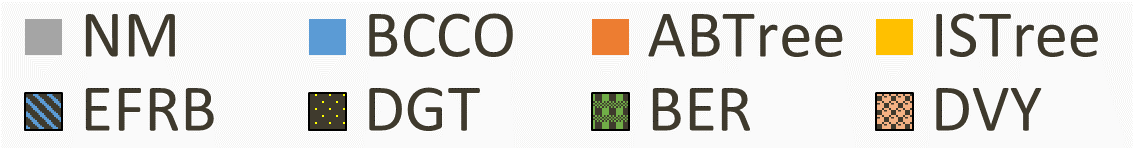}
\caption{Full comparison of basic operations for all (non-failing) data structures, higher is better}
\label{fig:appendix:experiments:excluding-data-structures}
\end{figure}

Unfortunately, the implementations of HJ and RM (which we obtained
from ASCYLIB~\cite{David:2015}) failed basic validation checksums.
Even after attempting to fix the implementations as much as possible,
the failures were persistent.
Since the implementations are not correct, it does not make sense to benchmark them.
The remaining implementations, EFRB, DVY, DGT, DER, RAVL and Chromatic
were all consistently outperformed by BCCO and, hence, merely obscured the results.
To give a sense of what the full data looks like,
Figure~\ref{fig:appendix:experiments:excluding-data-structures}
compares the implementations that do not fail checksum validation
using the same experimental setup as Section~\ref{sec:evaluation:basic-operations}.
(RAVL and Chromatic perform similarly to BER and are excluded.)
Thus, we decided not to include any of these data structures in our graphs.

\begin{figure*}[t]
\includegraphics[scale=0.165]{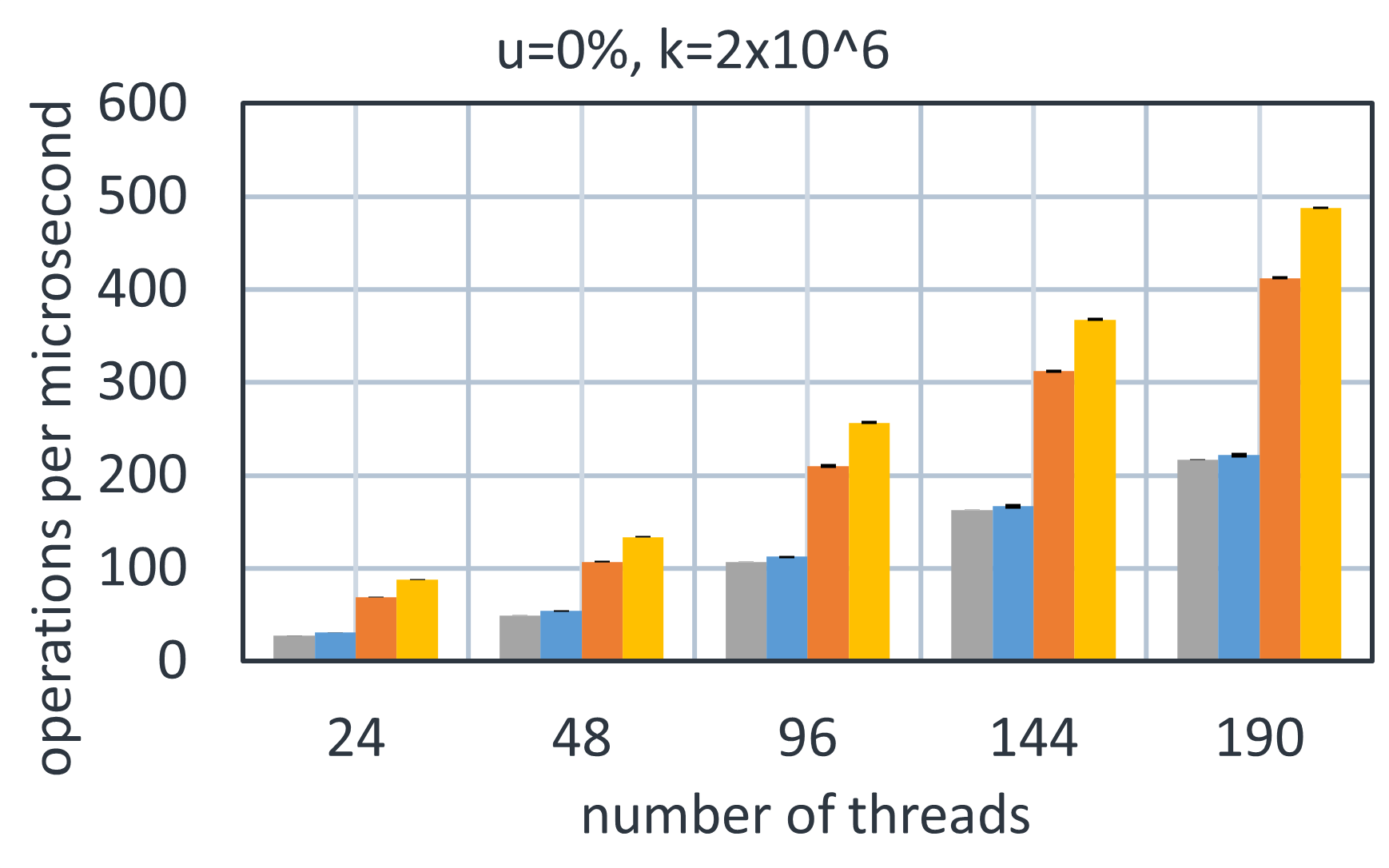}
\includegraphics[scale=0.165]{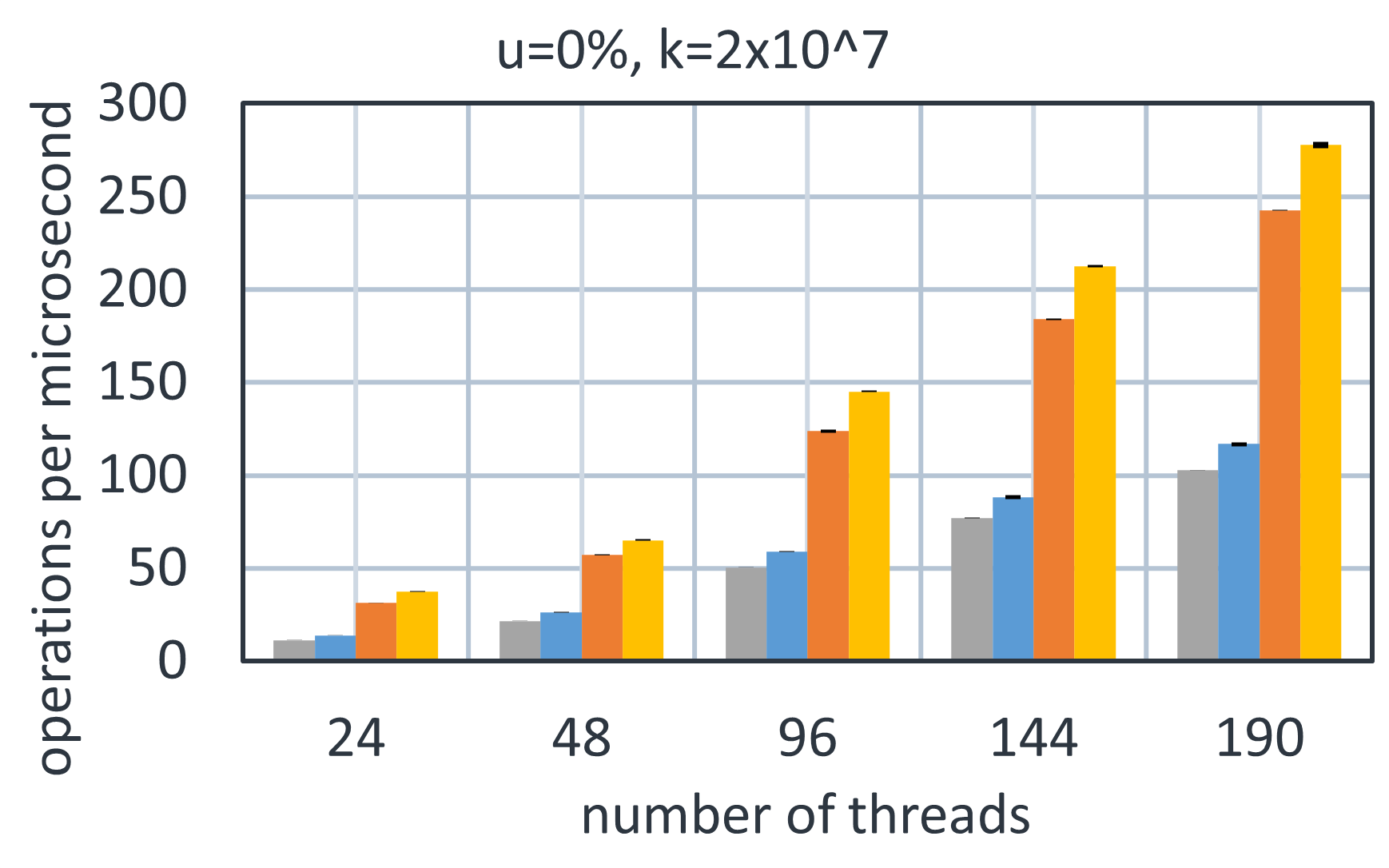}
\includegraphics[scale=0.165]{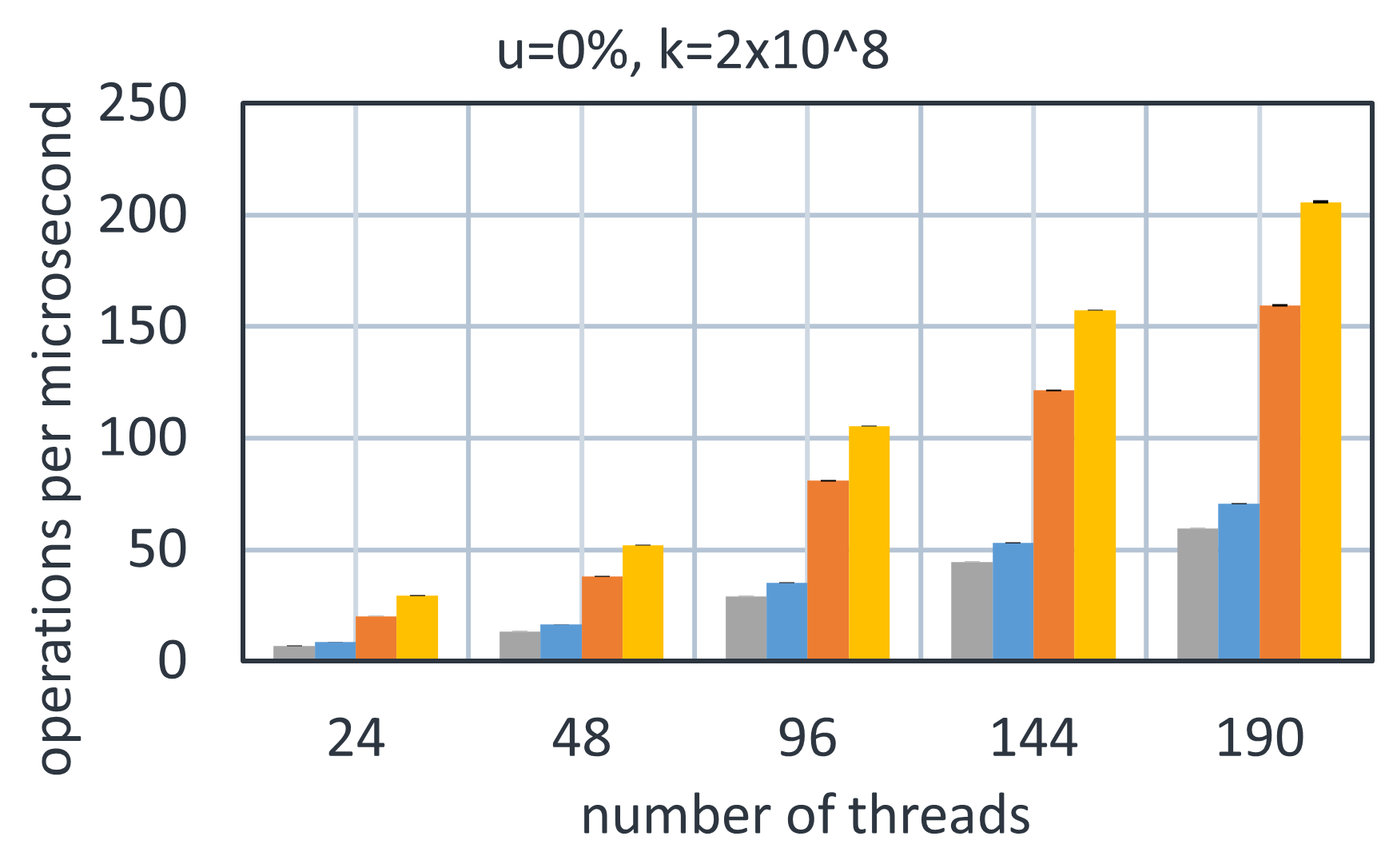}
\includegraphics[scale=0.165]{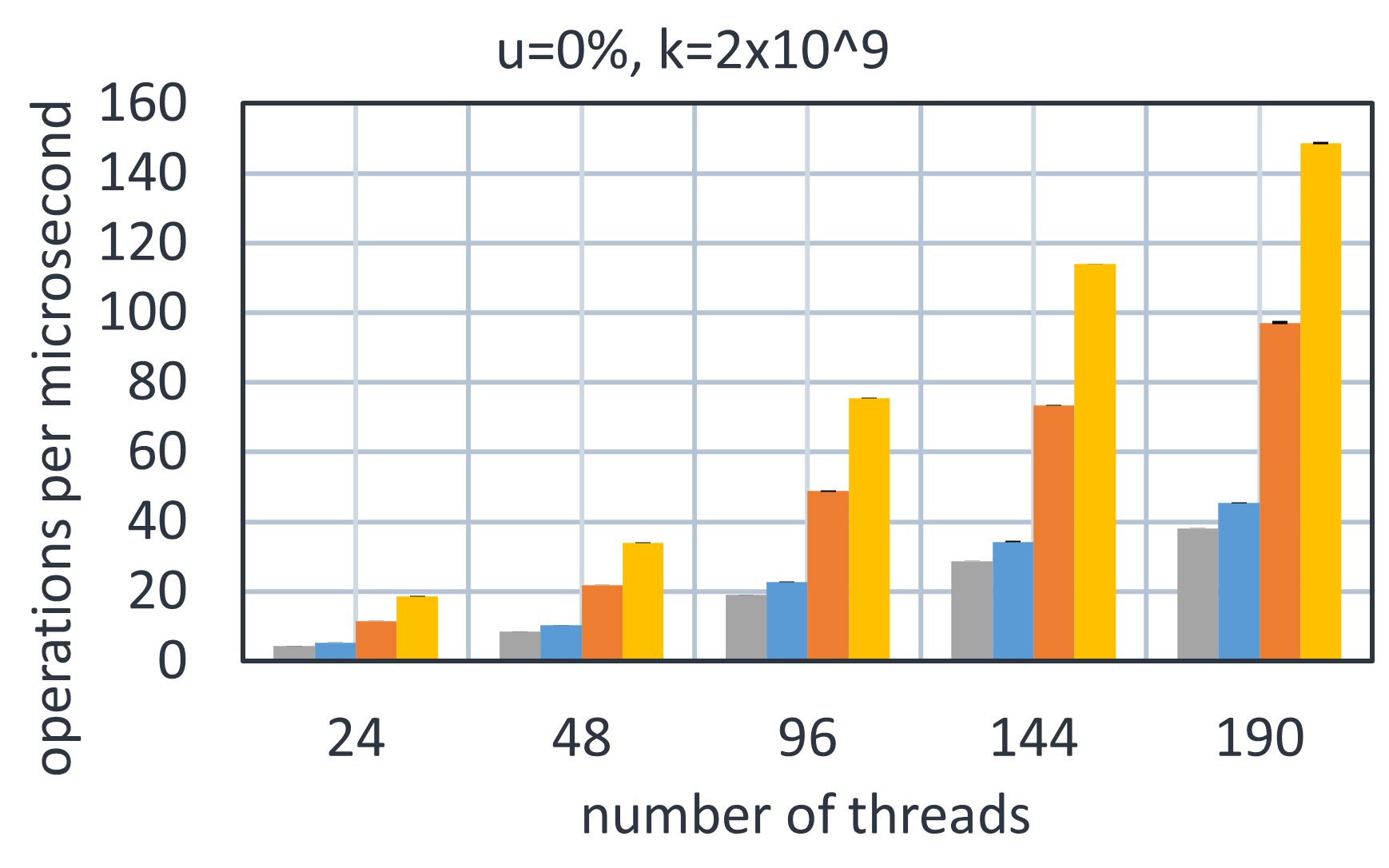}
\includegraphics[scale=0.165]{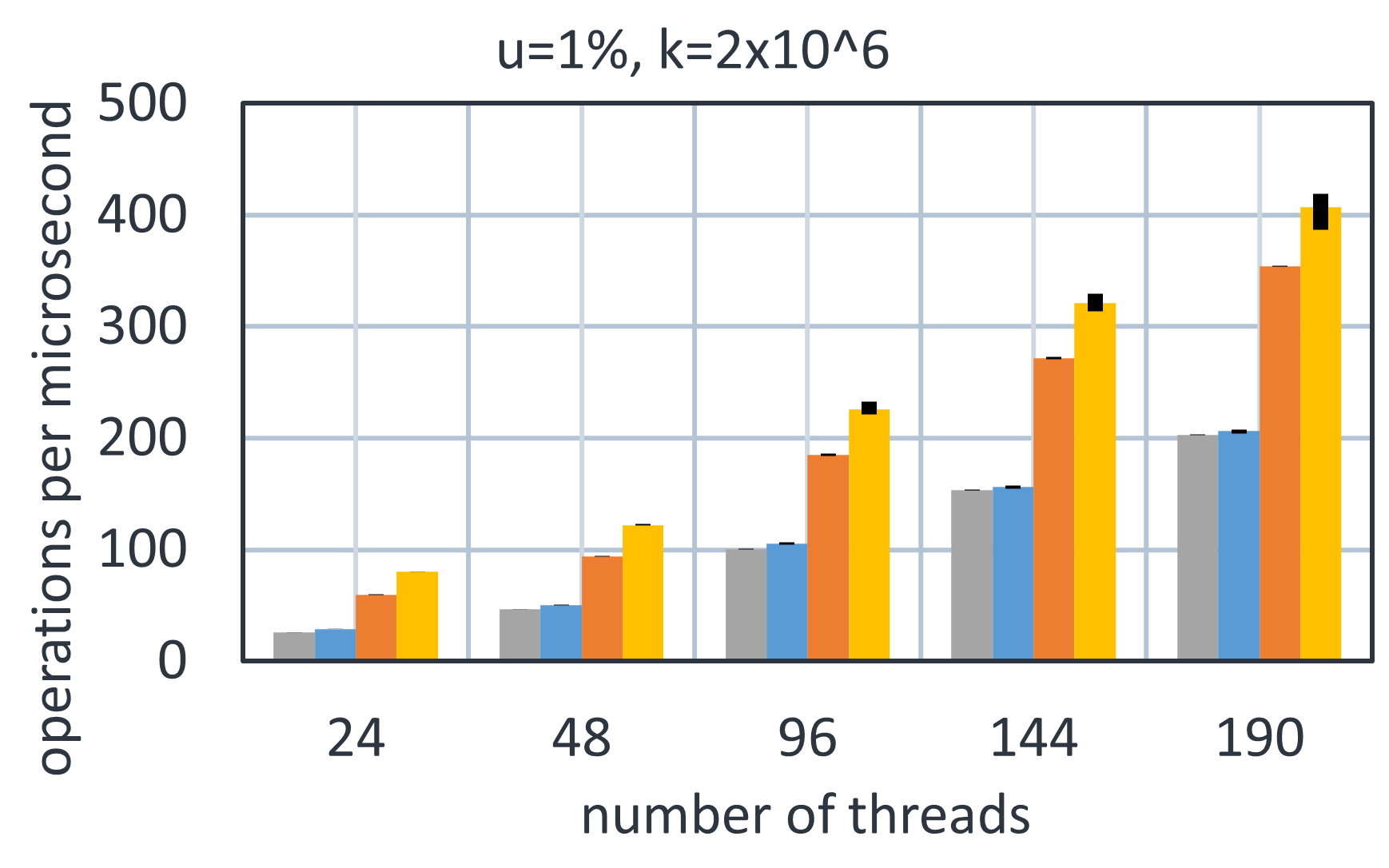}
\includegraphics[scale=0.165]{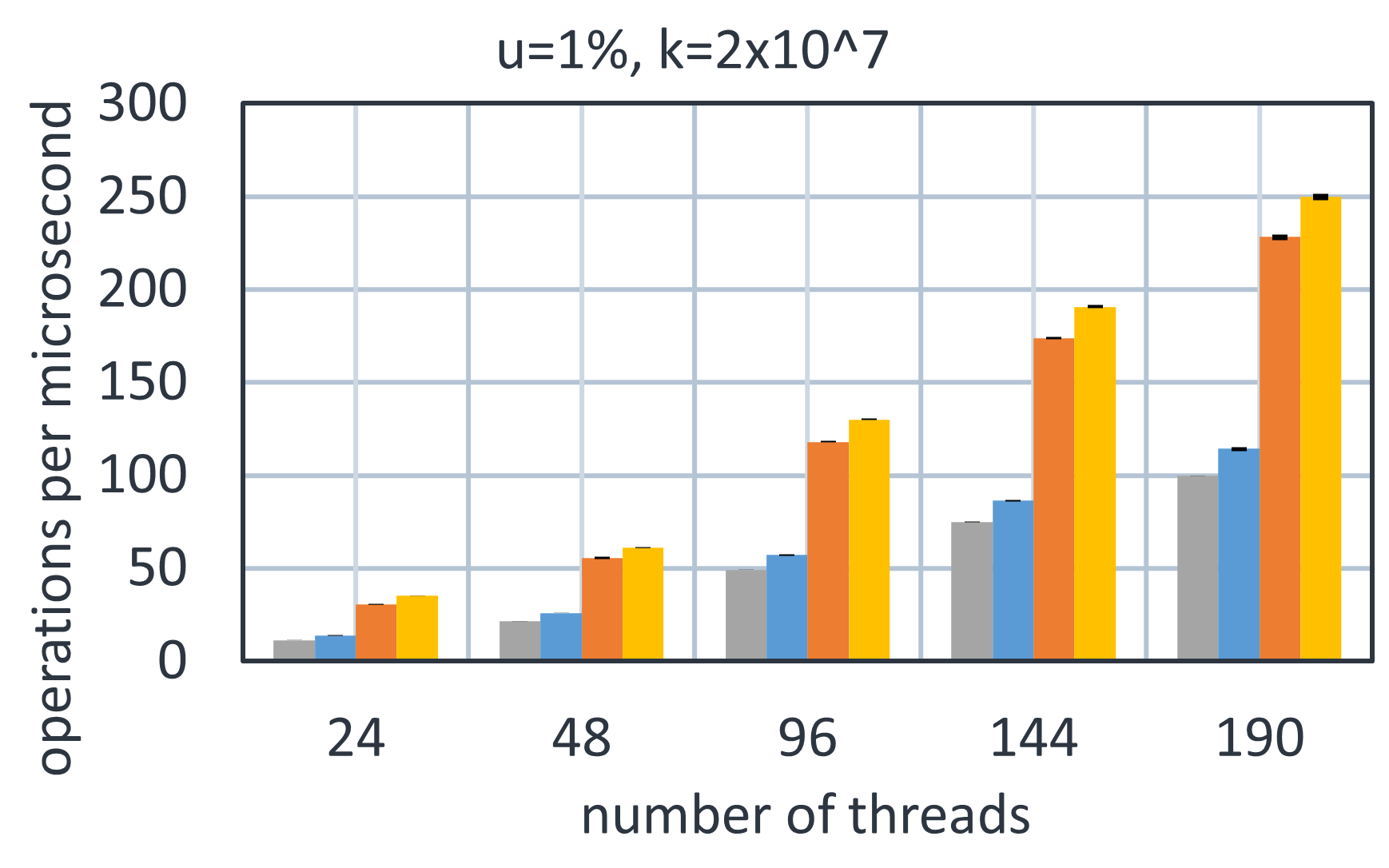}
\includegraphics[scale=0.165]{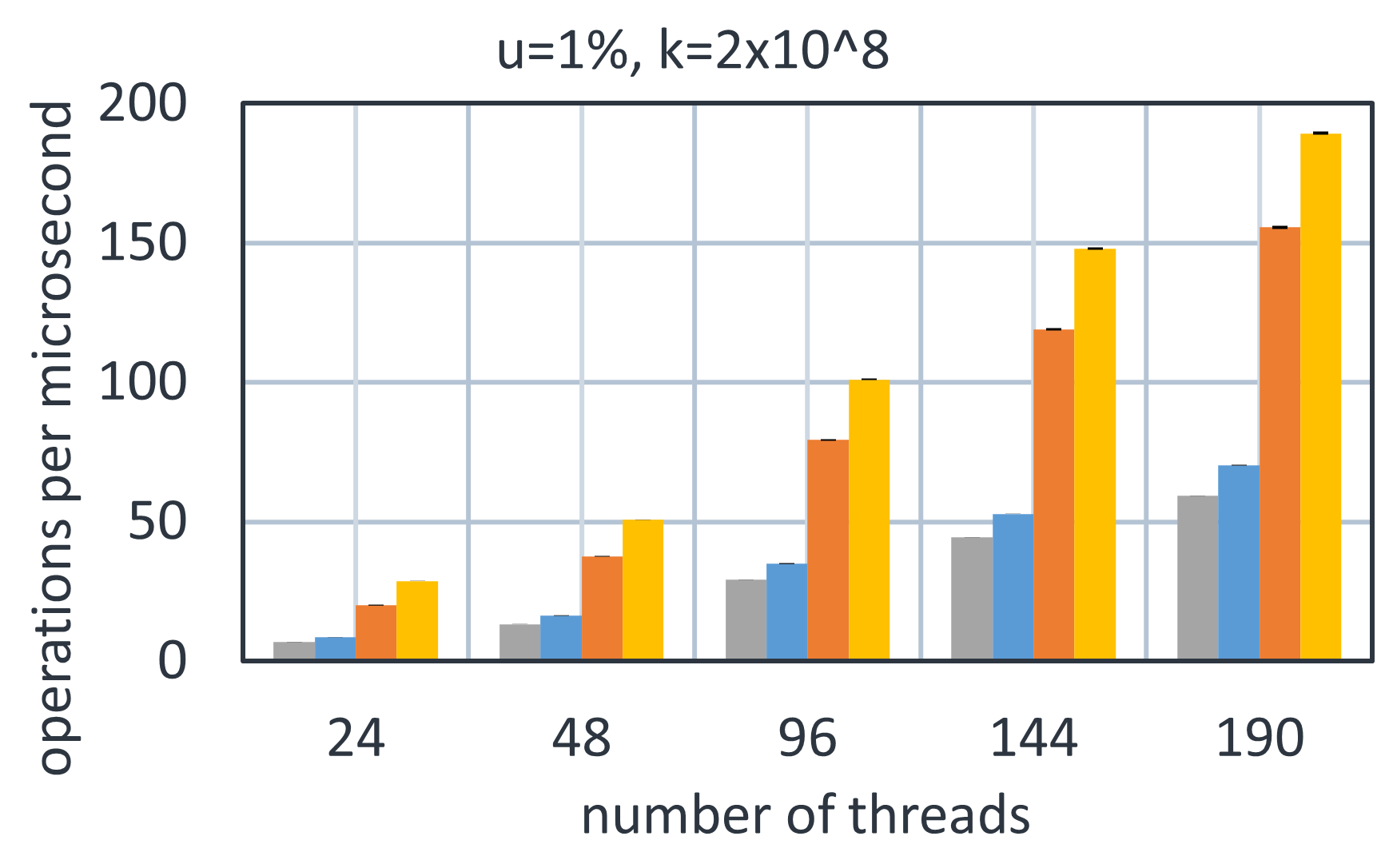}
\includegraphics[scale=0.165]{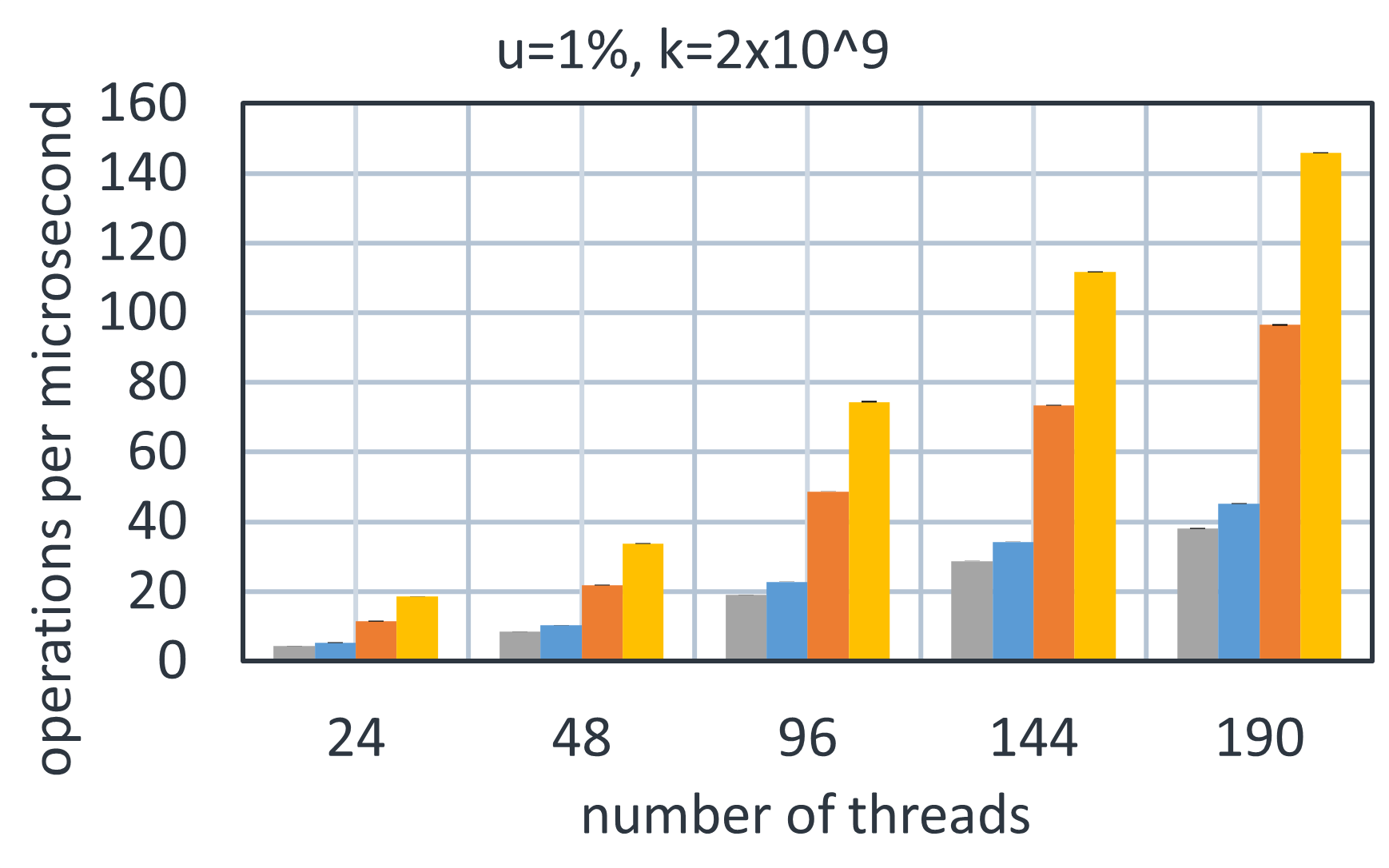}
\includegraphics[scale=0.165]{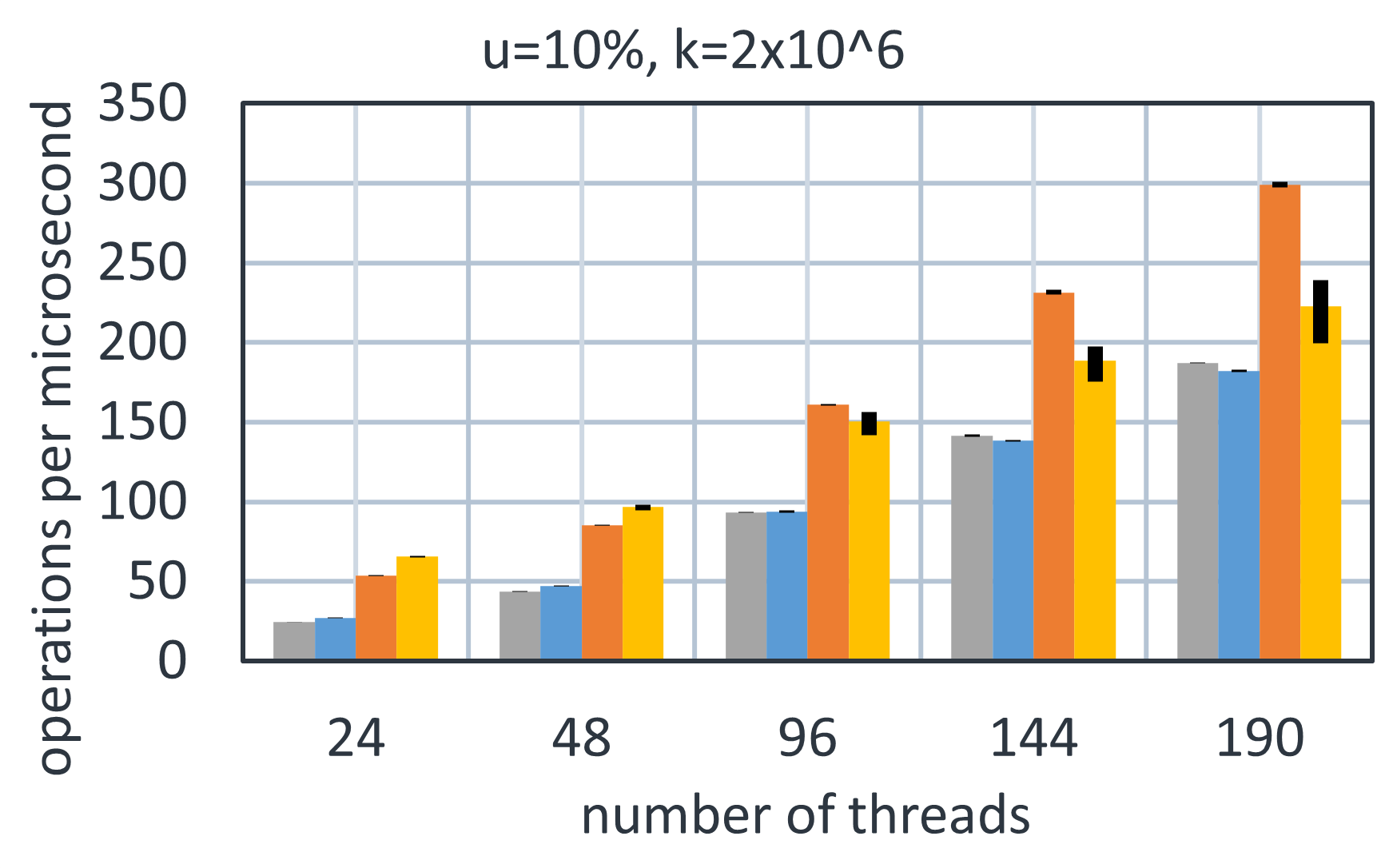}
\includegraphics[scale=0.165]{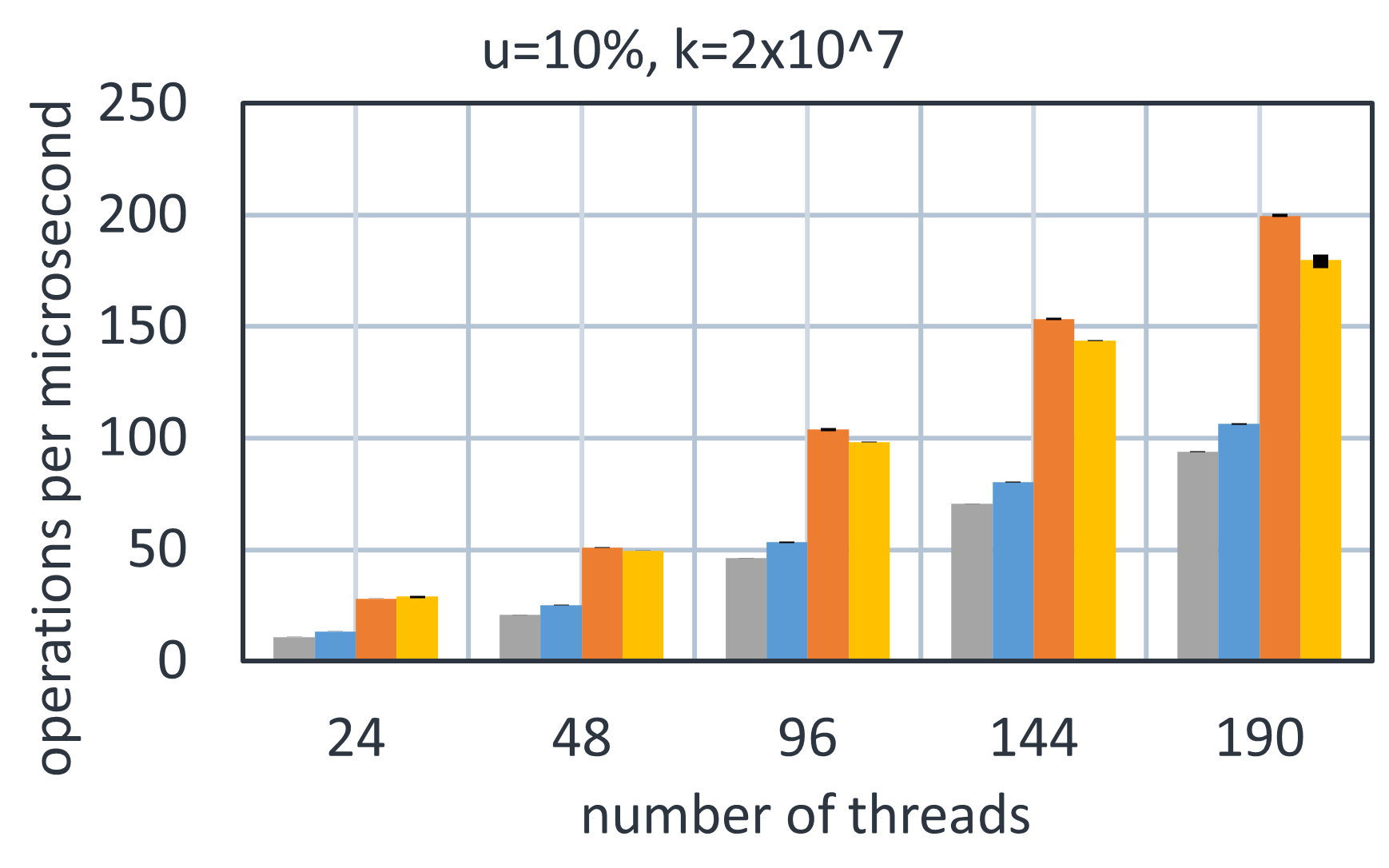}
\includegraphics[scale=0.165]{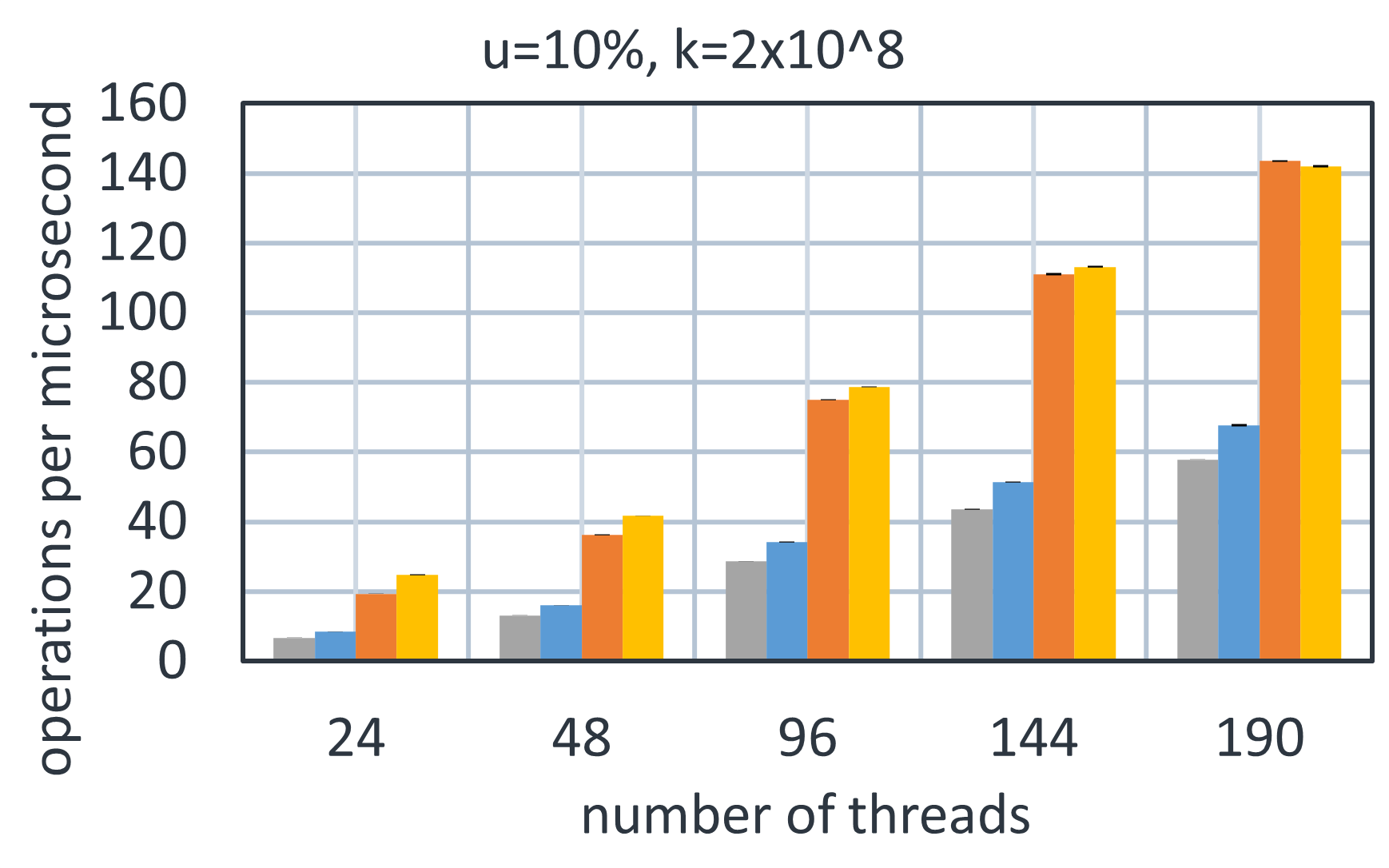}
\includegraphics[scale=0.165]{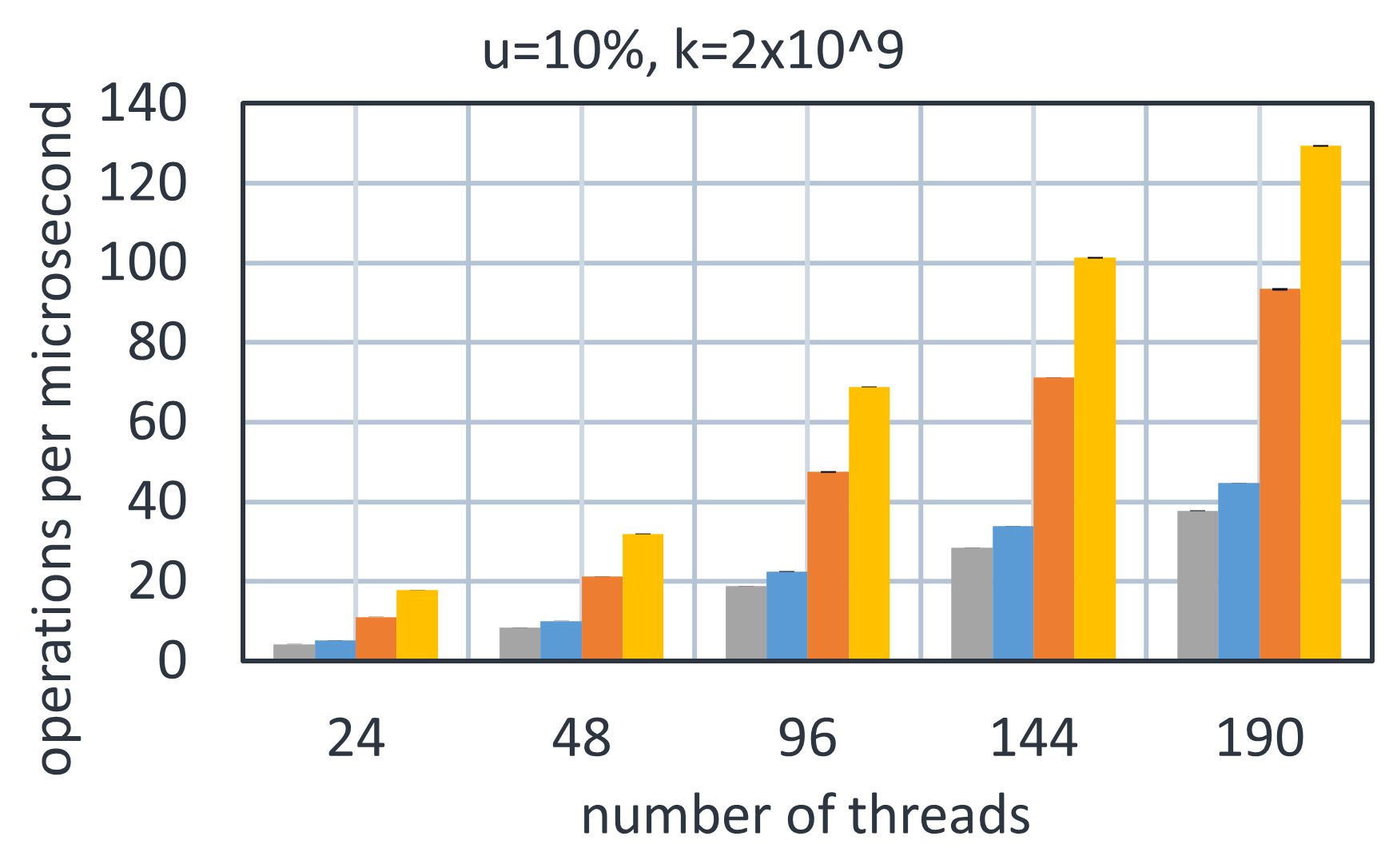}
\includegraphics[scale=0.165]{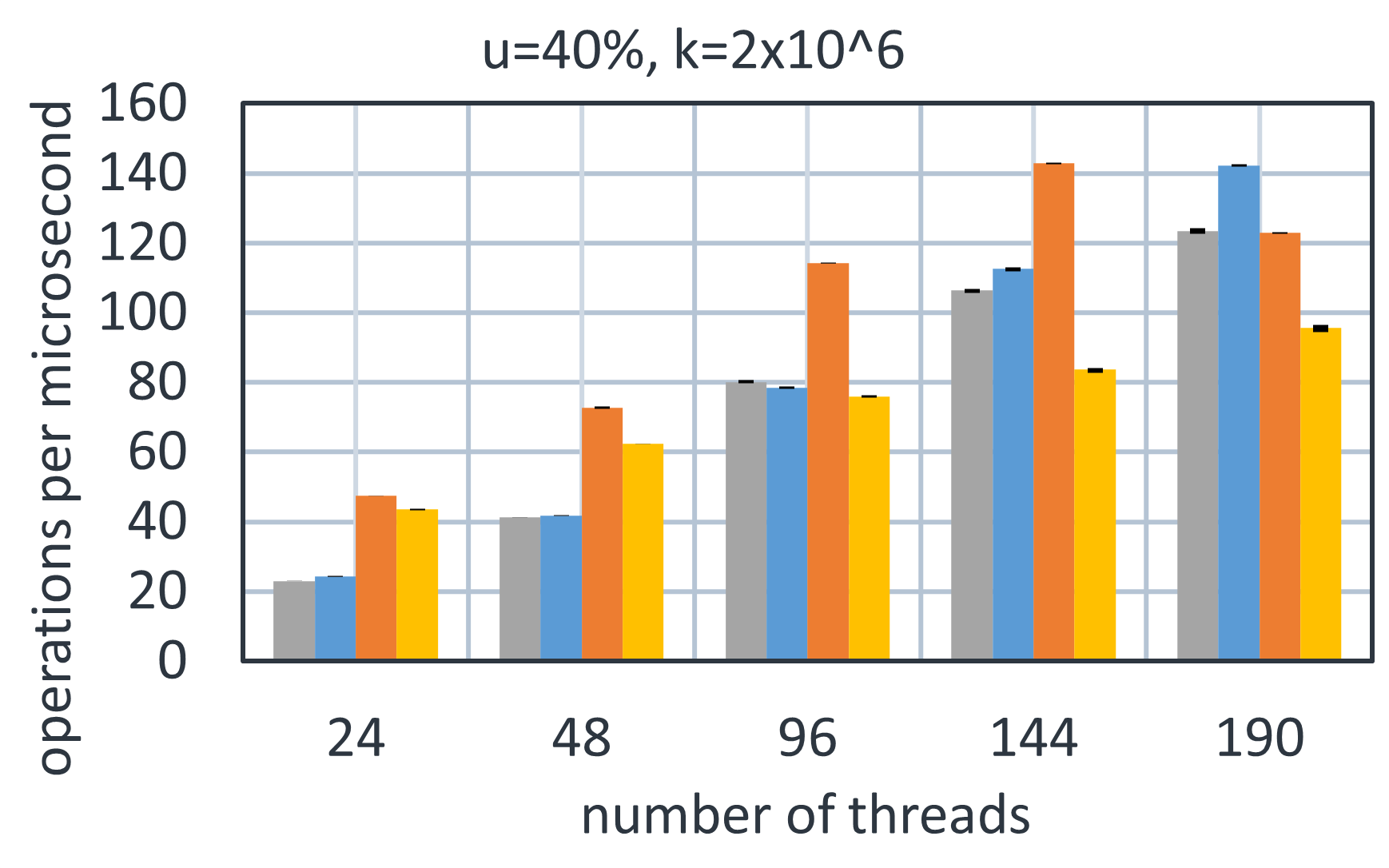}
\includegraphics[scale=0.165]{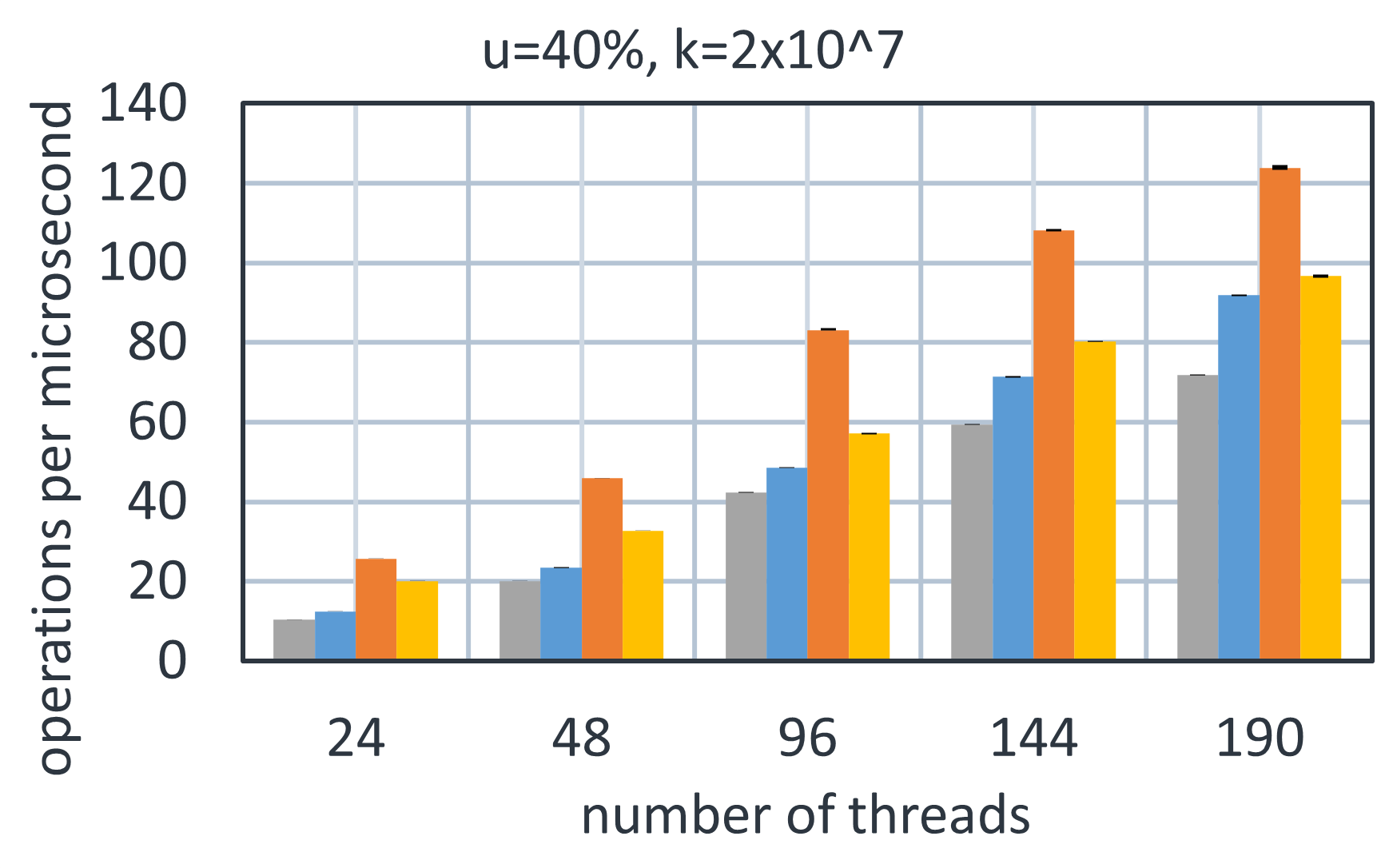}
\includegraphics[scale=0.165]{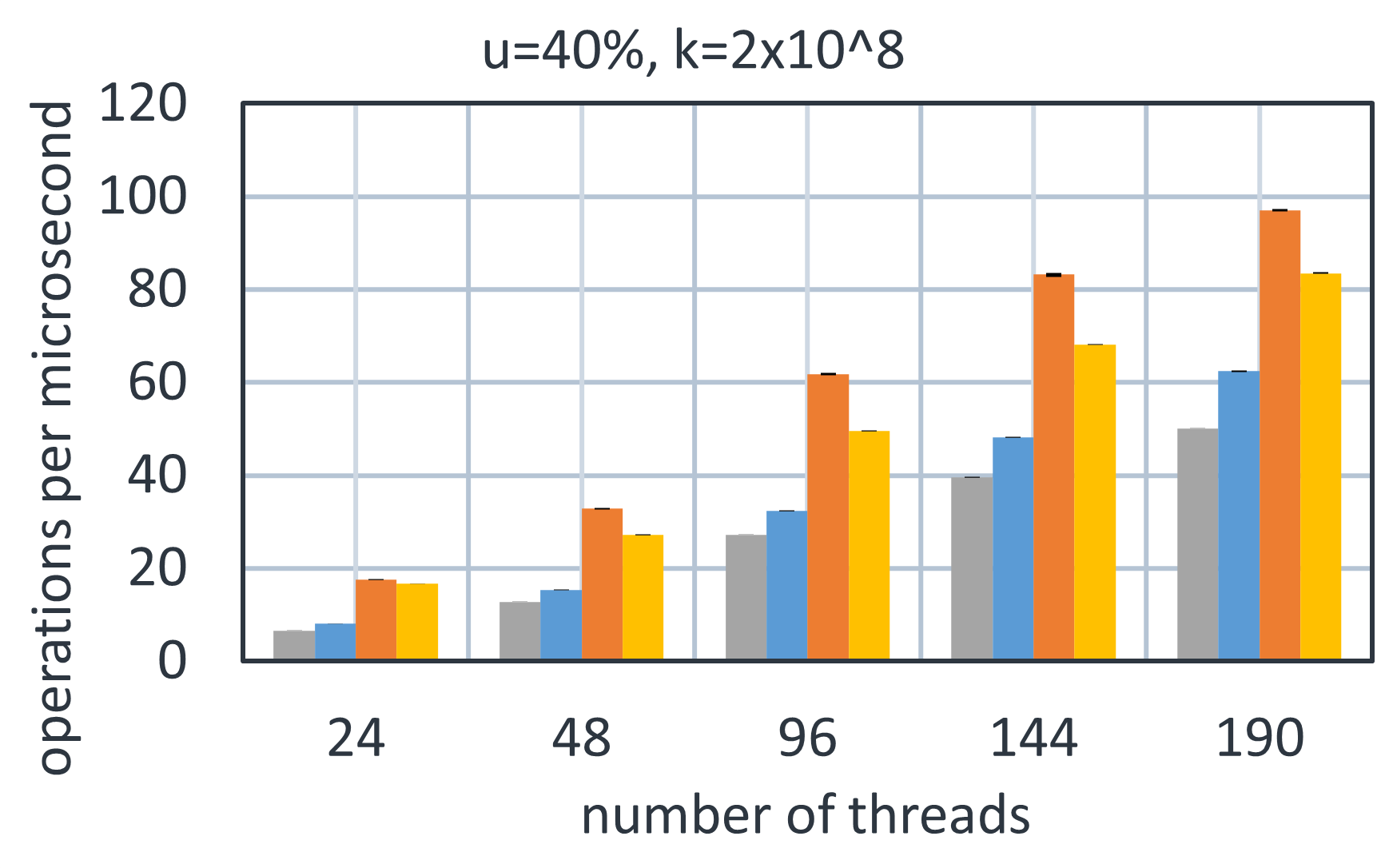}
\includegraphics[scale=0.165]{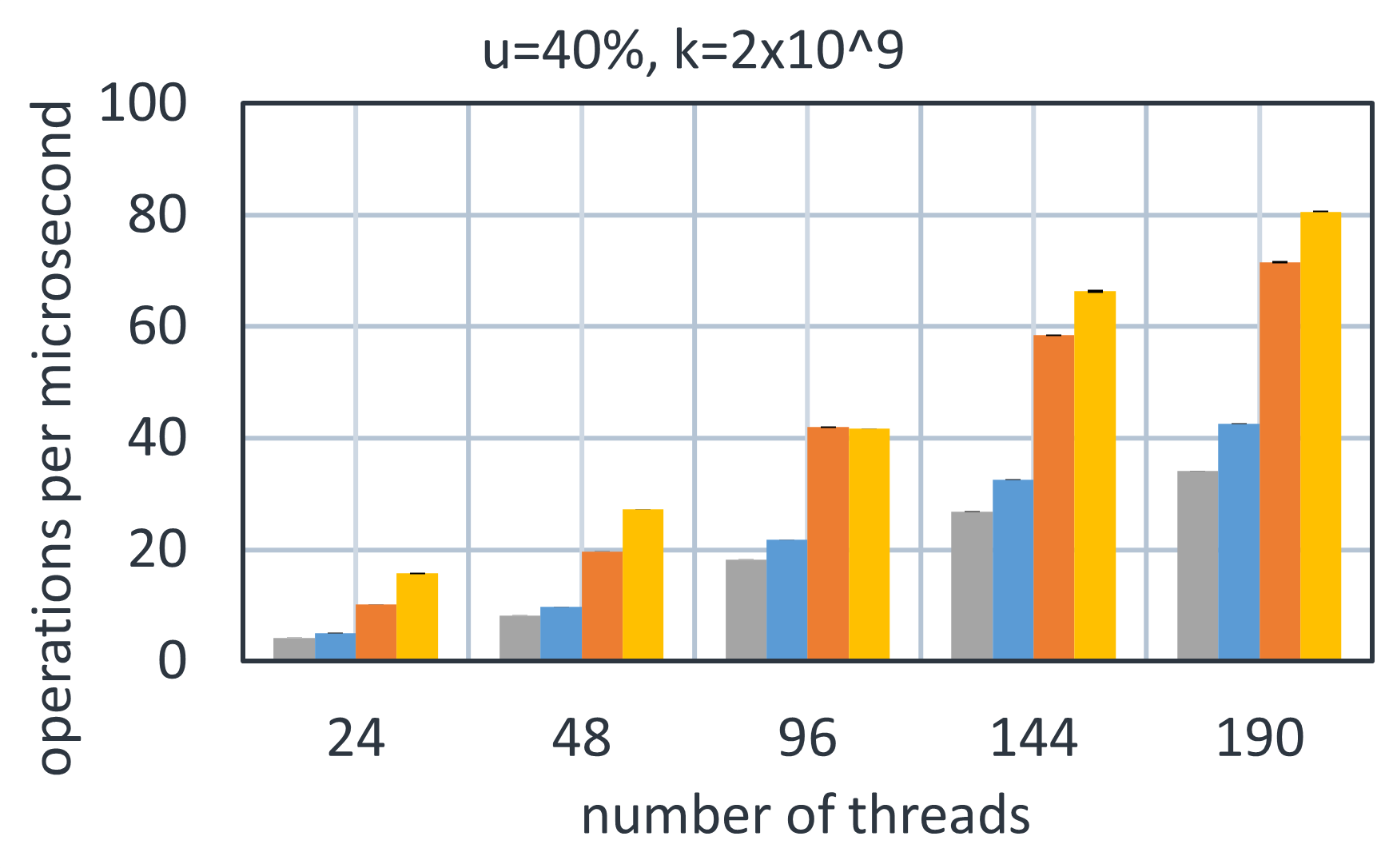}

\includegraphics[scale=0.25]{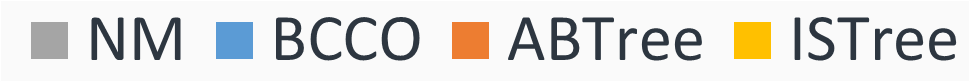}
\caption{Comparison of Basic Operations, Higher is Better}
\label{fig:appendix:experiments:basic-operations}
\end{figure*}

\paratitle{Throughput}
Figure~\ref{fig:appendix:experiments:basic-operations}
shows the throughput of the different concurrent data structures
for different update percentages ($0\%$, $1\%$, $10\%$ and $40\%$),
and dataset sizes ranging from $10^6$ to $10^9$.

\begin{figure*}[t]
\includegraphics[scale=0.165]{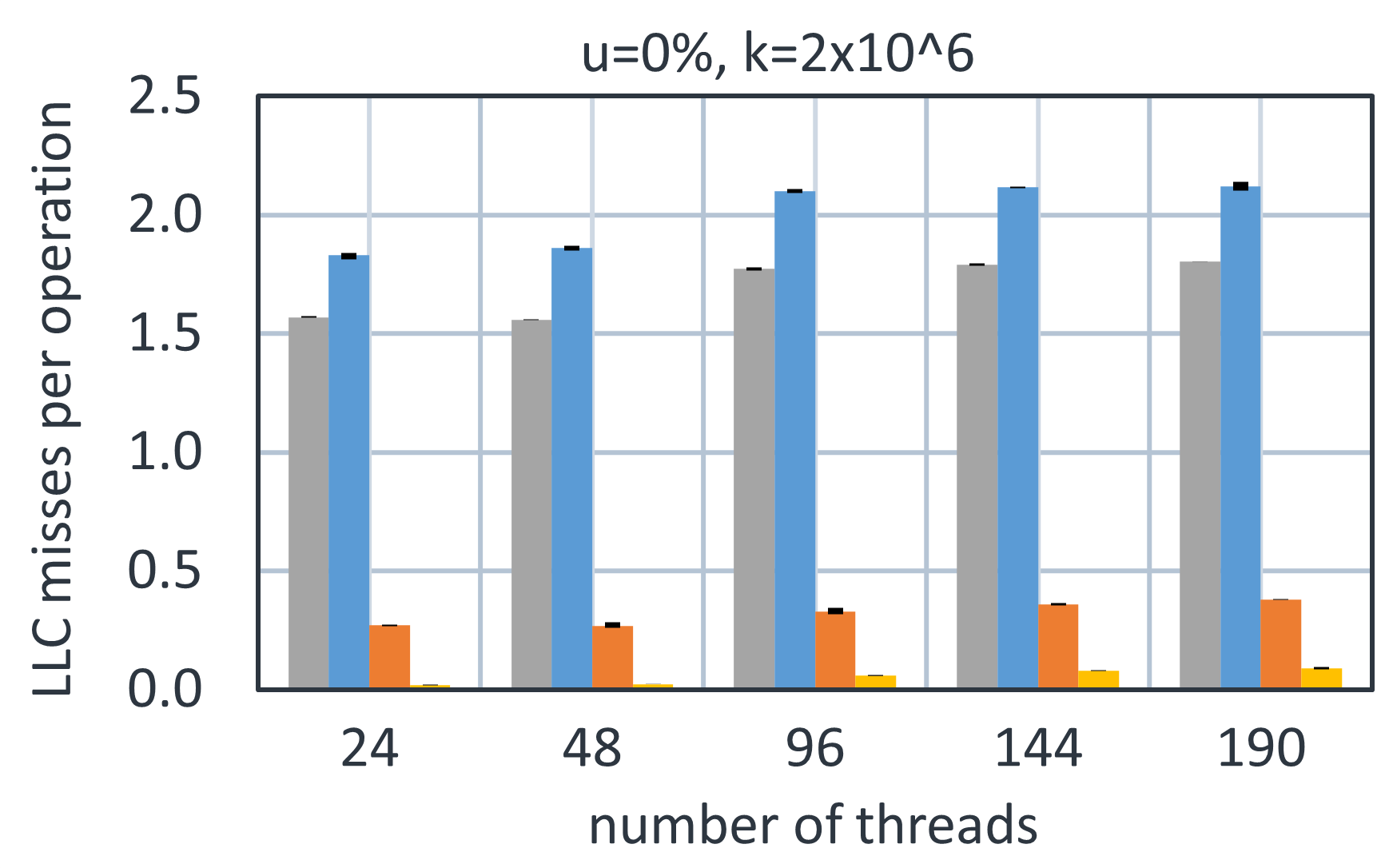}
\includegraphics[scale=0.165]{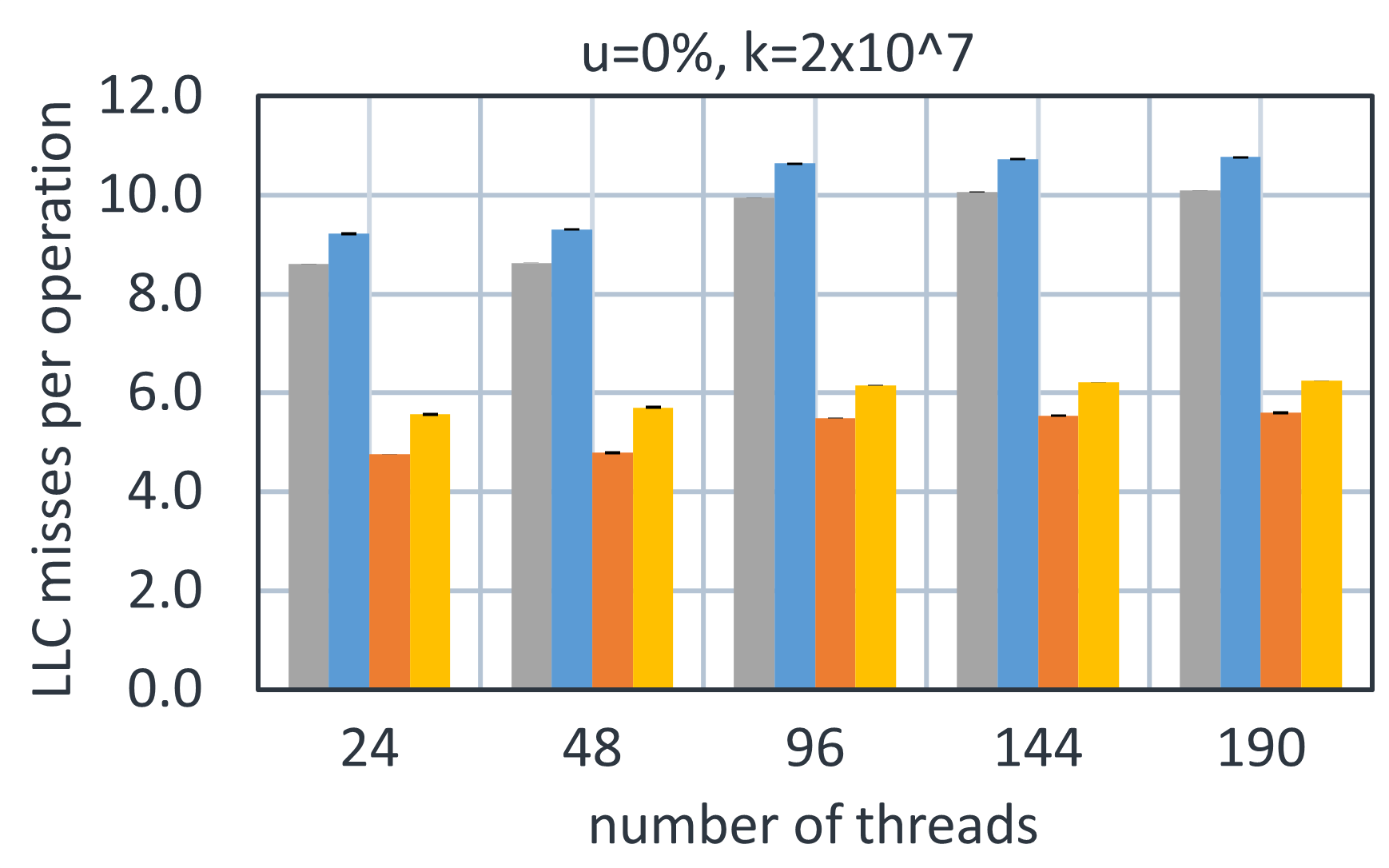}
\includegraphics[scale=0.165]{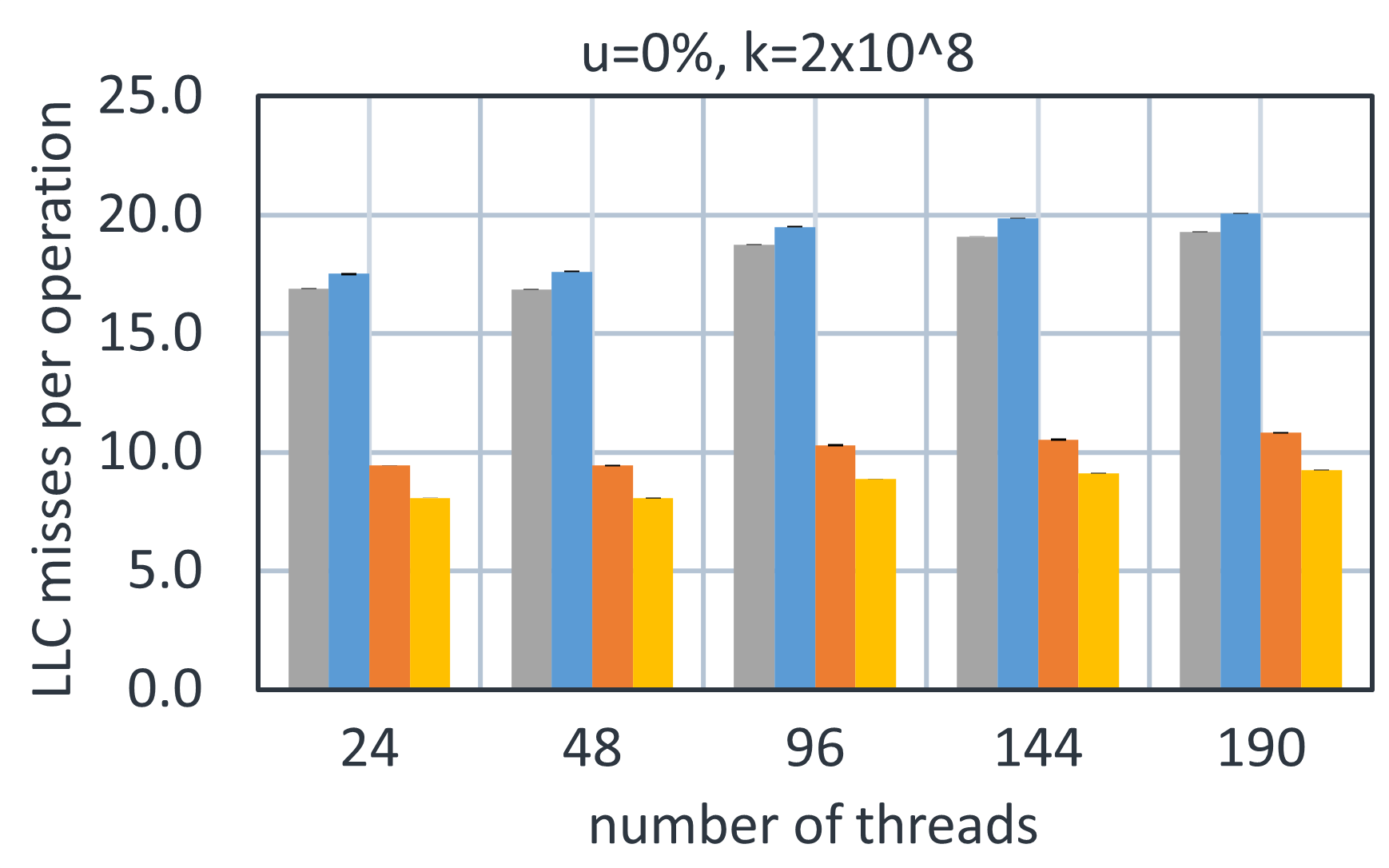}
\includegraphics[scale=0.165]{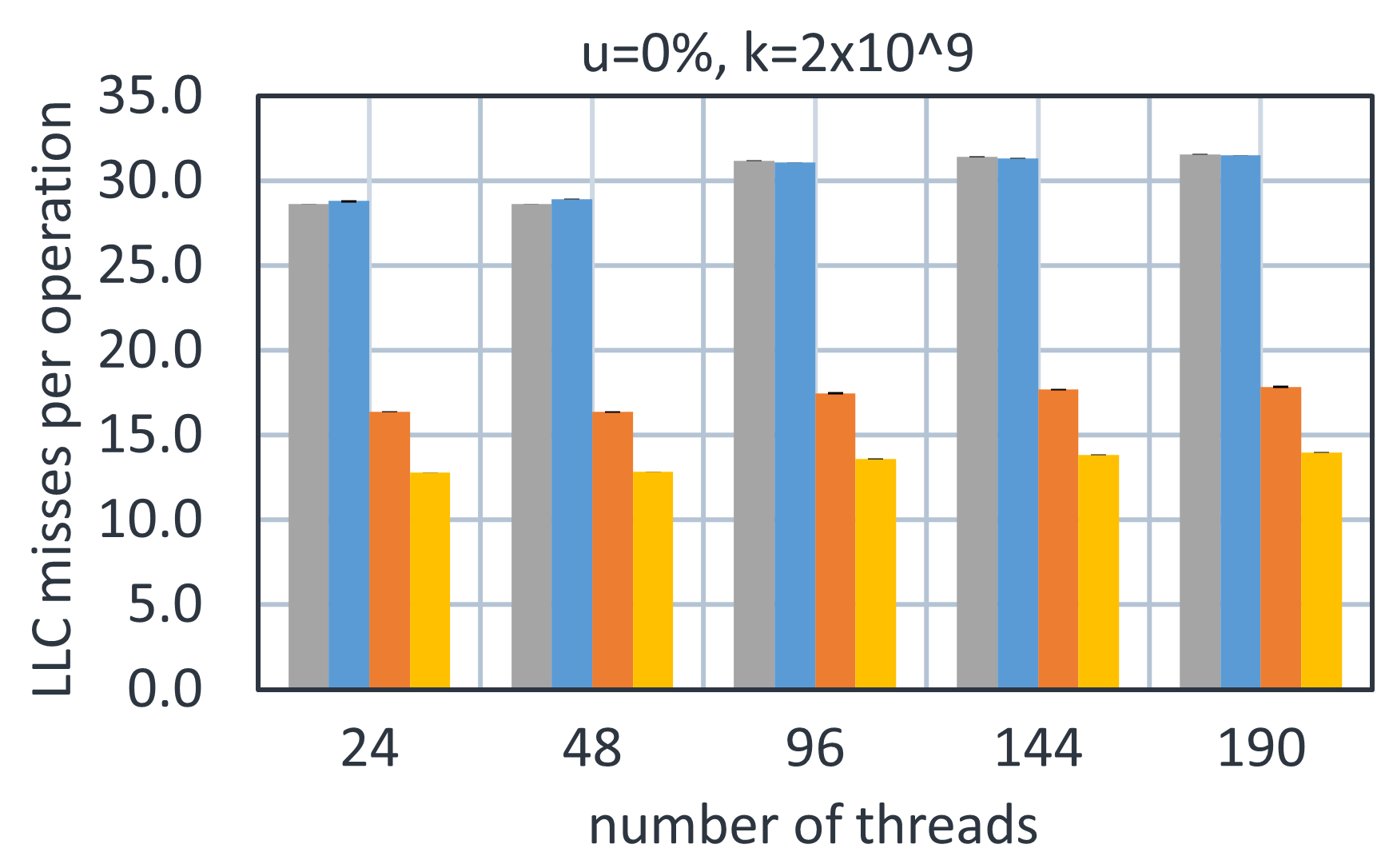}
\includegraphics[scale=0.165]{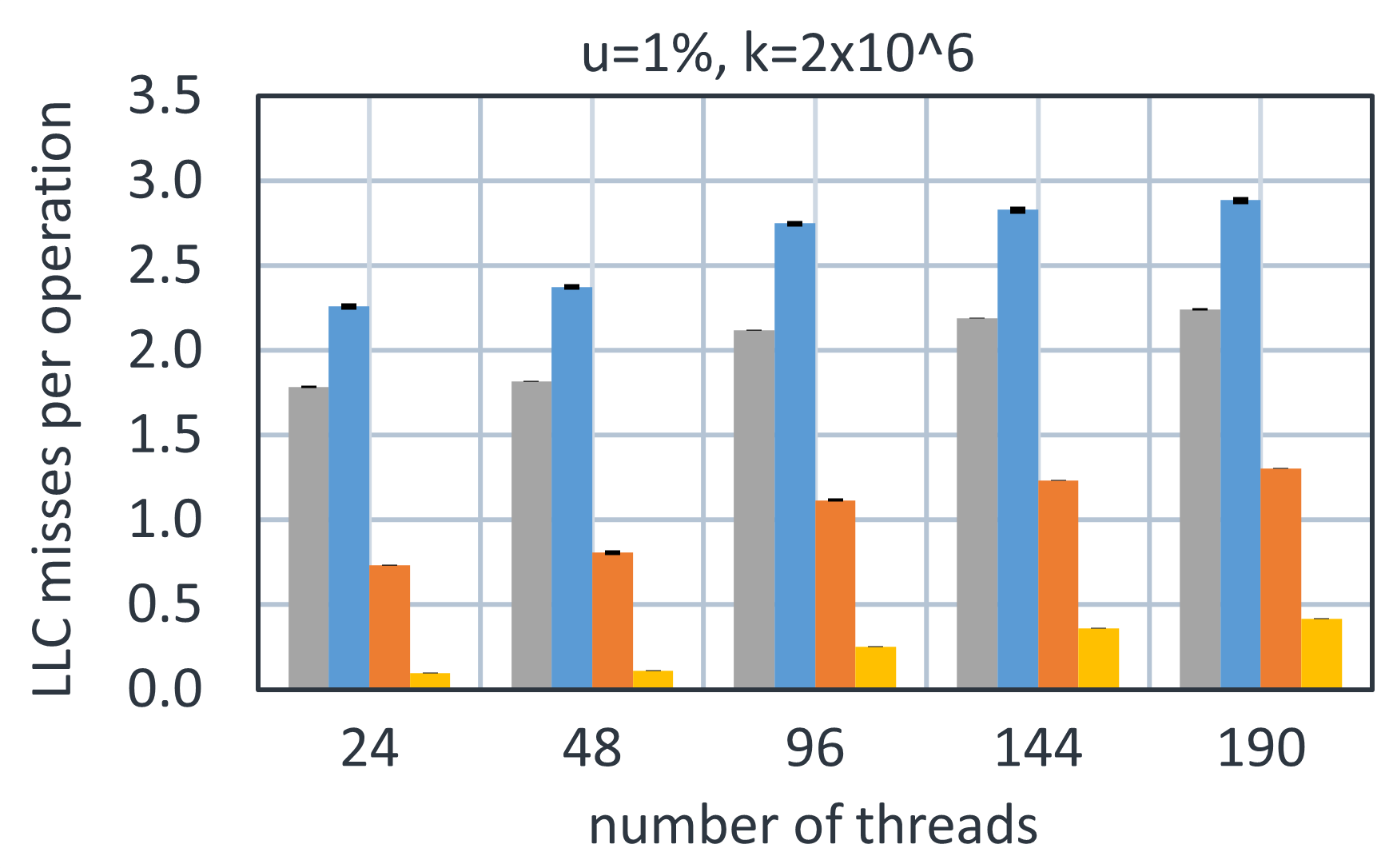}
\includegraphics[scale=0.165]{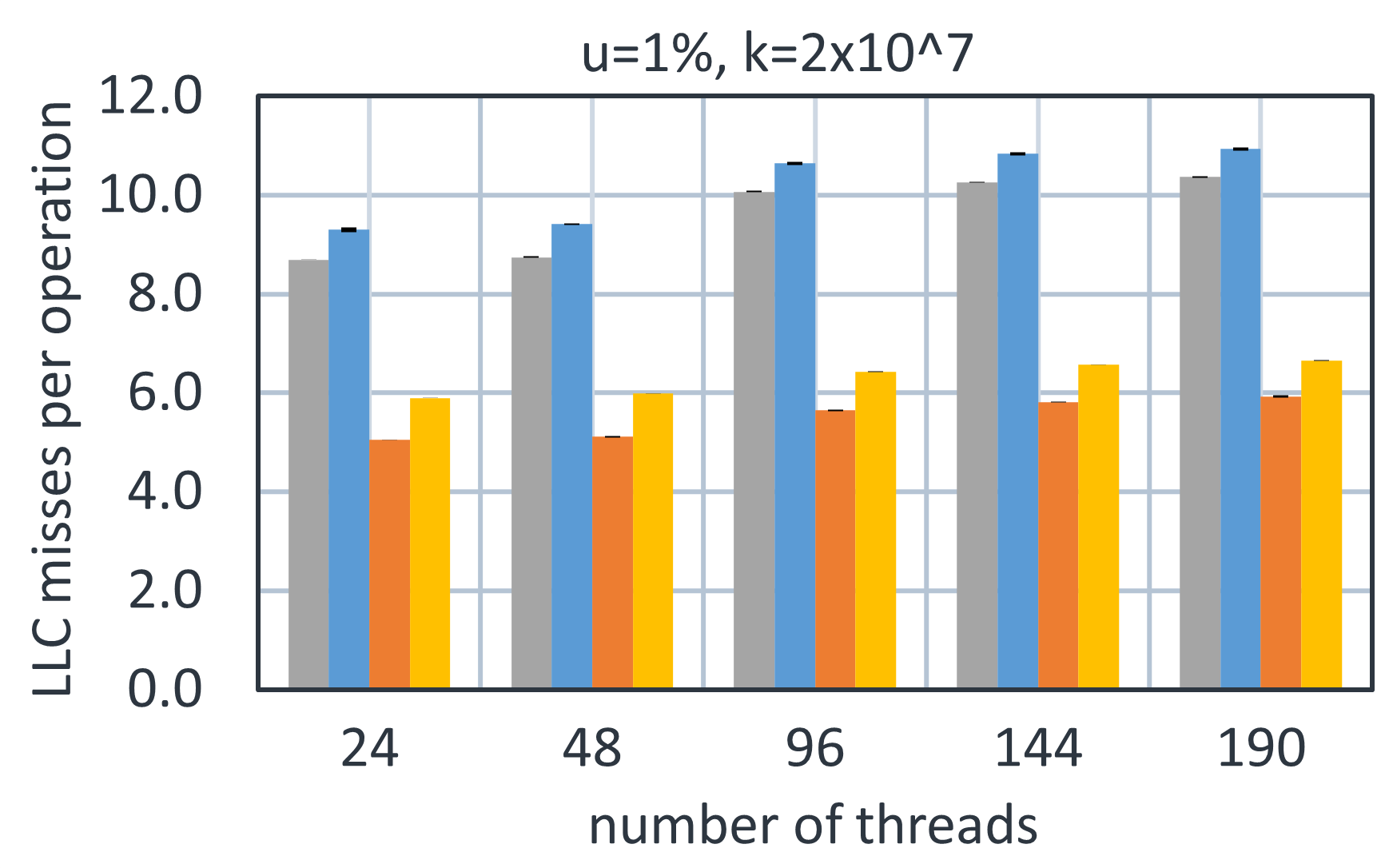}
\includegraphics[scale=0.165]{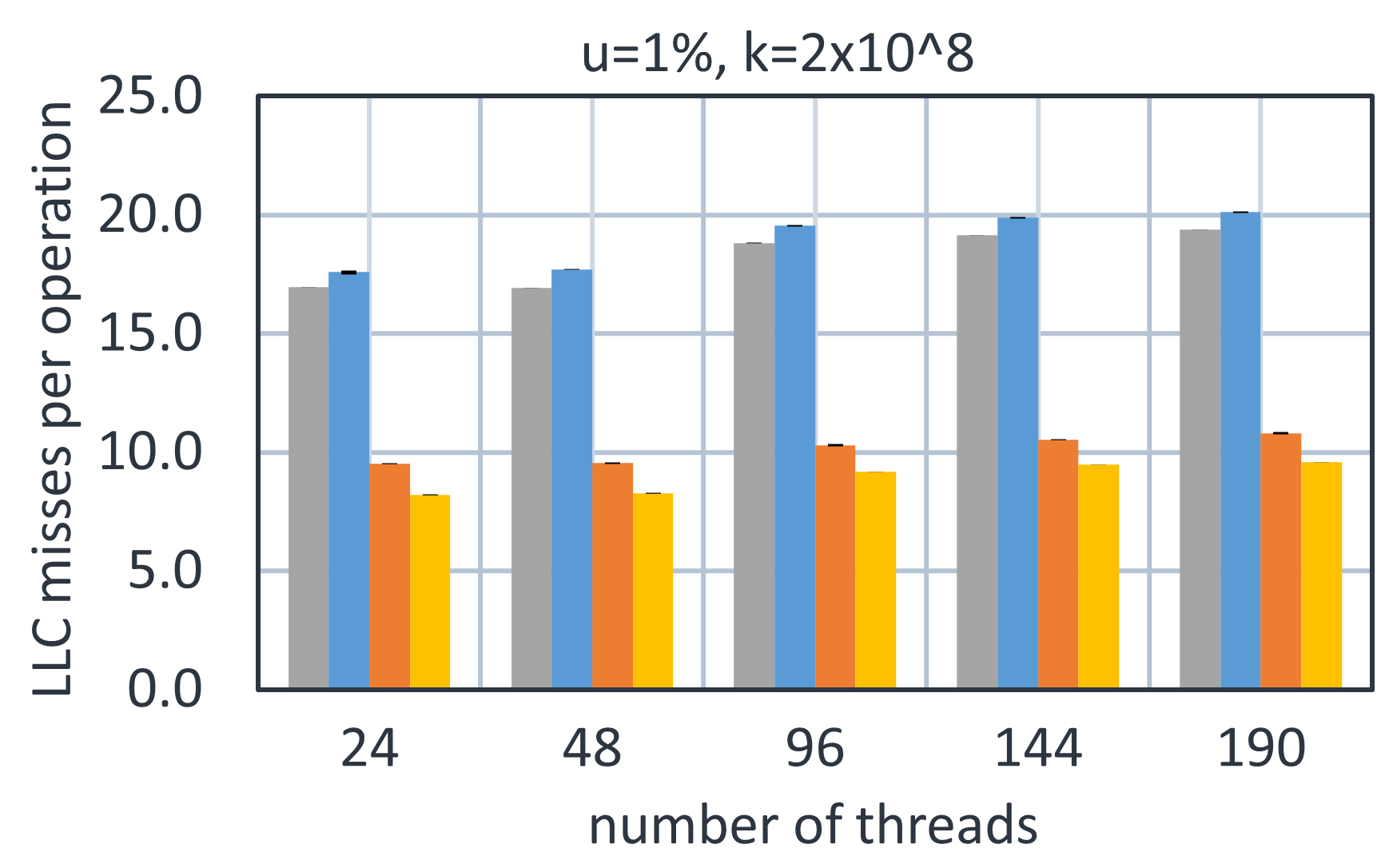}
\includegraphics[scale=0.165]{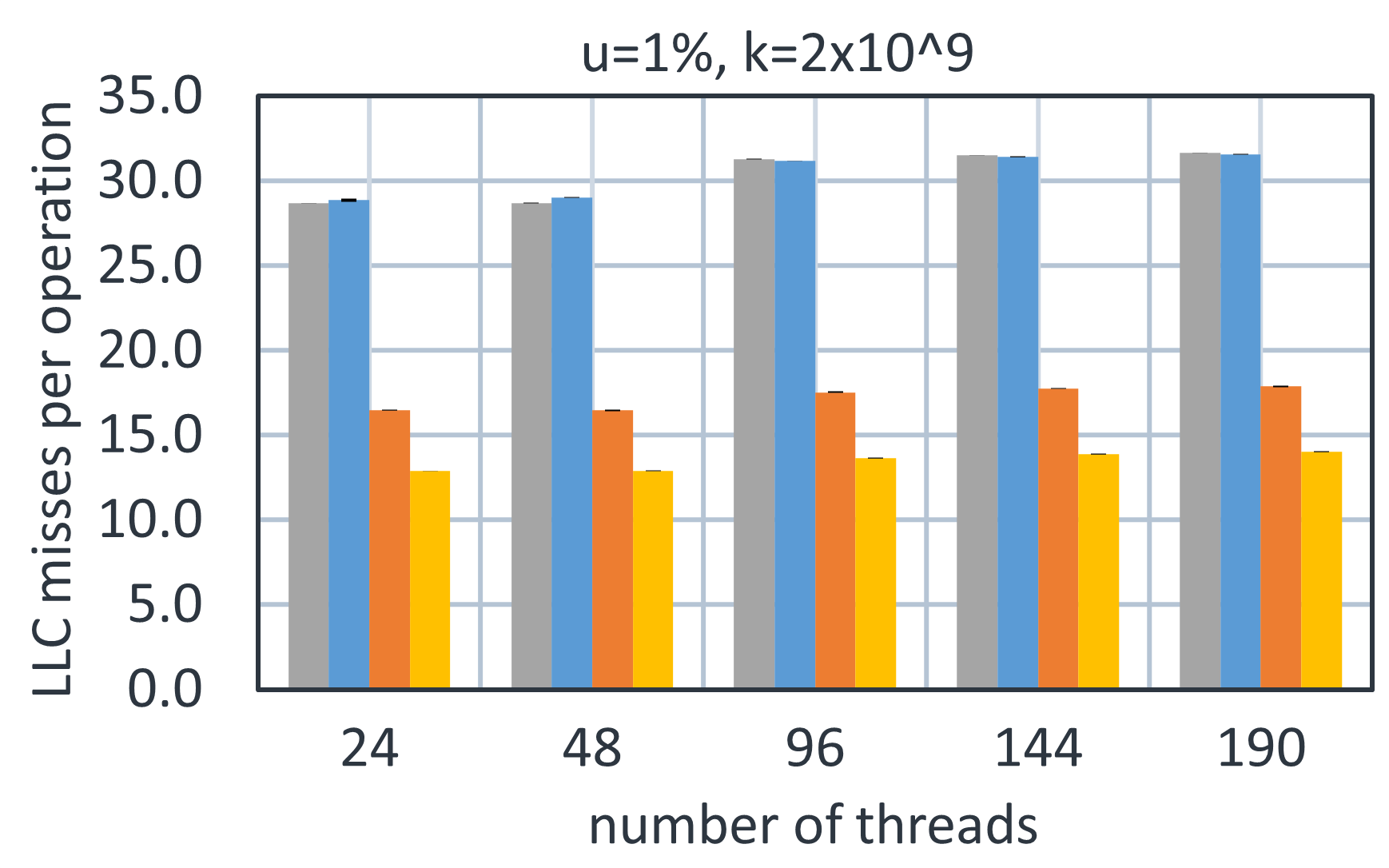}
\includegraphics[scale=0.165]{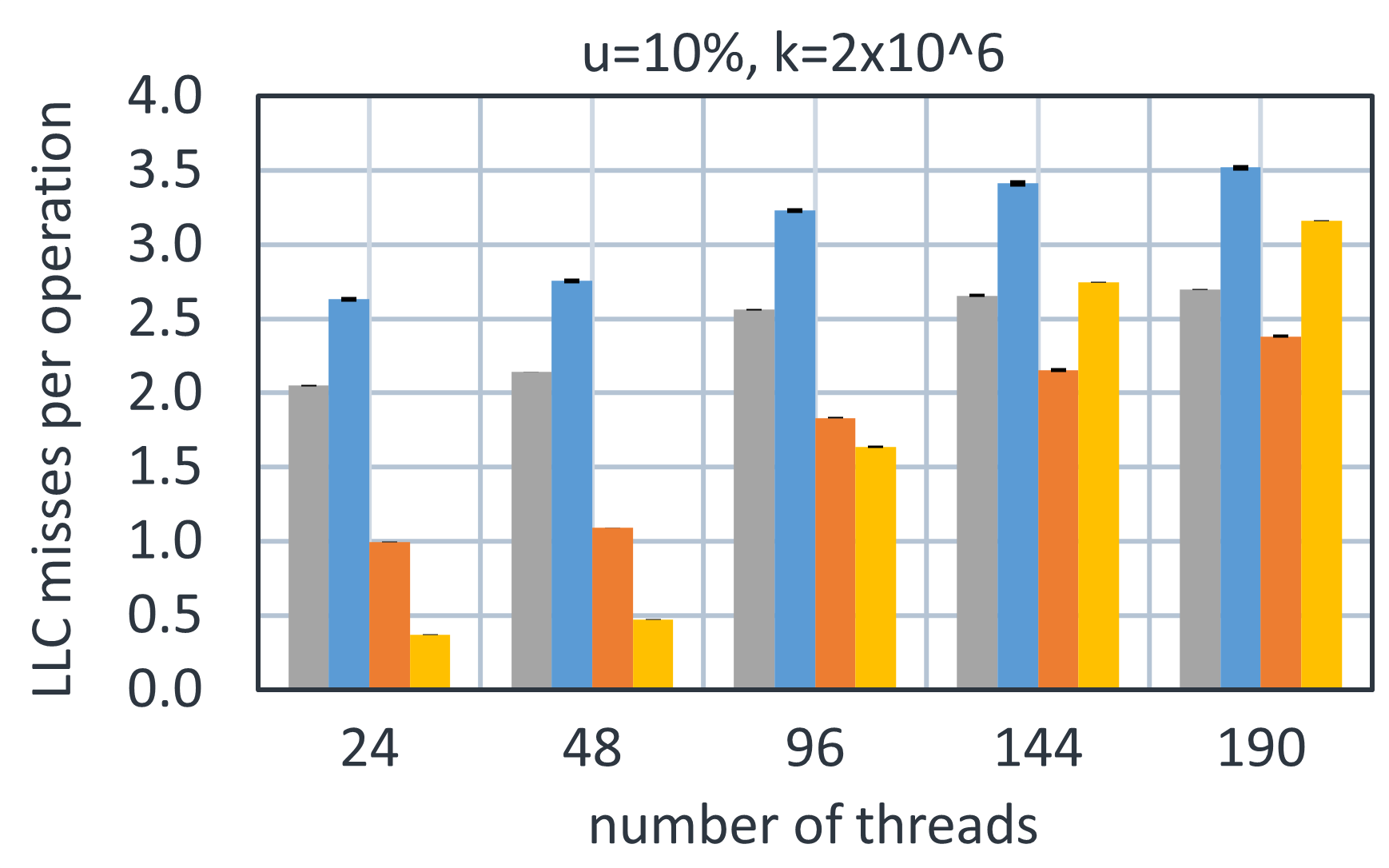}
\includegraphics[scale=0.165]{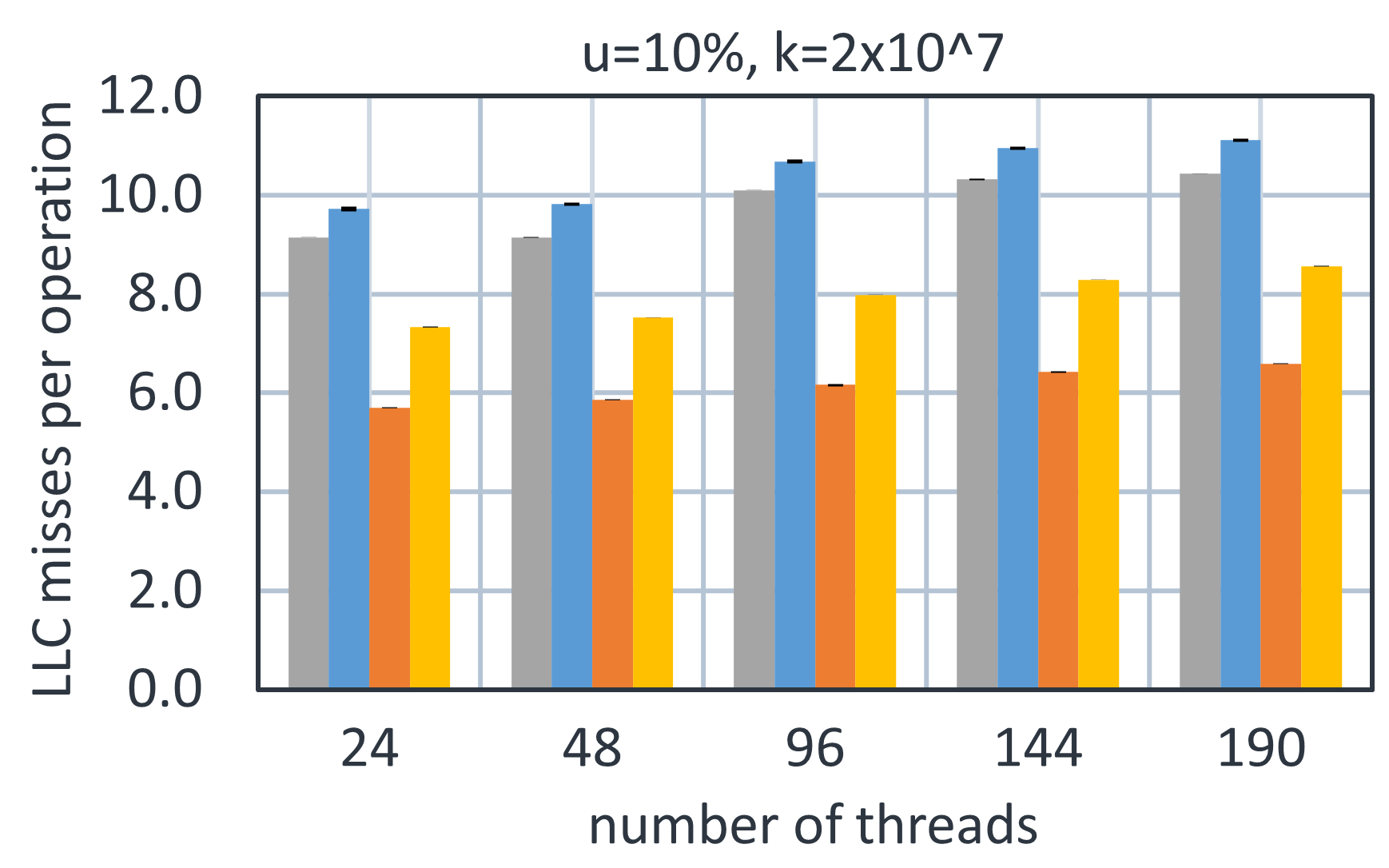}
\includegraphics[scale=0.165]{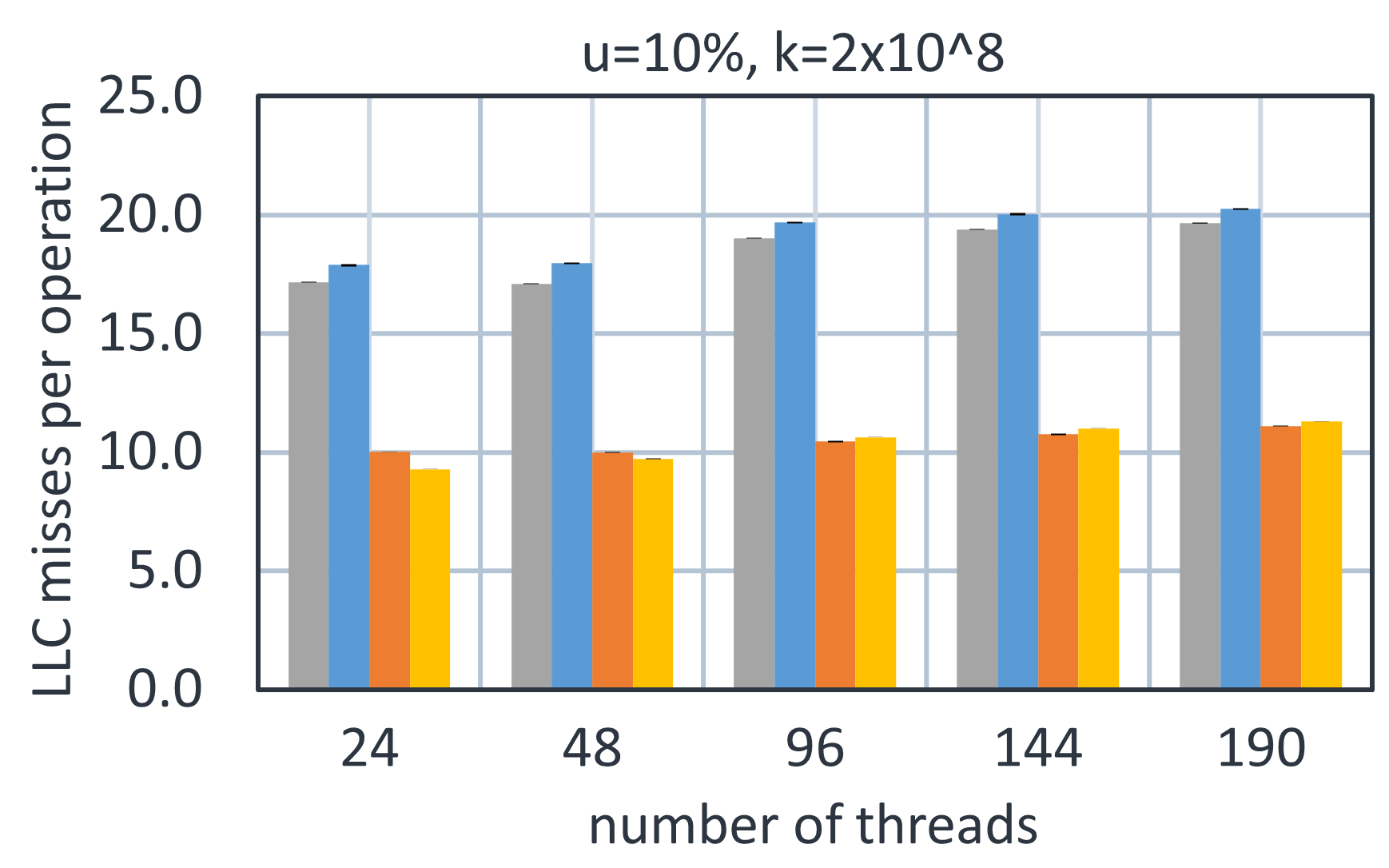}
\includegraphics[scale=0.165]{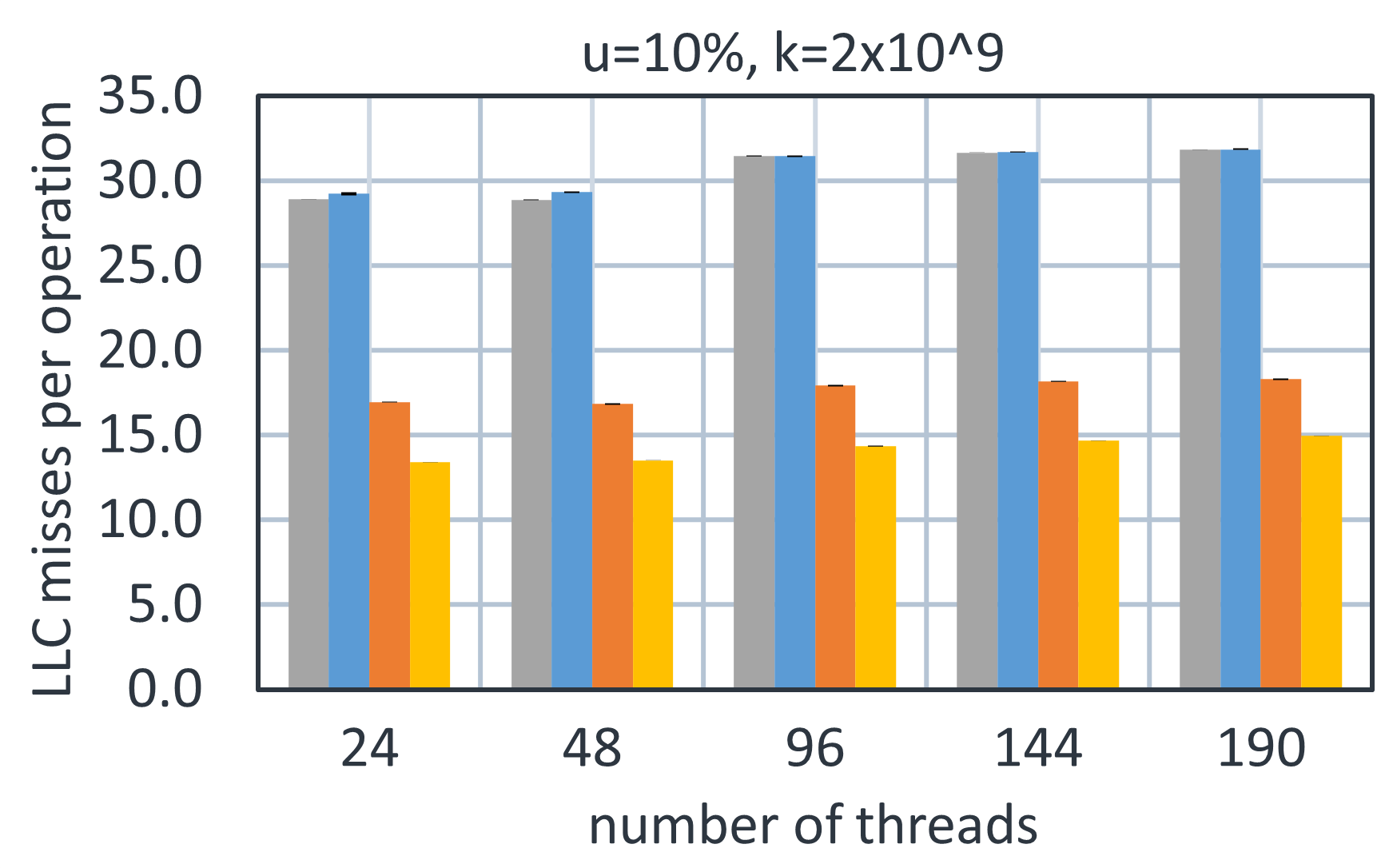}
\includegraphics[scale=0.165]{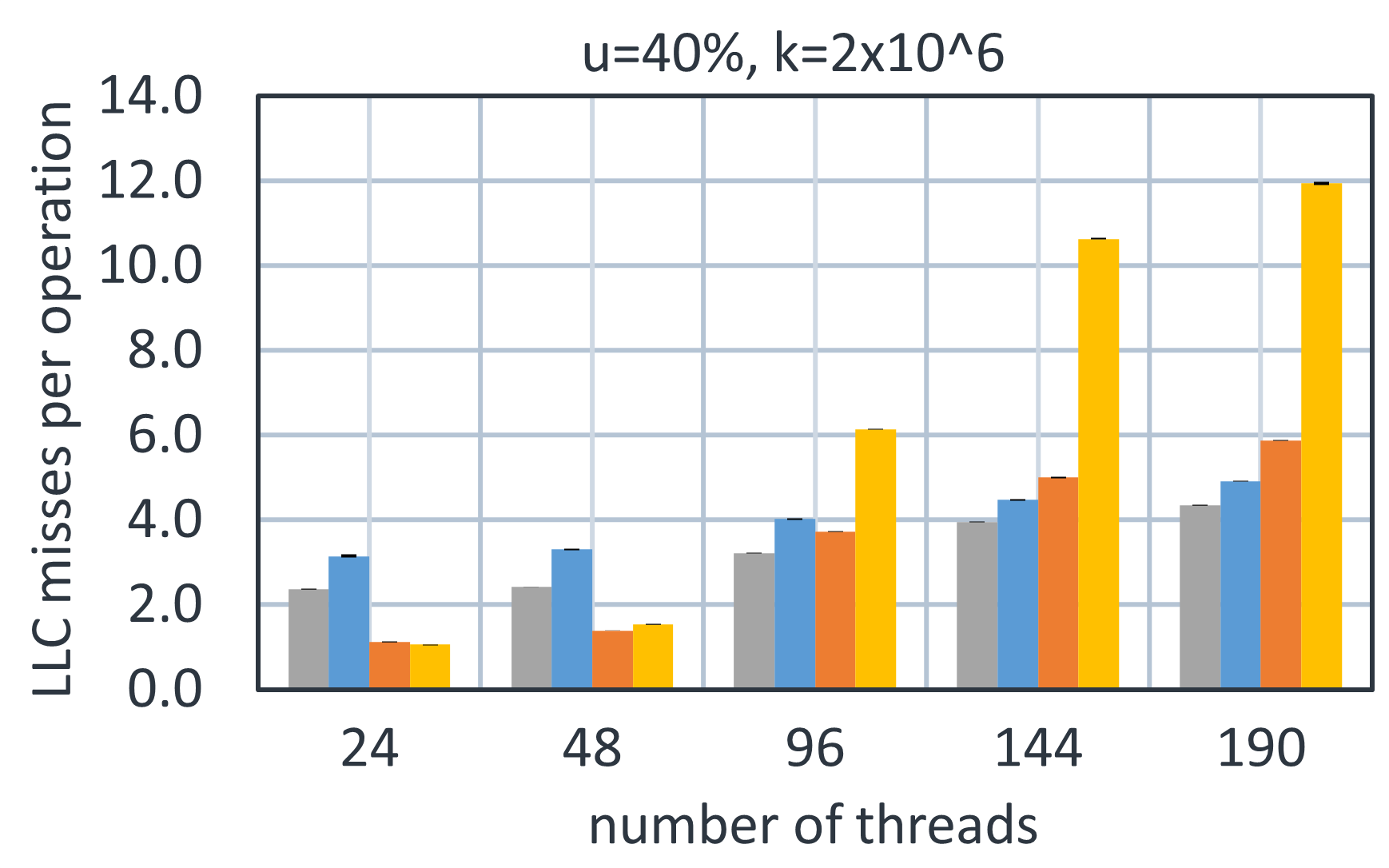}
\includegraphics[scale=0.165]{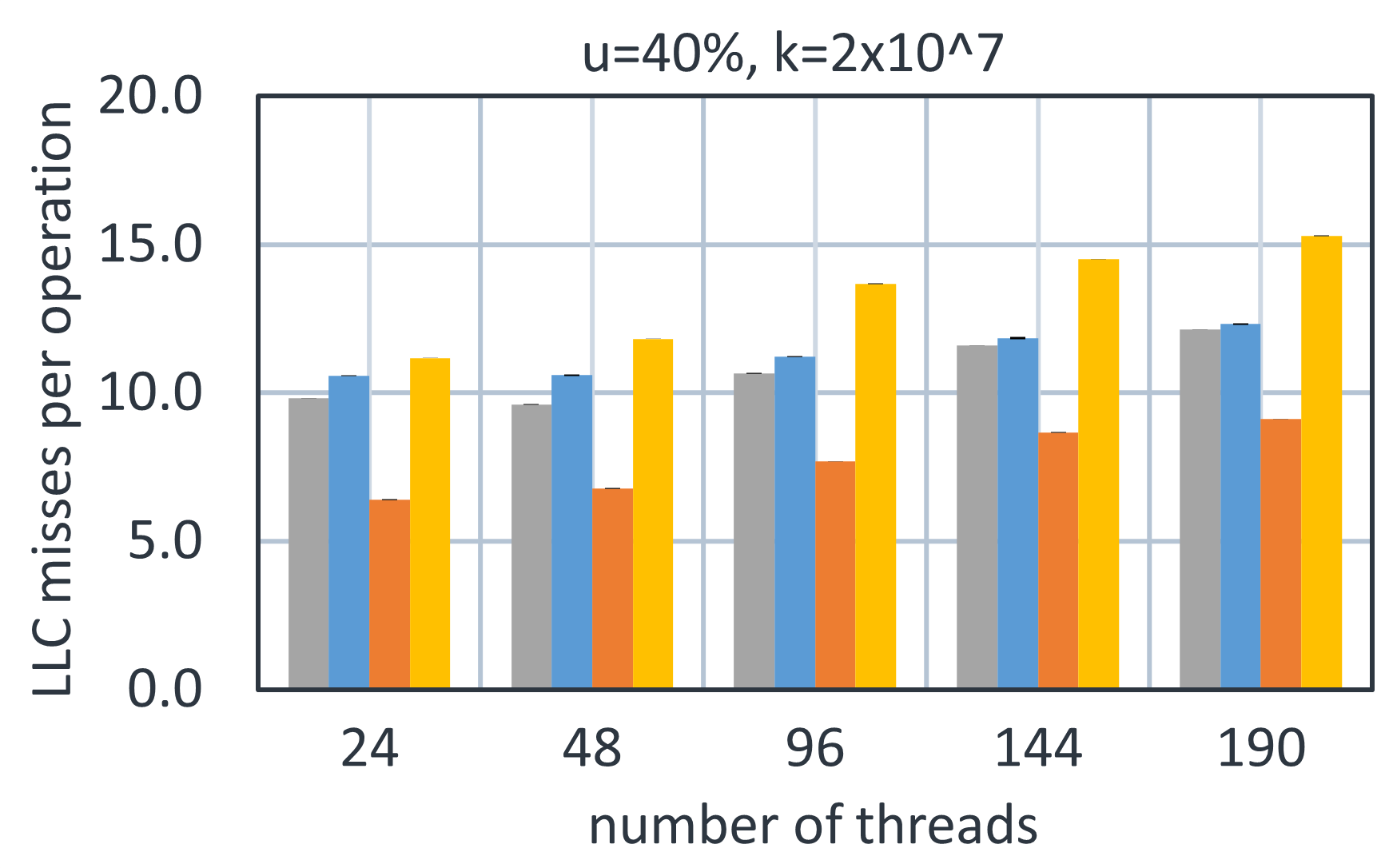}
\includegraphics[scale=0.165]{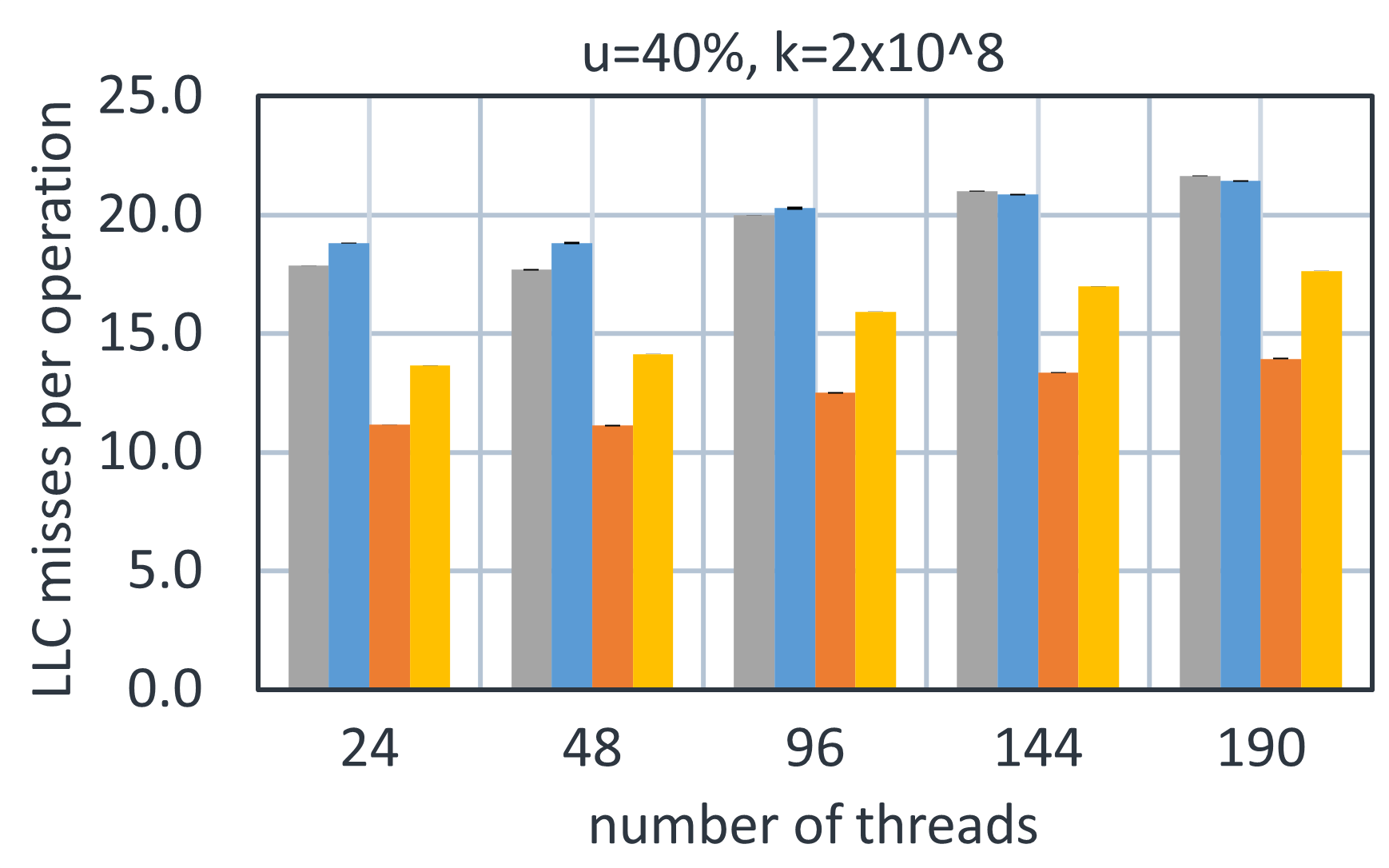}
\includegraphics[scale=0.165]{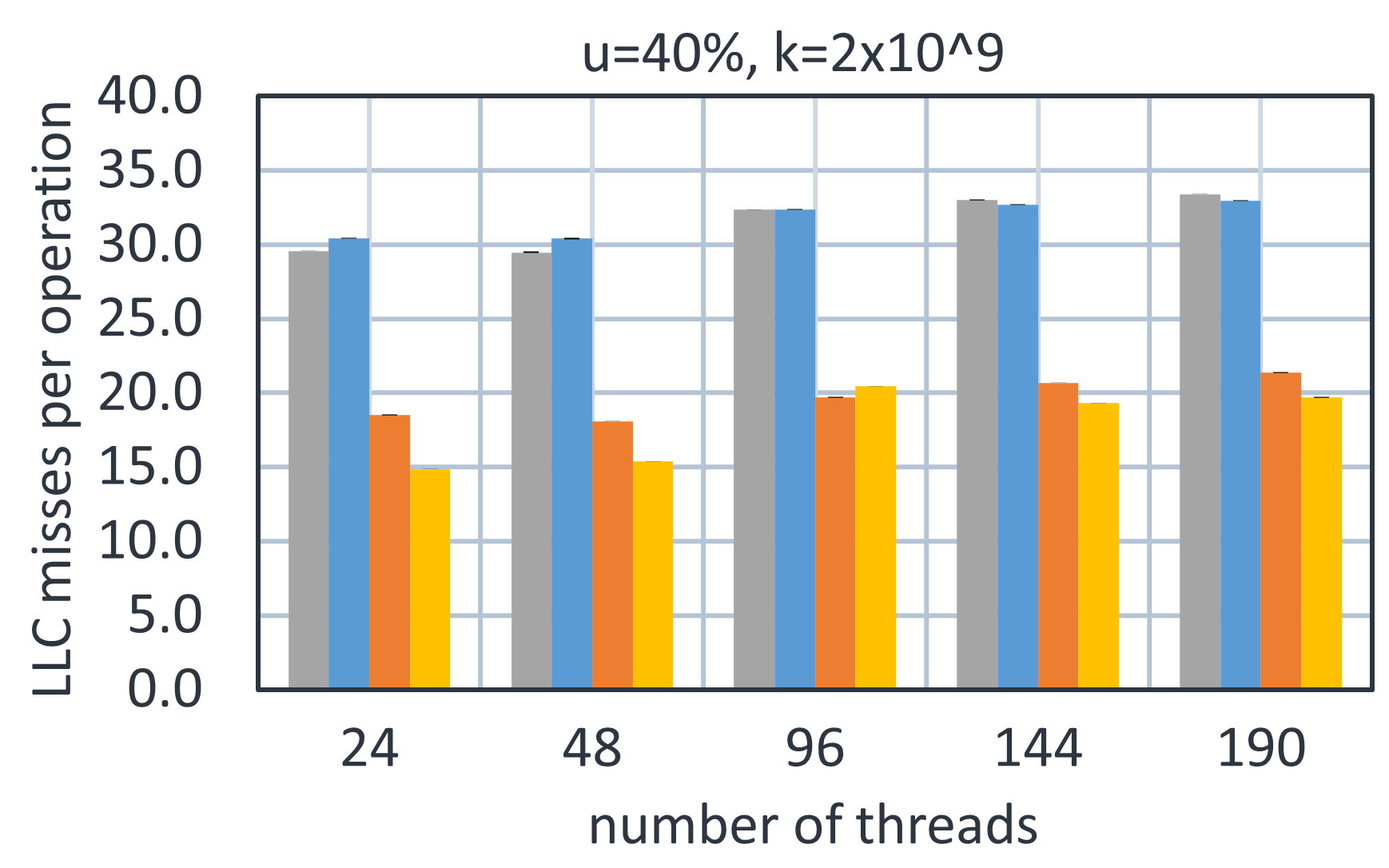}

\includegraphics[scale=0.25]{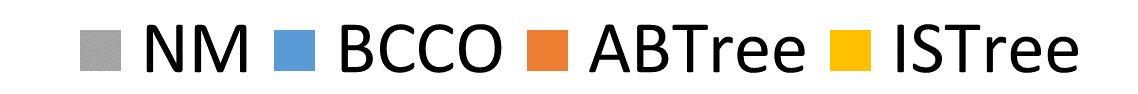}
\caption{Summary of Last-Level Cache Misses, Lower is Better}
\label{fig:appendix:experiments:llc}
\end{figure*}

\paratitle{LLC misses}
Figure~\ref{fig:appendix:experiments:llc}
shows the average count of last-level cache-misses
for different update percentages ($0\%$, $1\%$, $10\%$ and $40\%$),
and dataset sizes ranging from $10^6$ to $10^9$.

\begin{figure*}[ht]
\includegraphics[scale=0.20]{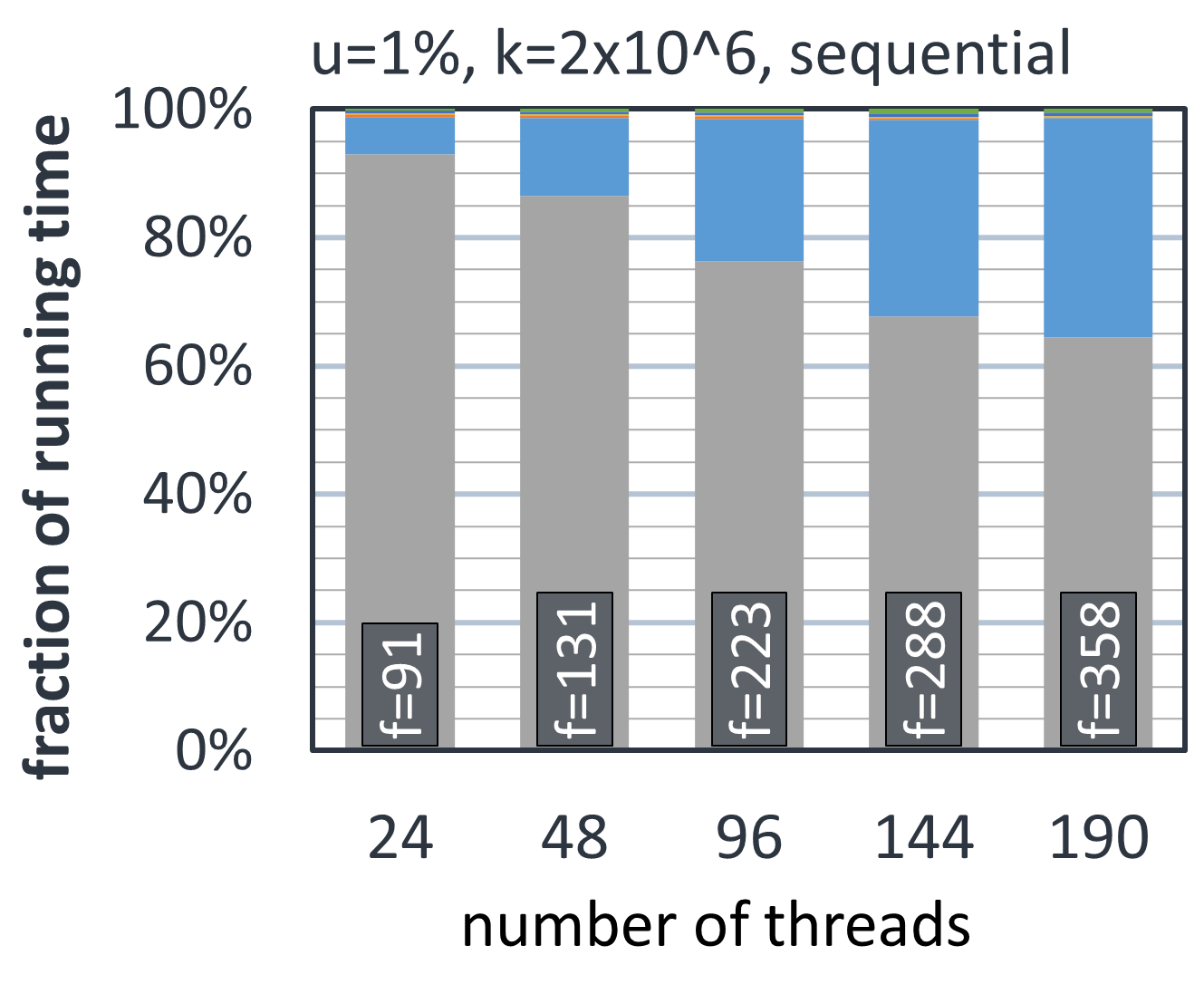}
\includegraphics[scale=0.20]{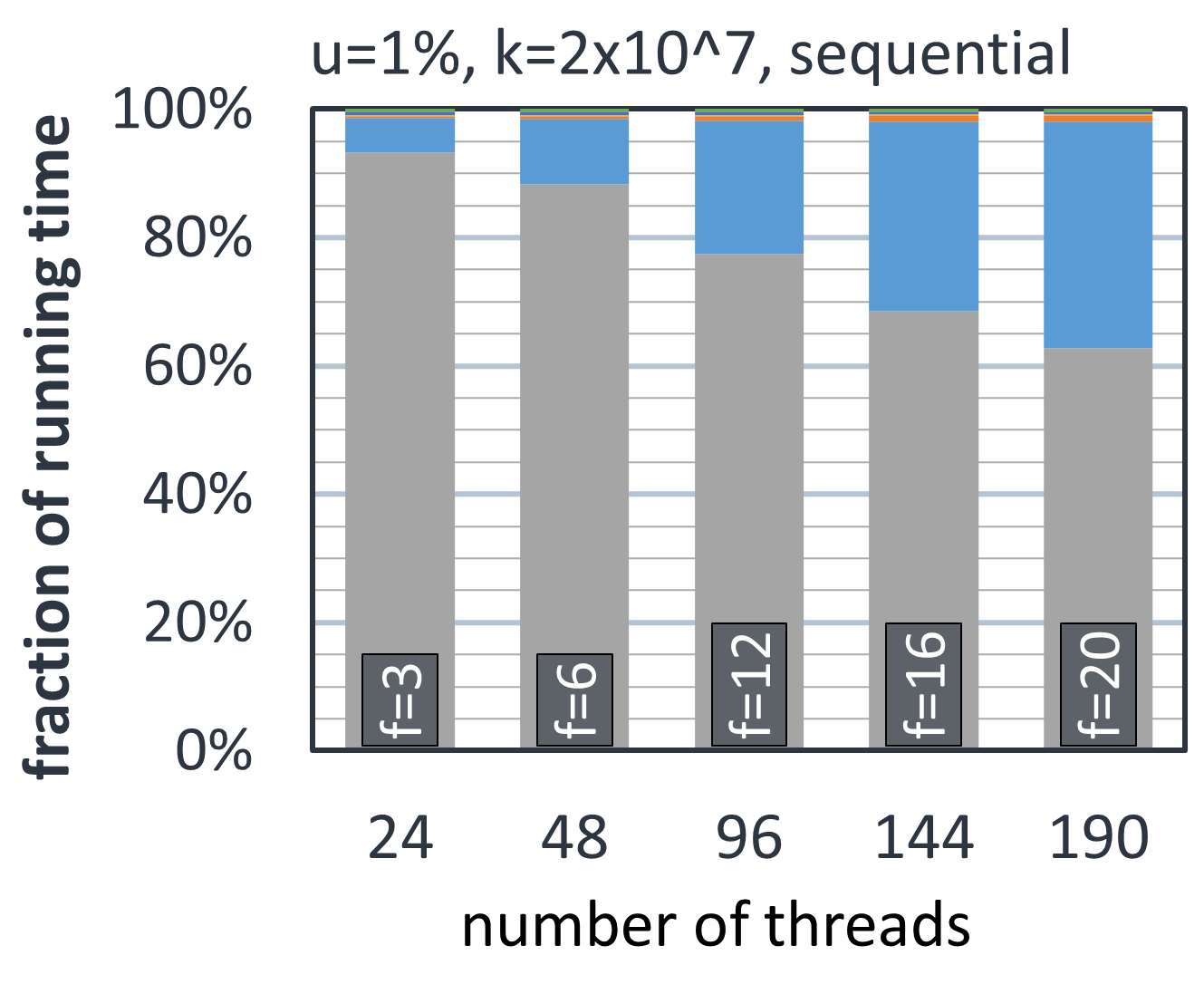}
\includegraphics[scale=0.20]{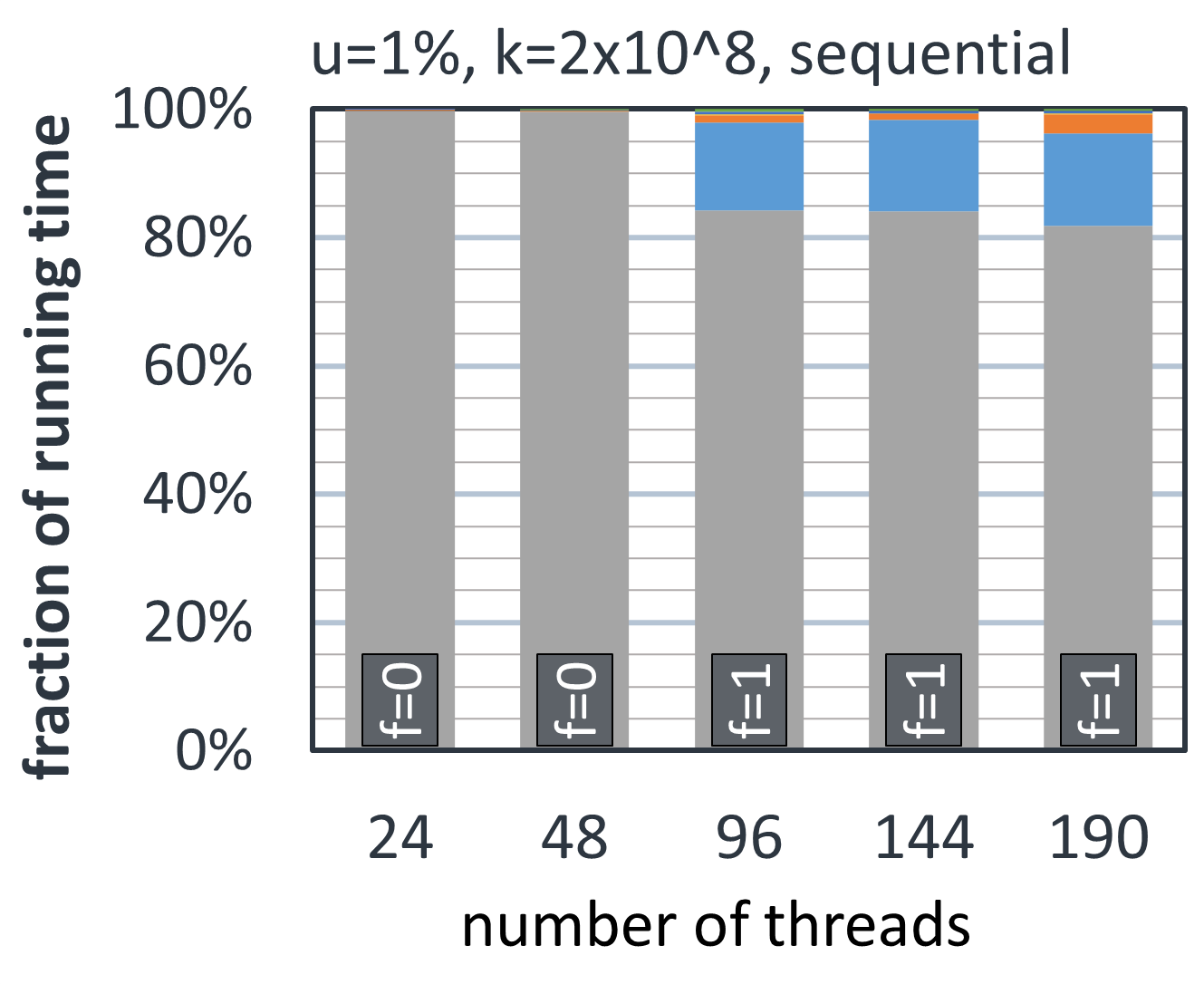}
\includegraphics[scale=0.20]{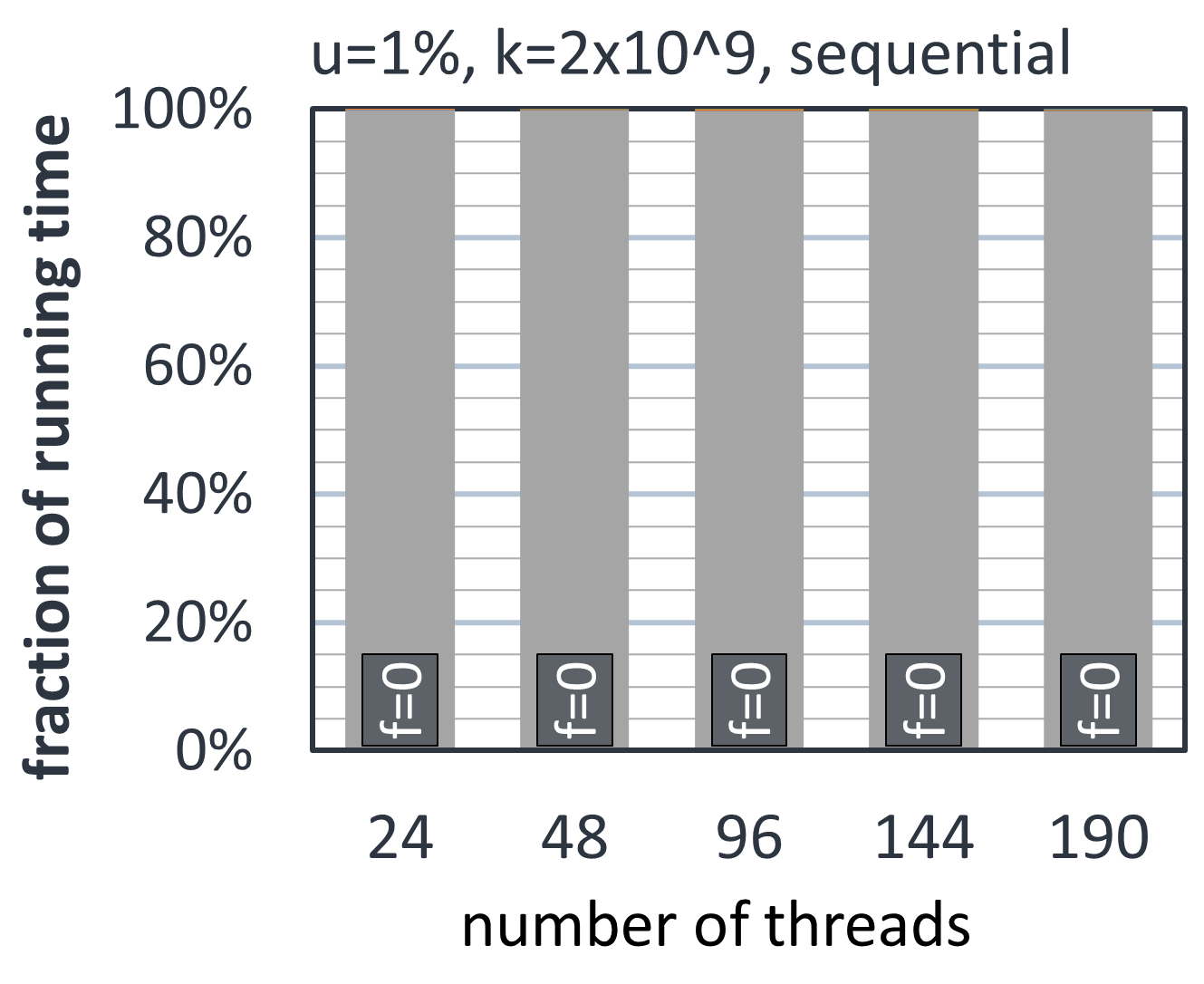}
\includegraphics[scale=0.20]{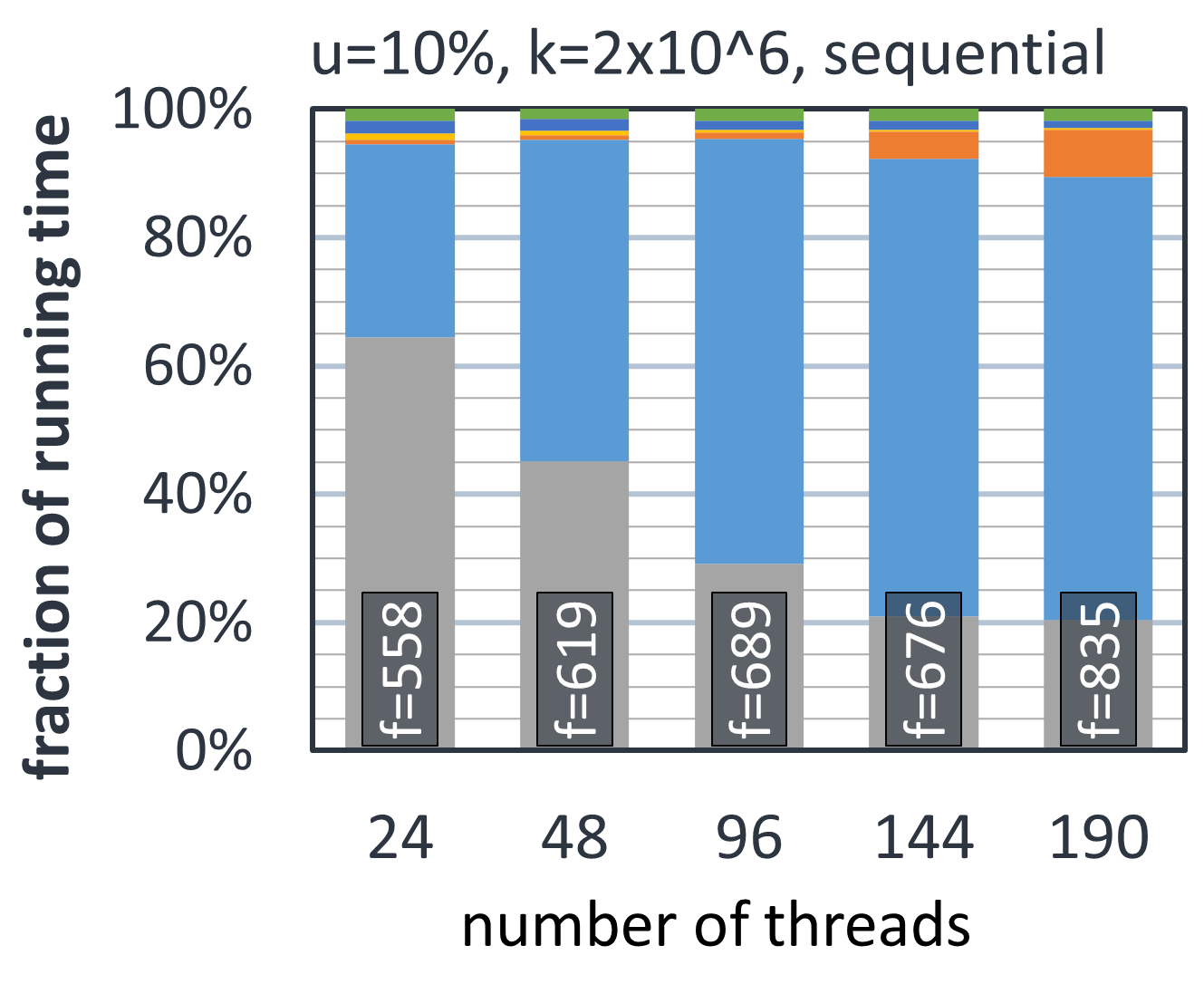}
\includegraphics[scale=0.20]{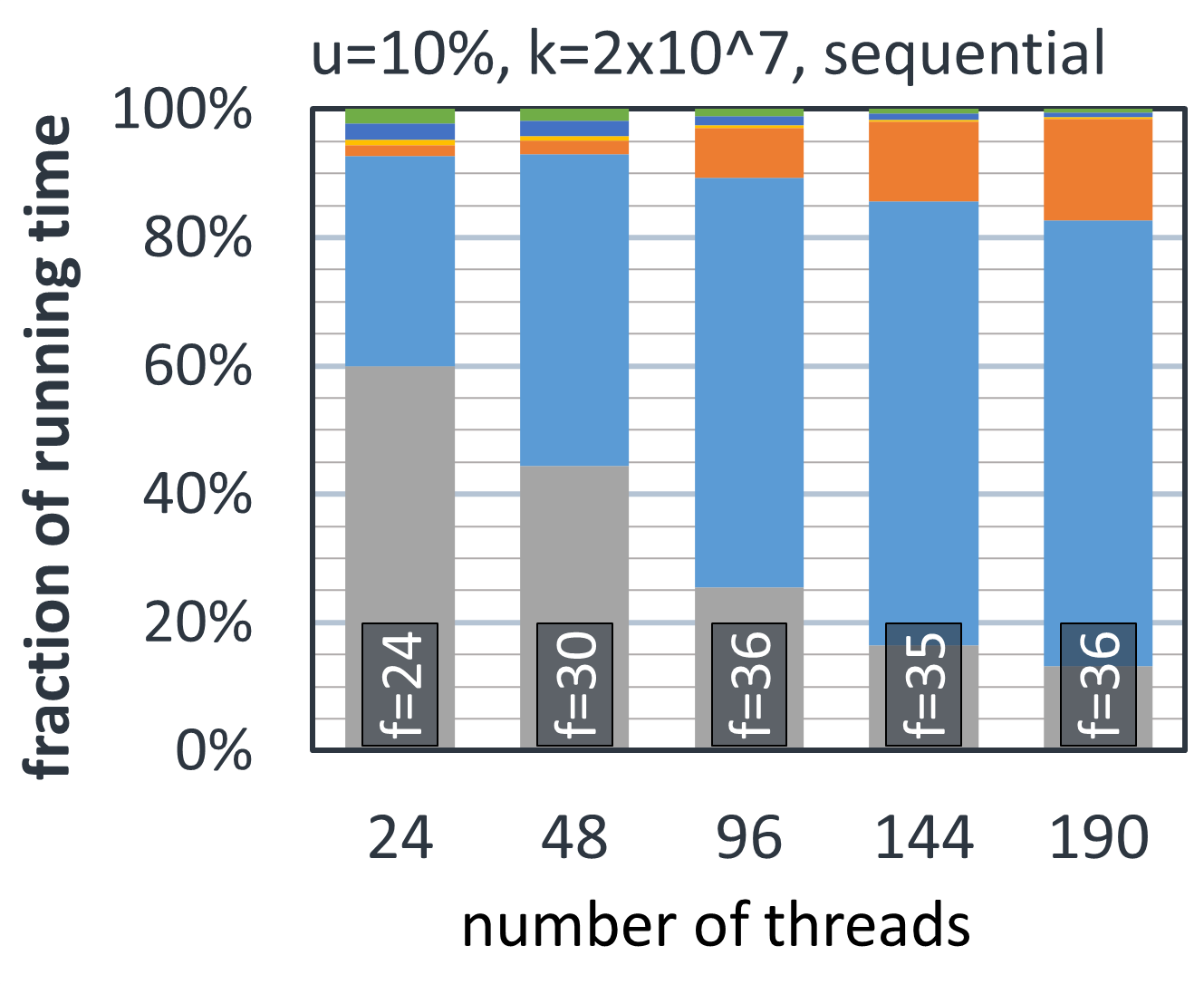}
\includegraphics[scale=0.20]{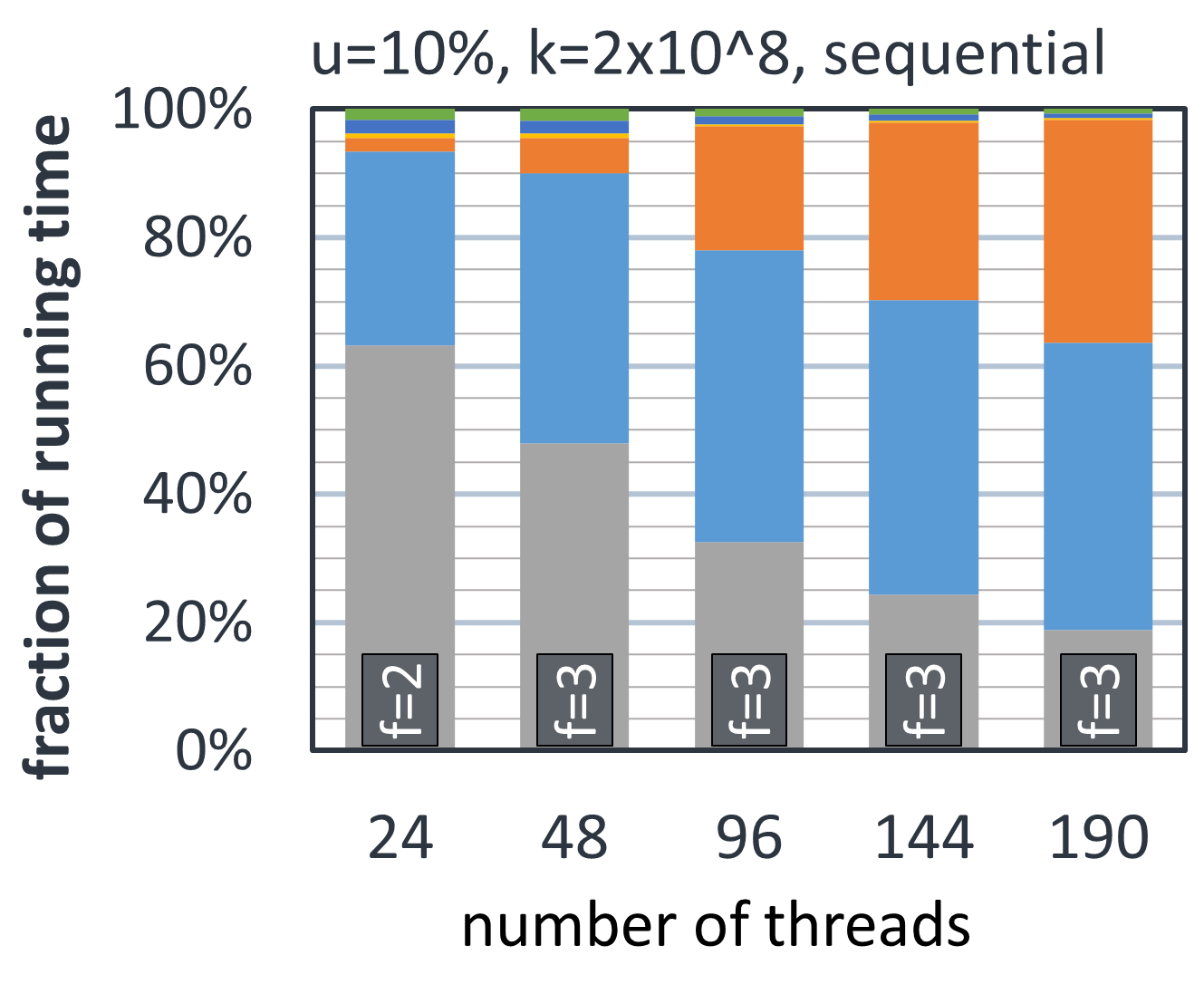}
\includegraphics[scale=0.20]{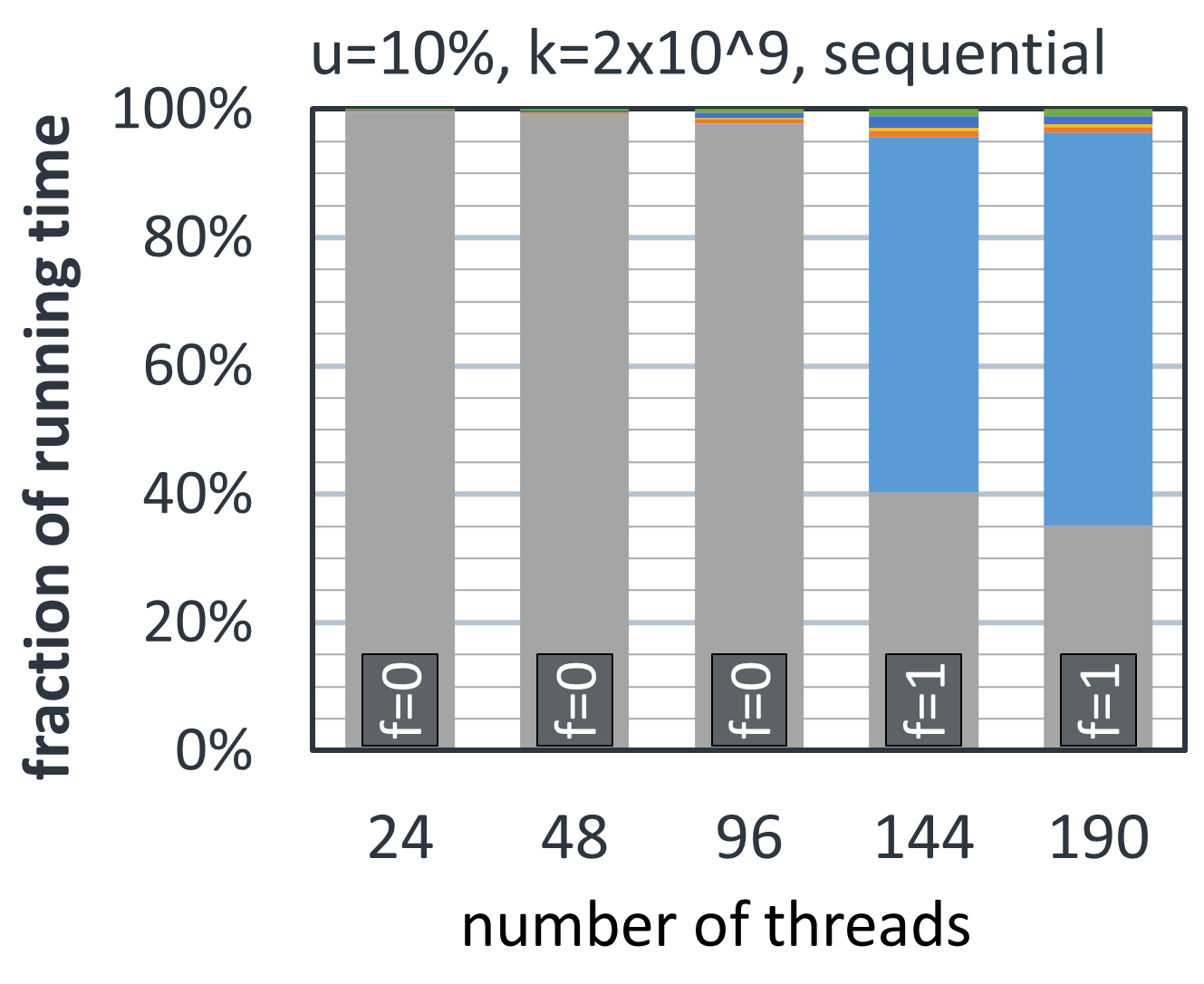}
\includegraphics[scale=0.20]{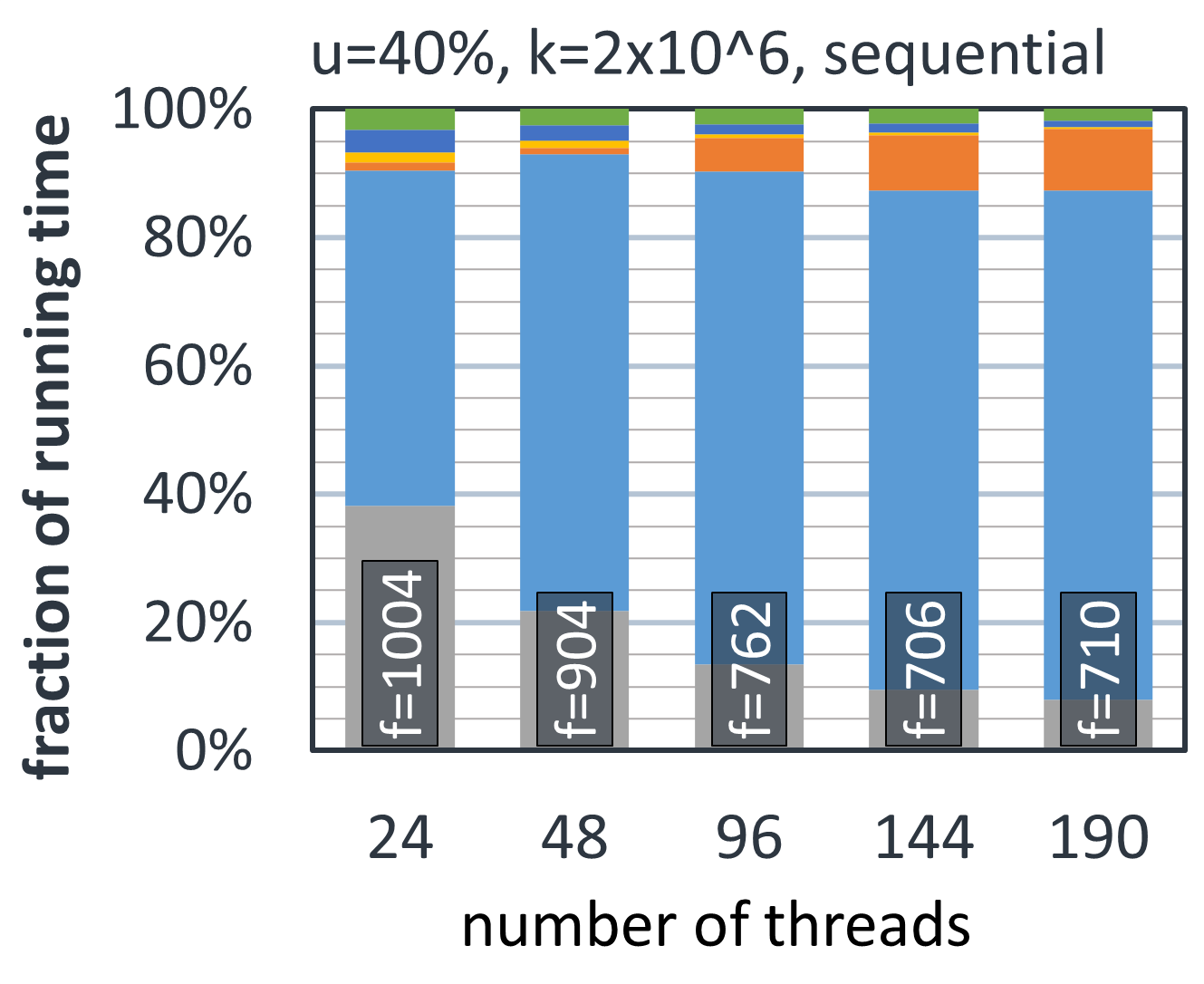}
\includegraphics[scale=0.20]{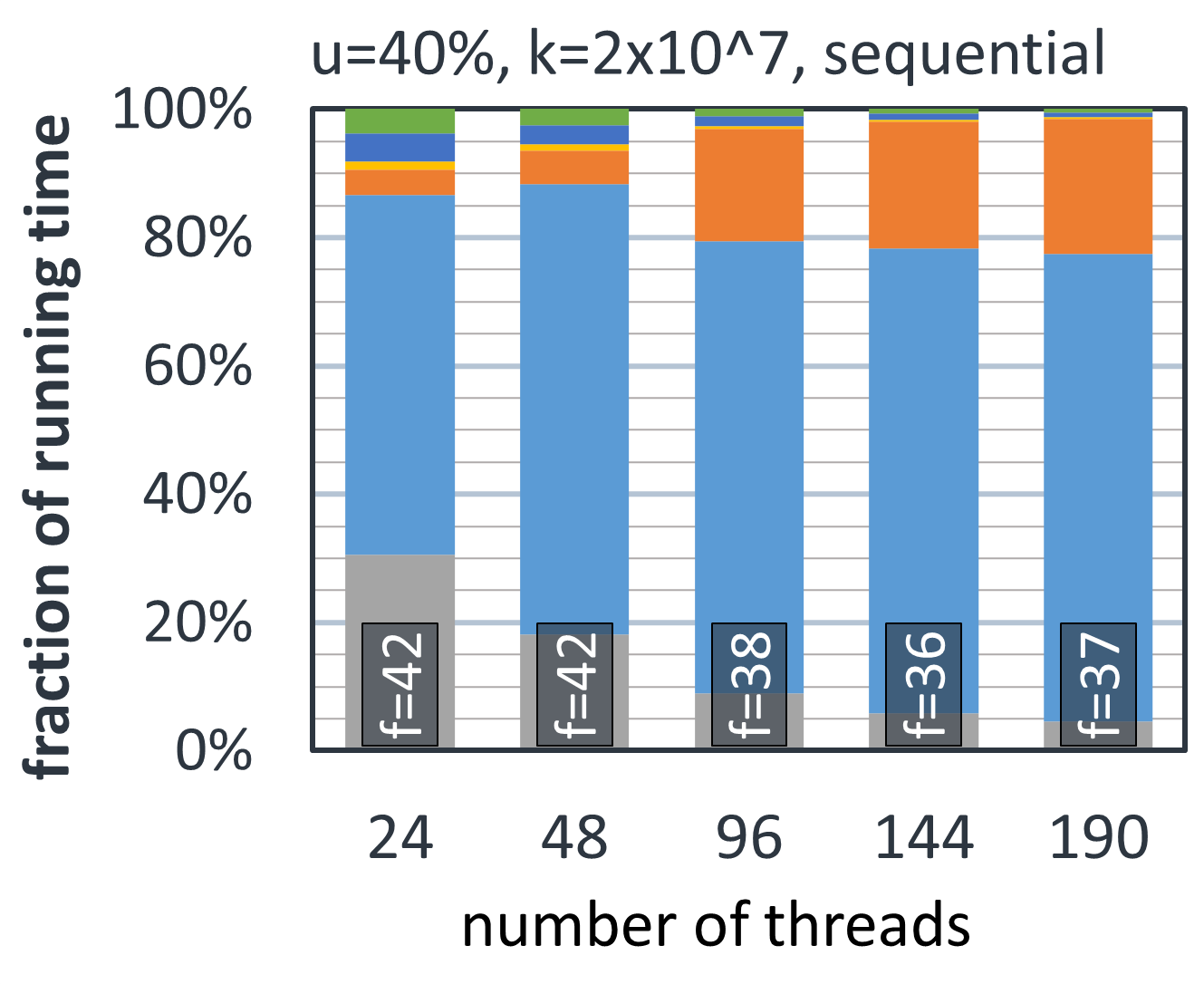}
\includegraphics[scale=0.20]{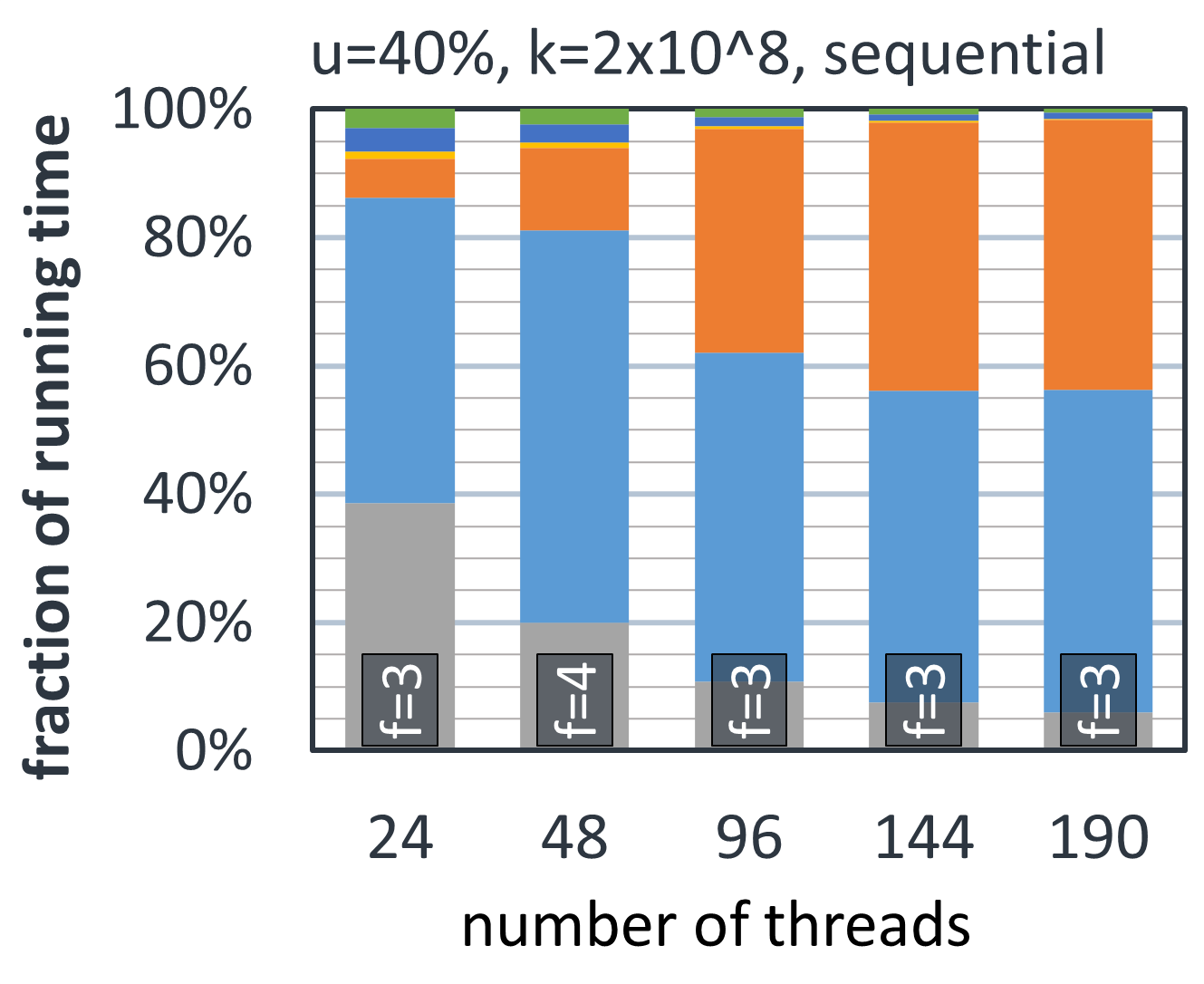}
\includegraphics[scale=0.20]{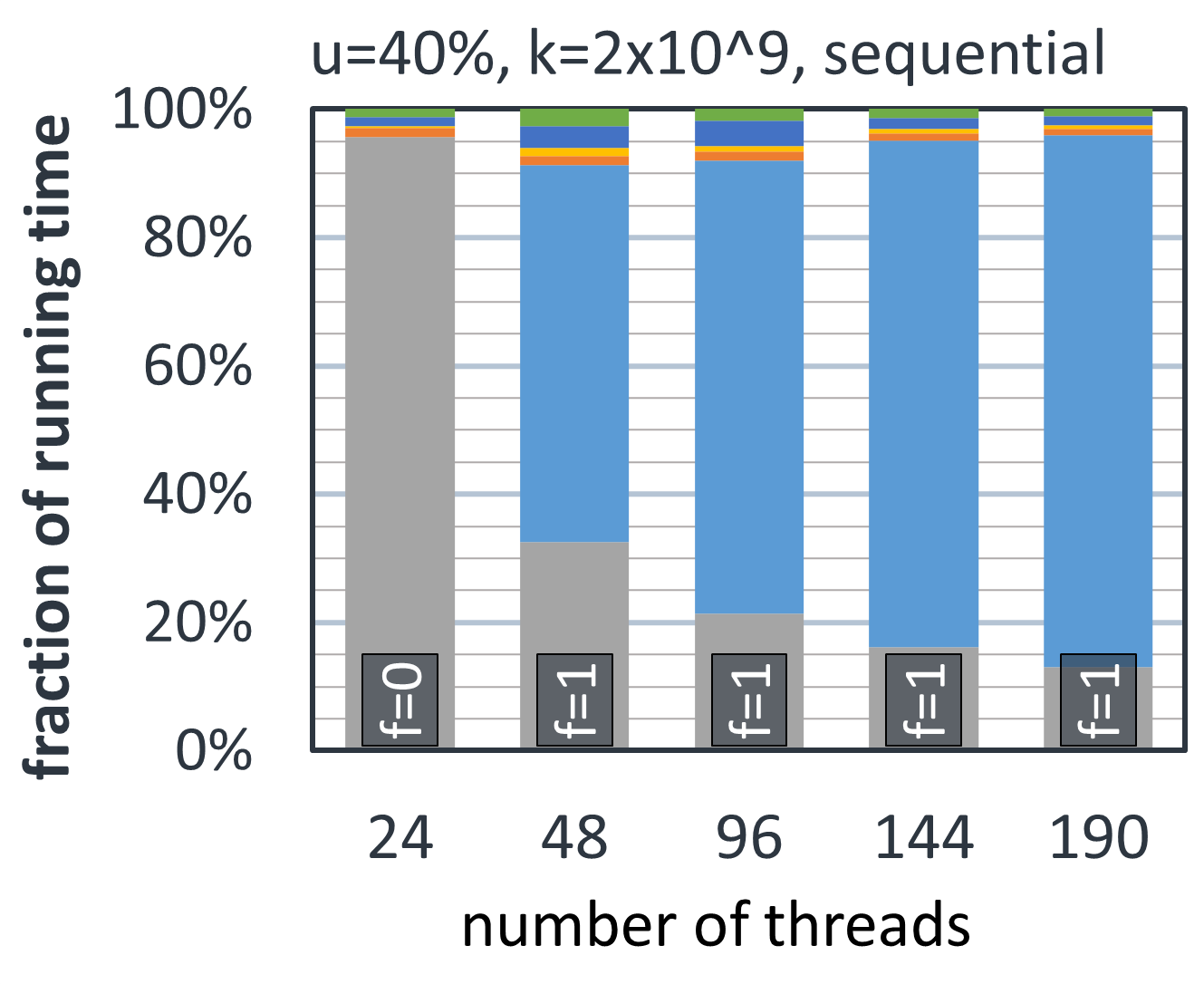}

\caption{
    Breakdown of Time Spent, when Using Non-Collaborative Rebuilding
    (\crule[DarkBluish]{2mm}{2mm}~creating,
    \crule[Greenish]{2mm}{2mm}~marking,
    \crule[Bluish]{2mm}{2mm}~useless helping,
    \crule[BloodOrange]{2mm}{2mm}~deallocation,
    \crule[SunsetYellow]{2mm}{2mm}~locating garbage,
    \crule[Gray]{2mm}{2mm}~other).
    Bars are annotated with $f$, the number of times the \textit{root} (entire tree) was rebuilt.
}
\label{fig:appendix:experiments:non-collaborative-rebuilding}
\end{figure*}

\begin{figure*}[ht]
\includegraphics[scale=0.20]{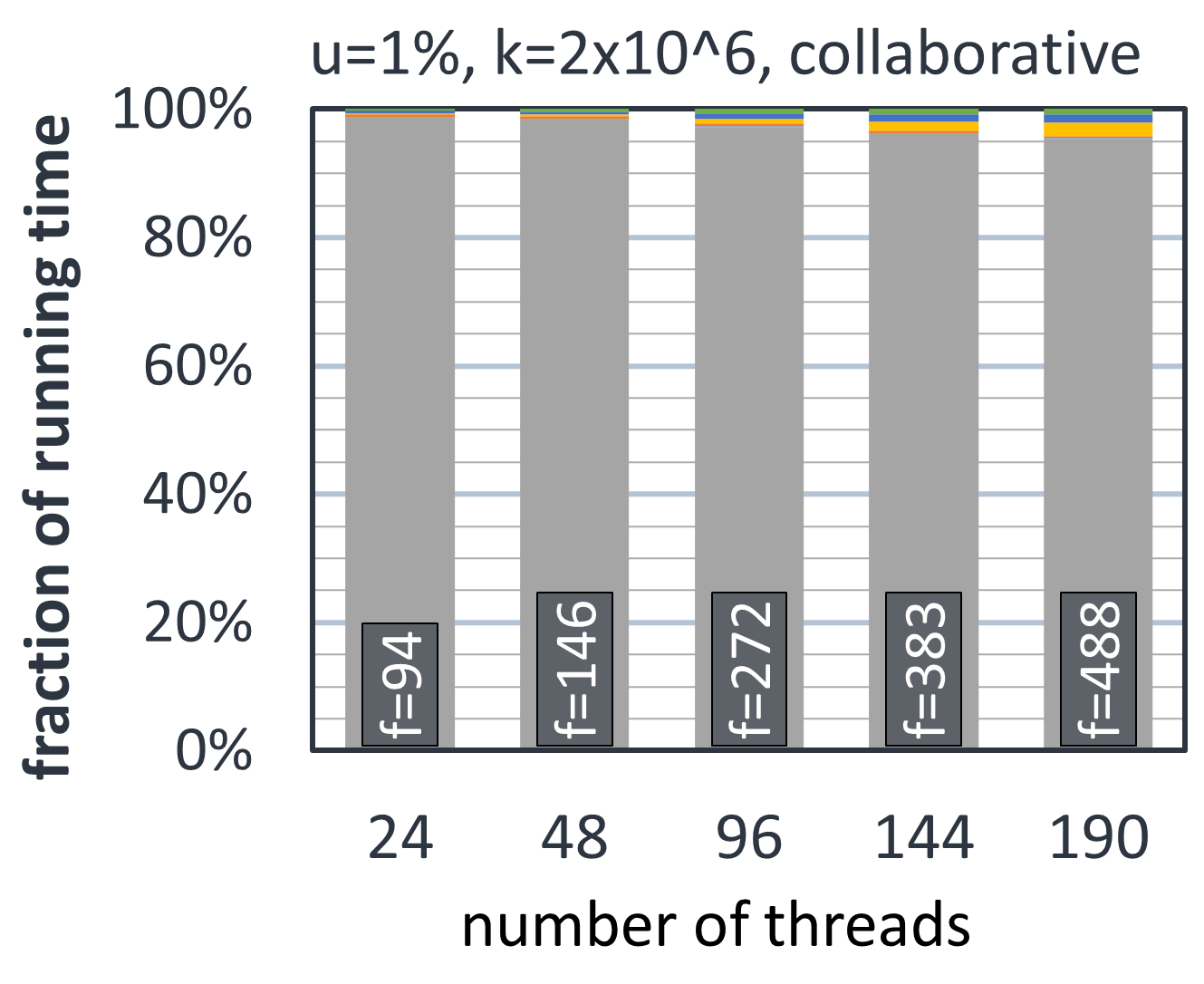}
\includegraphics[scale=0.20]{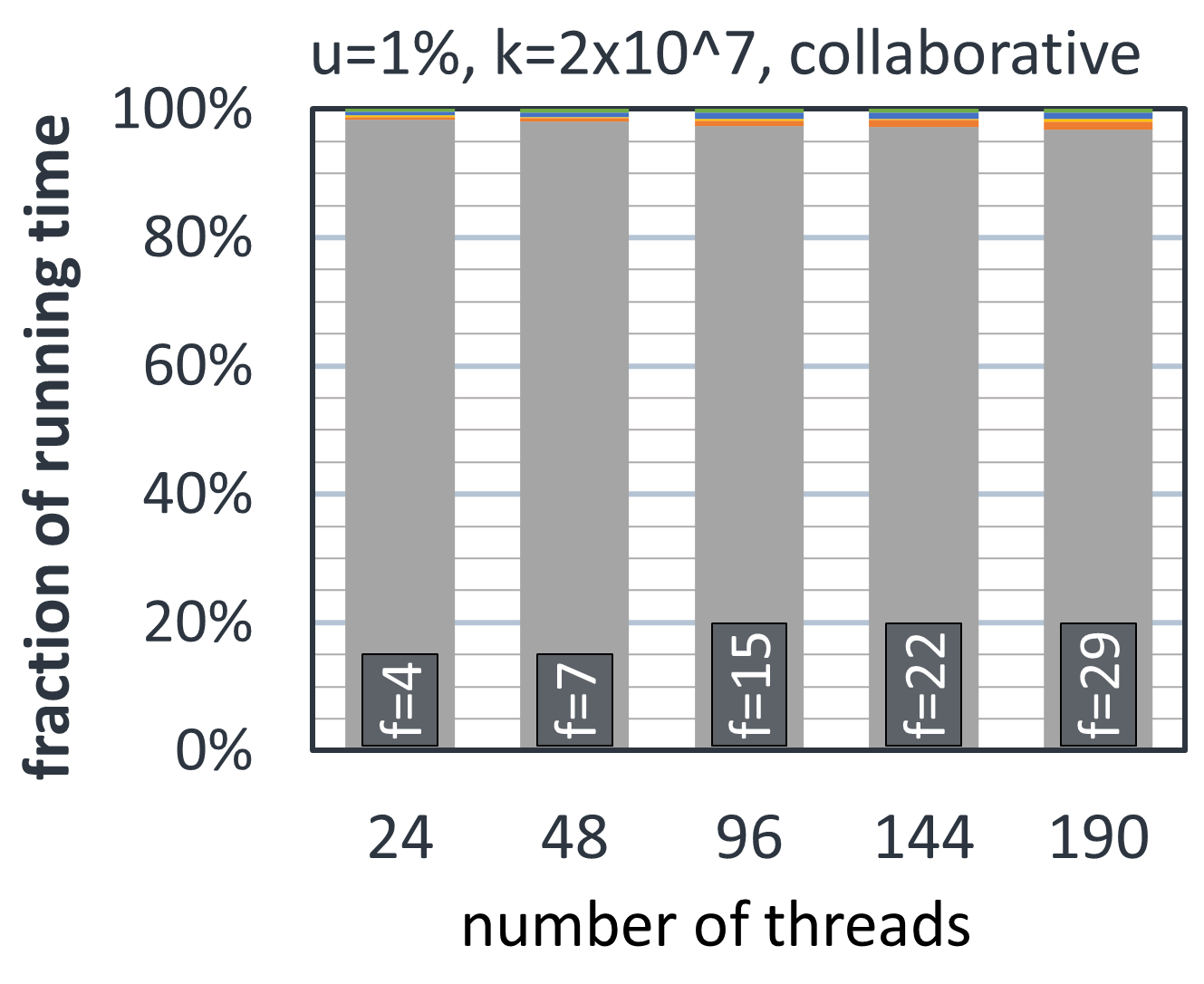}
\includegraphics[scale=0.20]{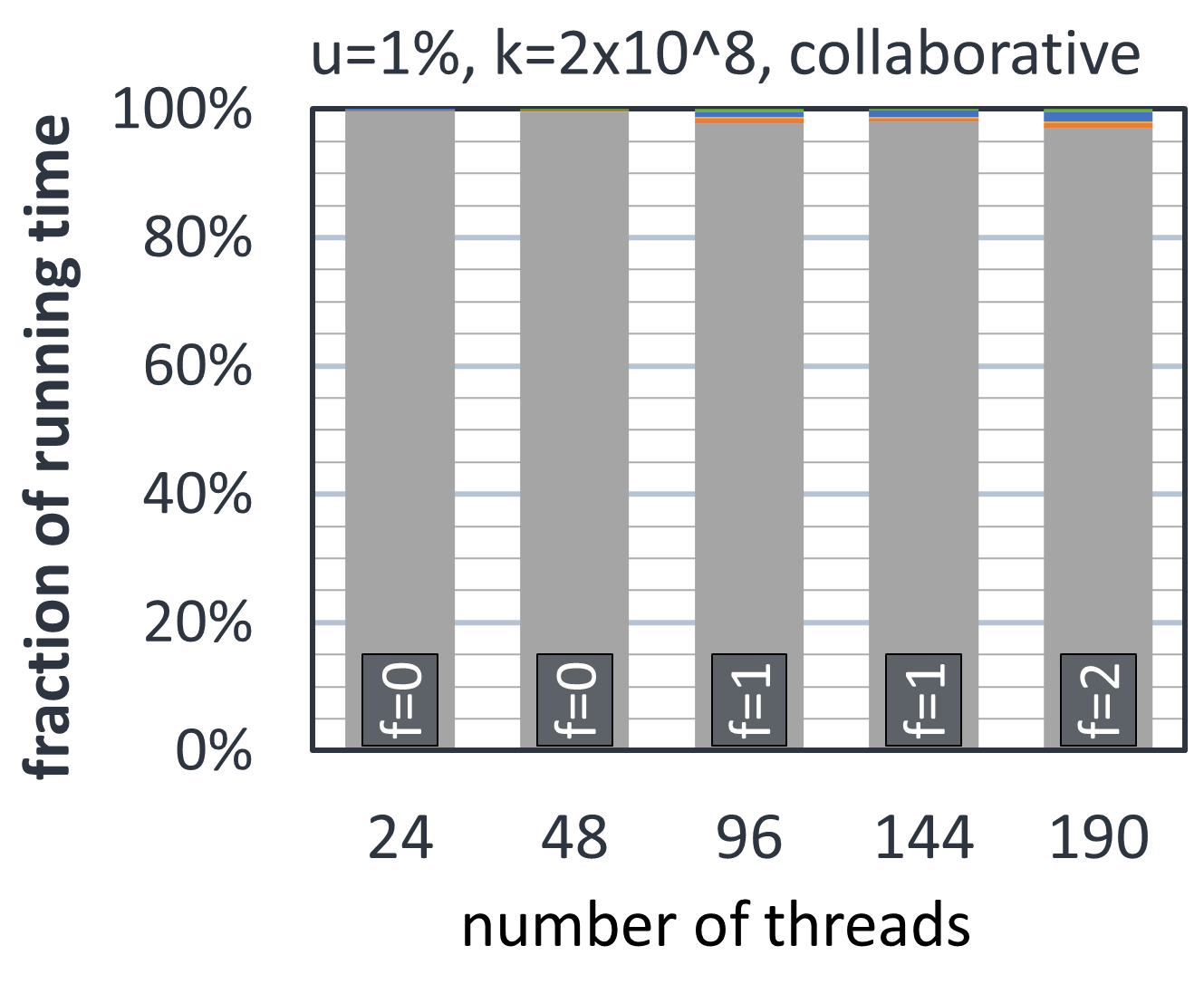}
\includegraphics[scale=0.20]{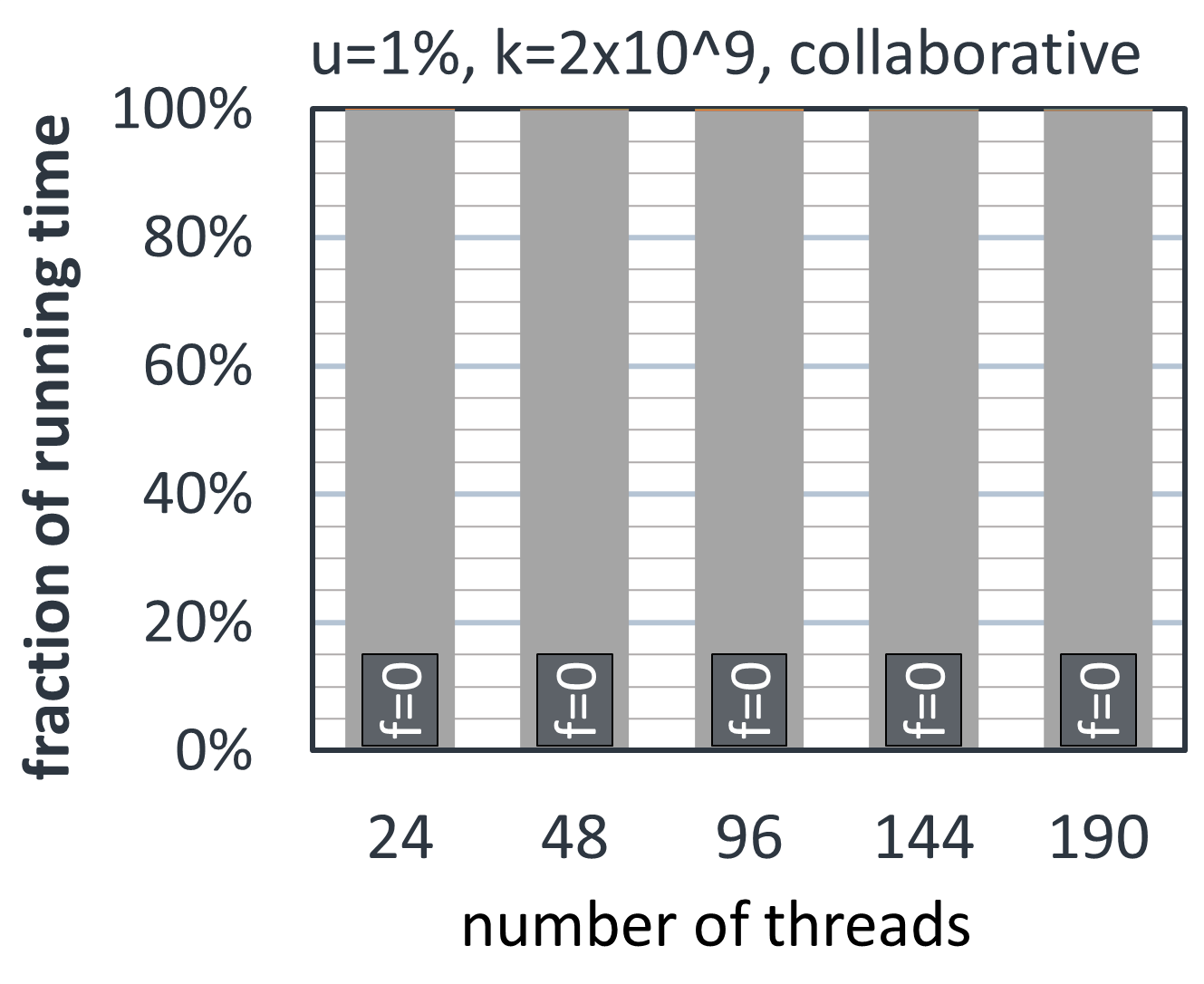}
\includegraphics[scale=0.20]{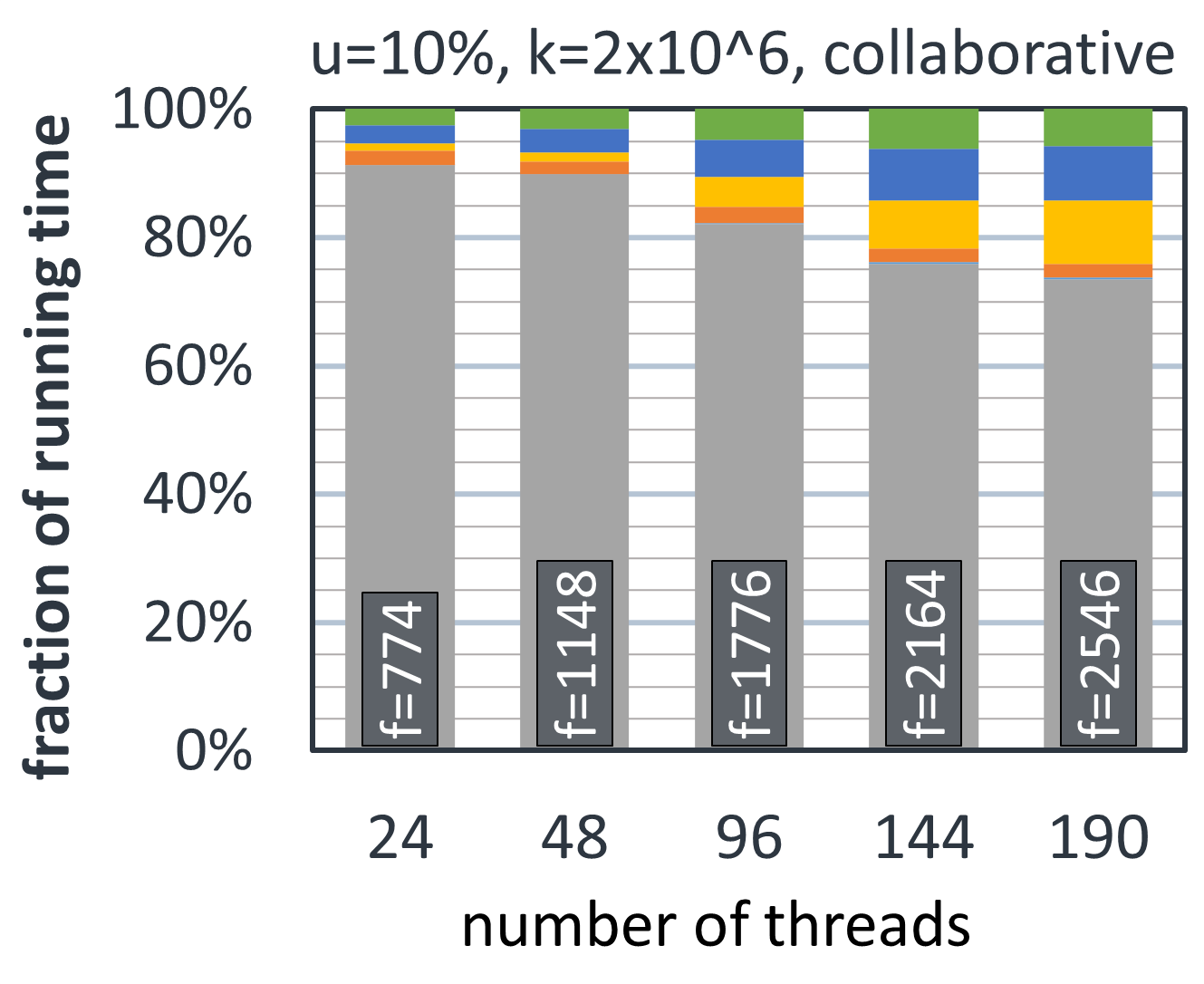}
\includegraphics[scale=0.20]{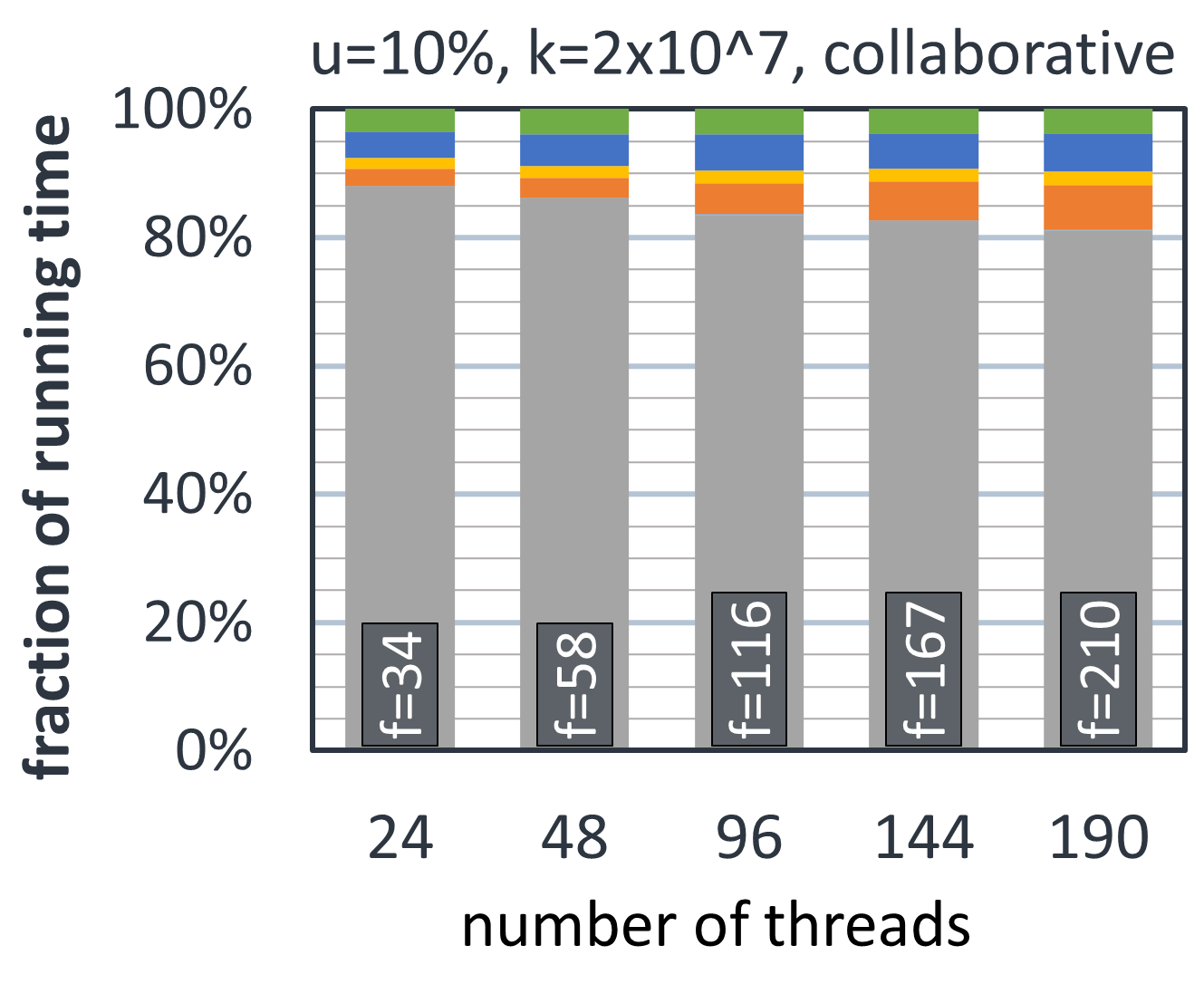}
\includegraphics[scale=0.20]{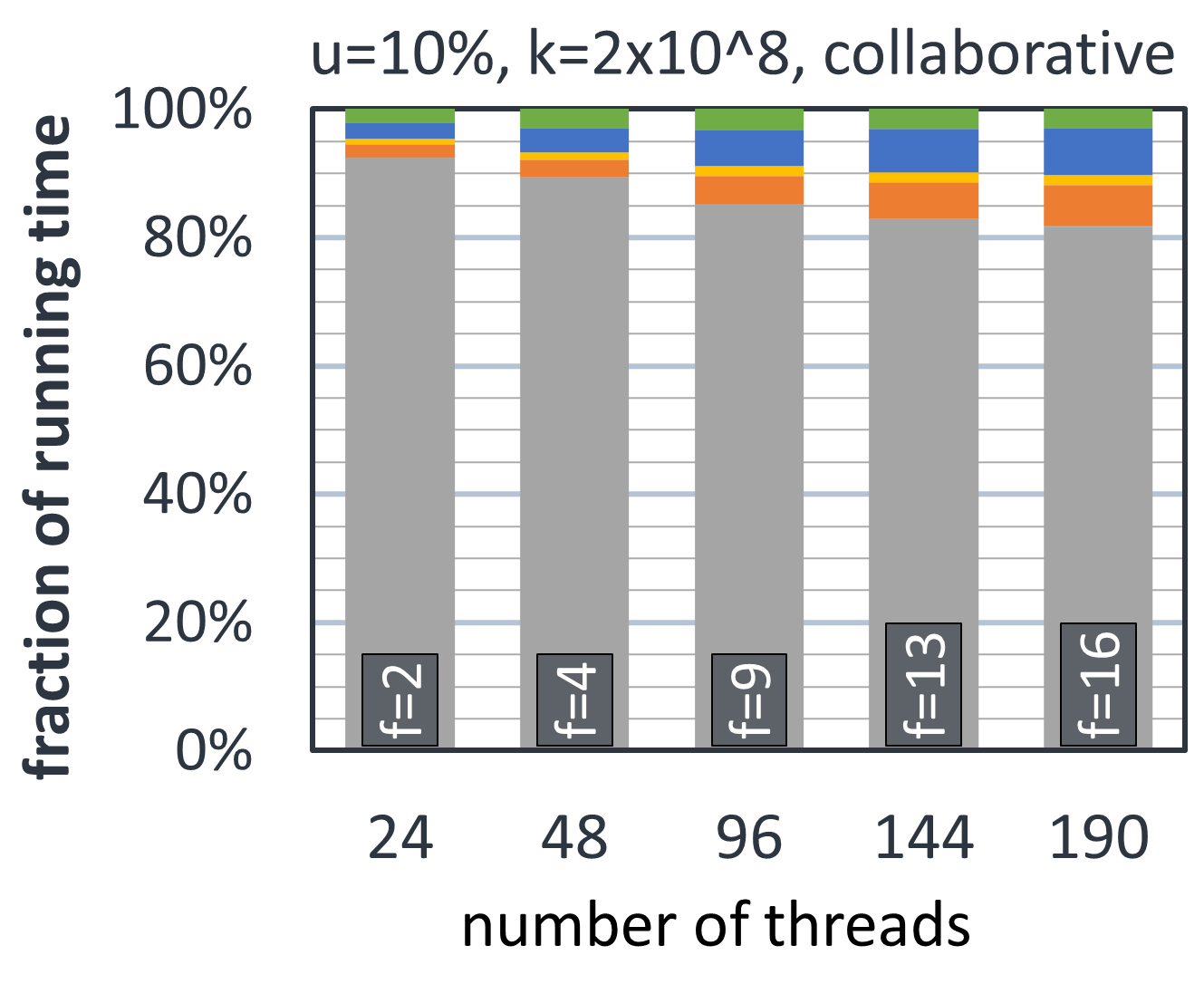}
\includegraphics[scale=0.20]{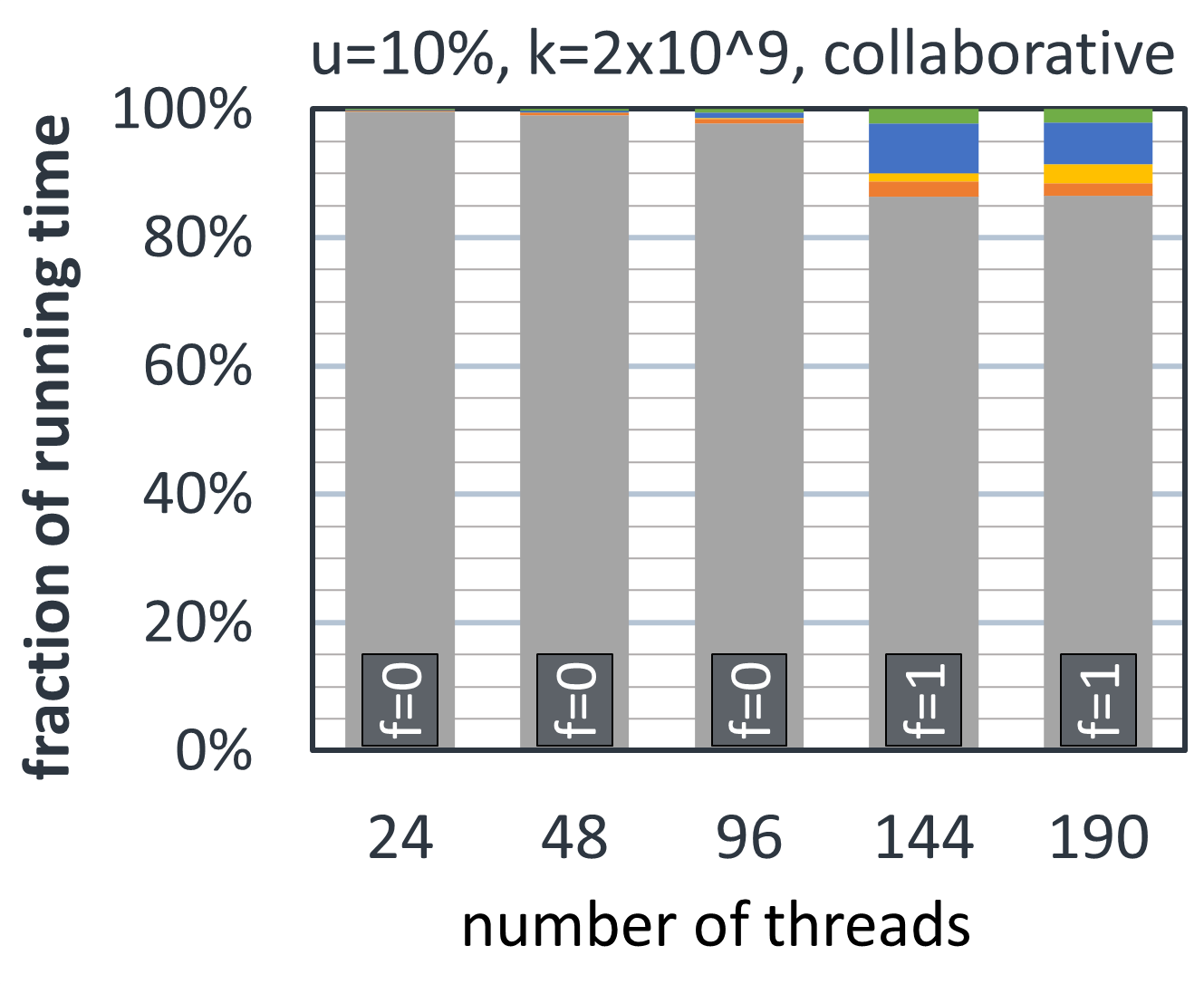}
\includegraphics[scale=0.20]{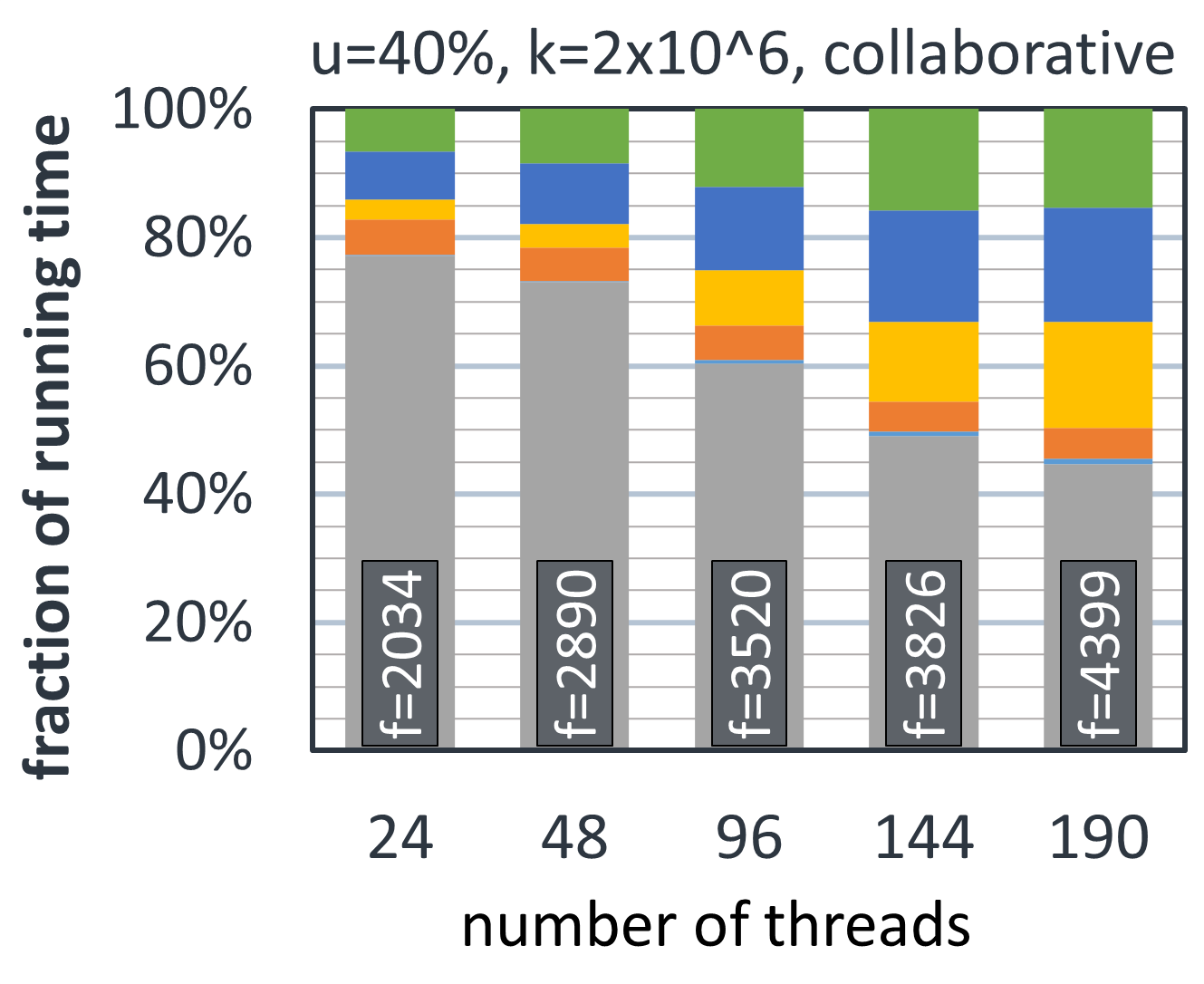}
\includegraphics[scale=0.20]{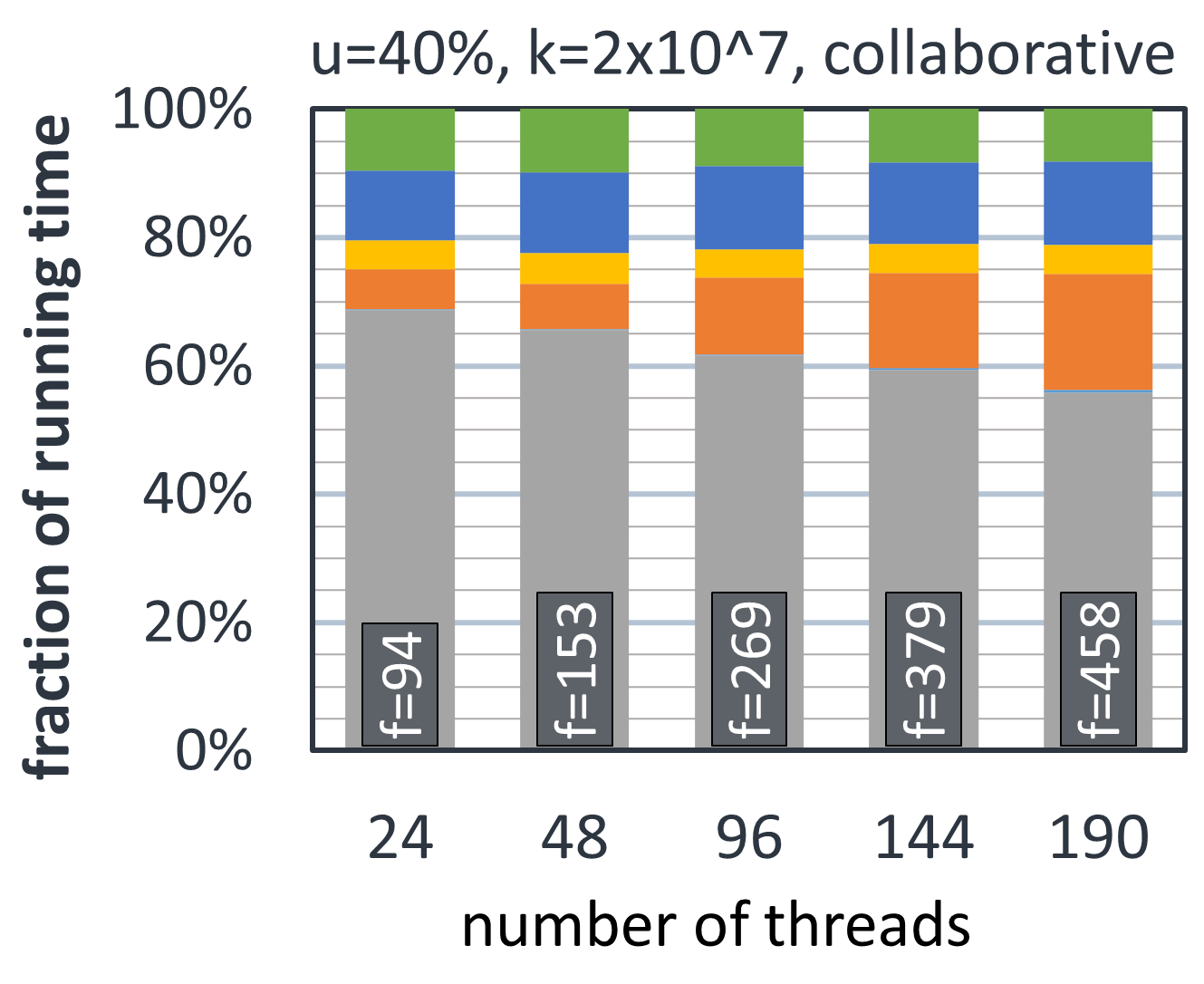}
\includegraphics[scale=0.20]{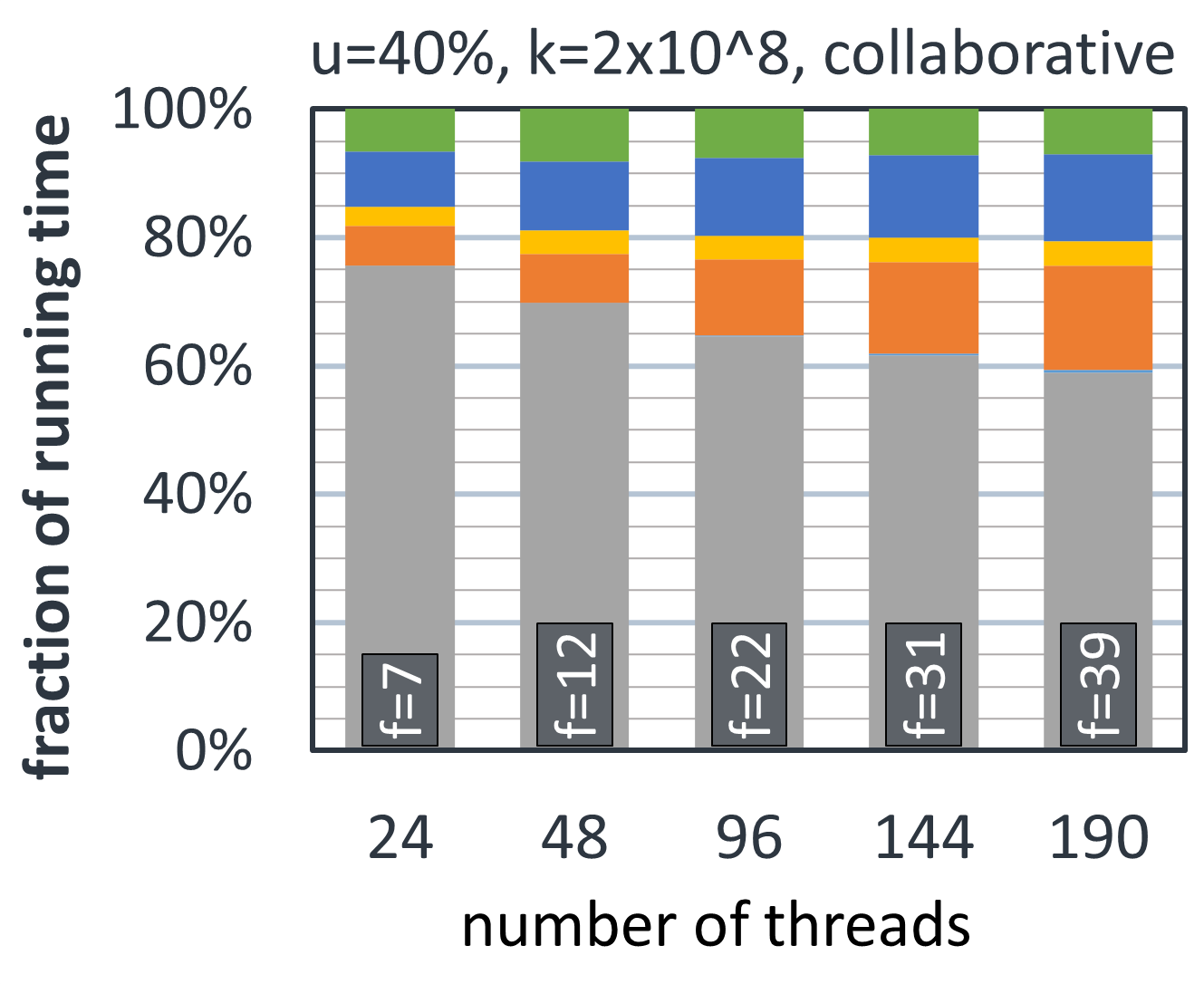}
\includegraphics[scale=0.20]{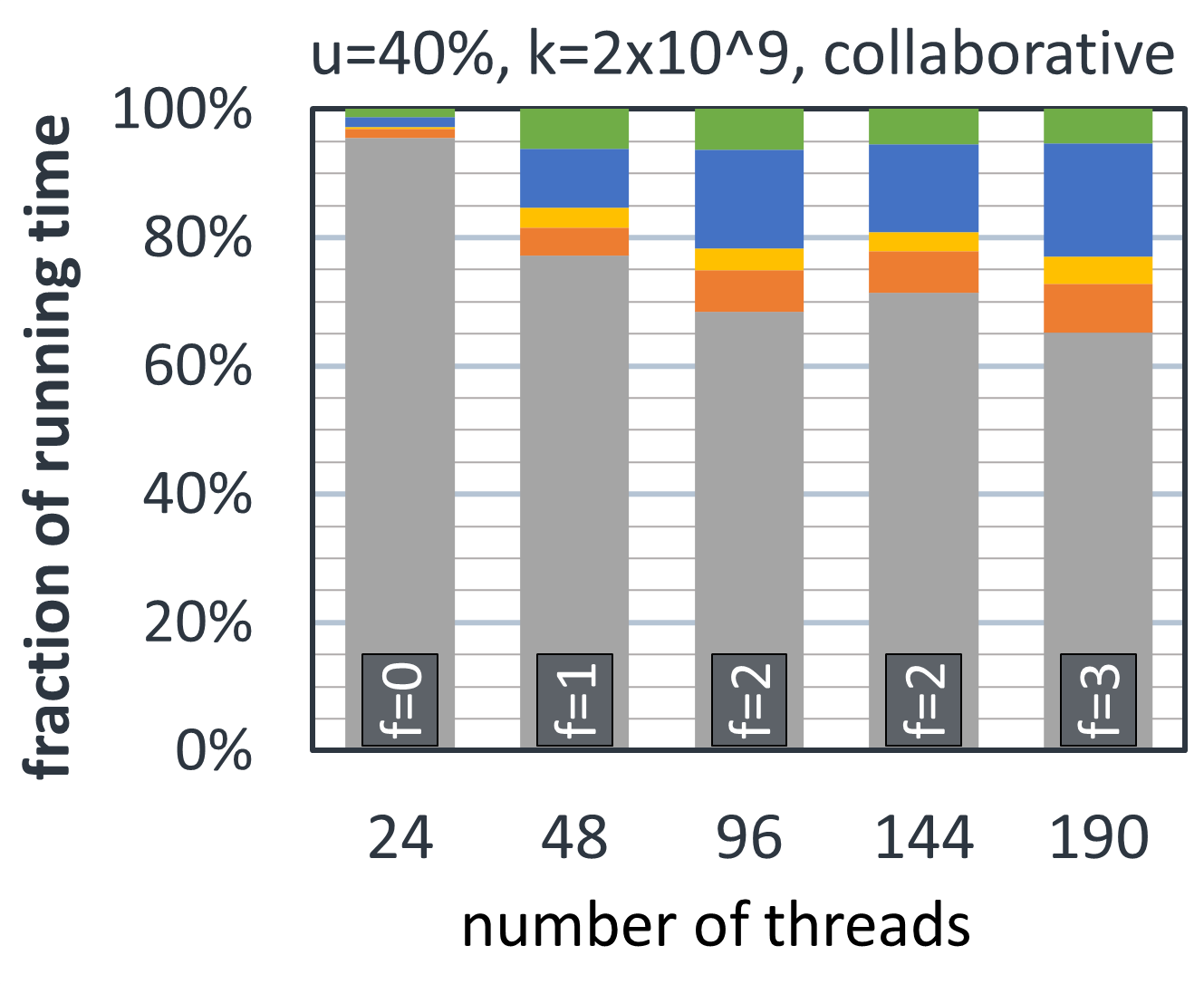}

\caption{
    Breakdown of Time Spent, when Using Collaborative Rebuilding
    (\crule[DarkBluish]{2mm}{2mm}~creating,
    \crule[Greenish]{2mm}{2mm}~marking,
    \crule[Bluish]{2mm}{2mm}~useless helping,
    \crule[BloodOrange]{2mm}{2mm}~deallocation,
    \crule[SunsetYellow]{2mm}{2mm}~locating garbage,
    \crule[Gray]{2mm}{2mm}~other).
    Bars are annotated with $f$, the number of times the \textit{root} (entire tree) was rebuilt.
}
\label{fig:appendix:experiments:collaborative-rebuilding}
\end{figure*}

\paratitle{Collaborative rebuilding}
Figures~\ref{fig:appendix:experiments:non-collaborative-rebuilding}
and \ref{fig:appendix:experiments:collaborative-rebuilding}
show the breakdown of the execution time,
with non-collaborative and collaborative rebuilding, respectively.
Plots in Figure~\ref{fig:appendix:experiments:non-collaborative-rebuilding}
show that, as the ratio of updates grows beyond $10\%$,
the execution time quickly becomes dominated by rebuilding.
In fact, most of the rebuilding time is spent in helping operations
in which a created subtree ends up being discarded.
The proportion of discarded subtrees grows with the number of threads.
Plots in Figure~\ref{fig:appendix:experiments:collaborative-rebuilding}
show that, when using collaborative rebuilding,
almost no time is spent in discarding the subtrees.

\begin{figure}
\includegraphics[scale=0.3]{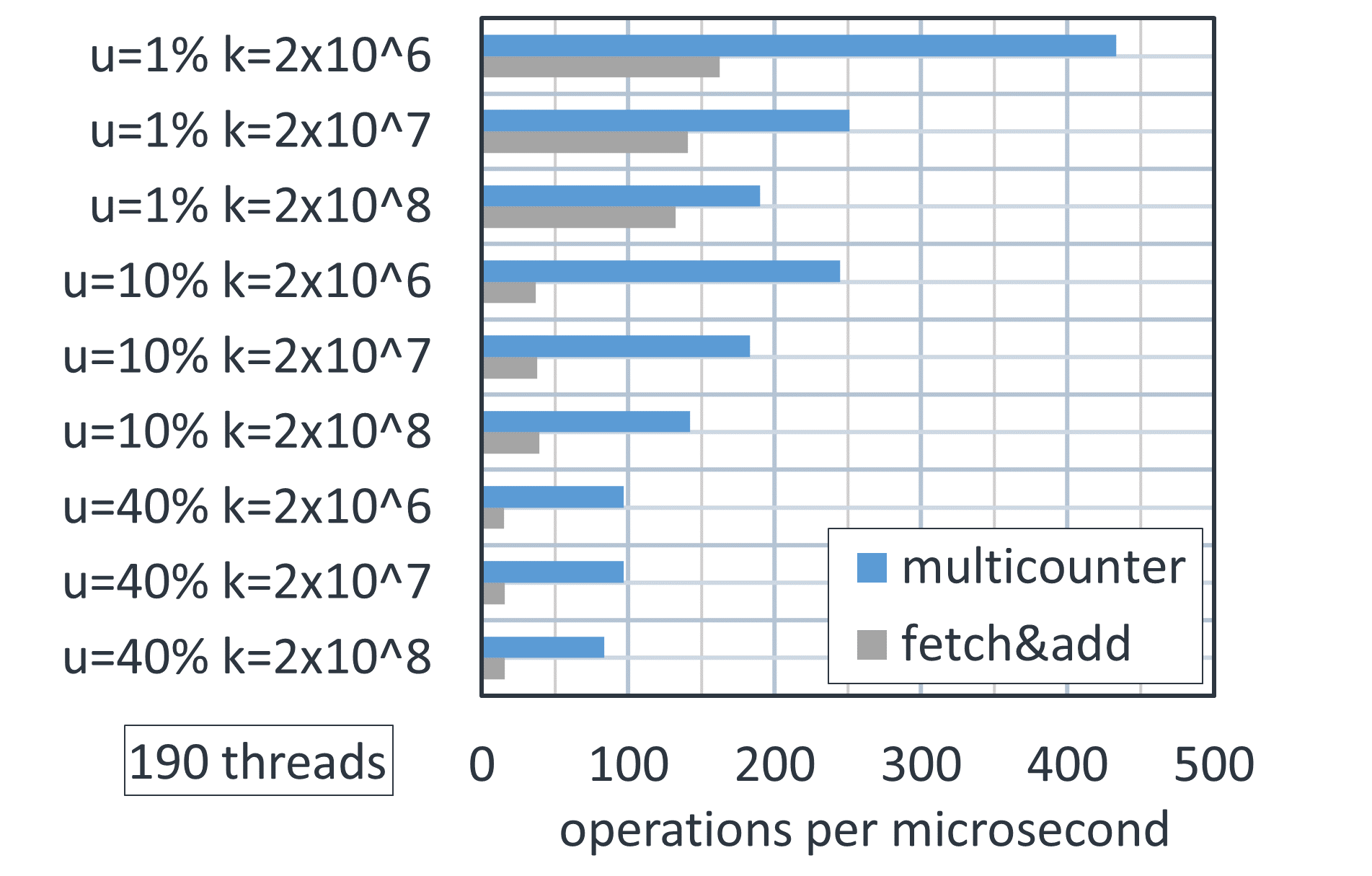}

\caption{Effect of Using a Multicounter in the Root Node}
\label{fig:appendix:experiments:multicounter}
\end{figure}

\paratitle{Using a multicounter in the root node}
Our concurrent IST implementation uses atomic fetch-and-add instructions
to implement update counters in inner nodes,
but uses a multicounter in the root node to reduce contention.
Figure~\ref{fig:appendix:experiments:multicounter}
compares the throughput of a concurrent IST with a multicounter
against an IST with a simple fetch-and-add-counter in the root node,
when using $190$ threads.
In the variant that uses a simple fetch-and-add-counter in the root,
throughput is reduced to $15\%-33\%$ compared to the multicounter variant.

\paratitle{No-SQL database workload}
We study a simple \textit{in-memory database management system} called DBx1000~\cite{dbx1000},
which is used in multi-core database research. DBx implements a simple relational database,
which contains one or more \textit{tables}. 
Each table can have one or more \textit{key fields} and associated \textit{indexes}. Each index
allows processes to query a specific key field, quickly locating any rows in which the key field
contains a desired value. We replace the default index implementation in DBx with each of the
BSTs that we study.


\begin{figure*}
\includegraphics[width=0.45\textwidth]{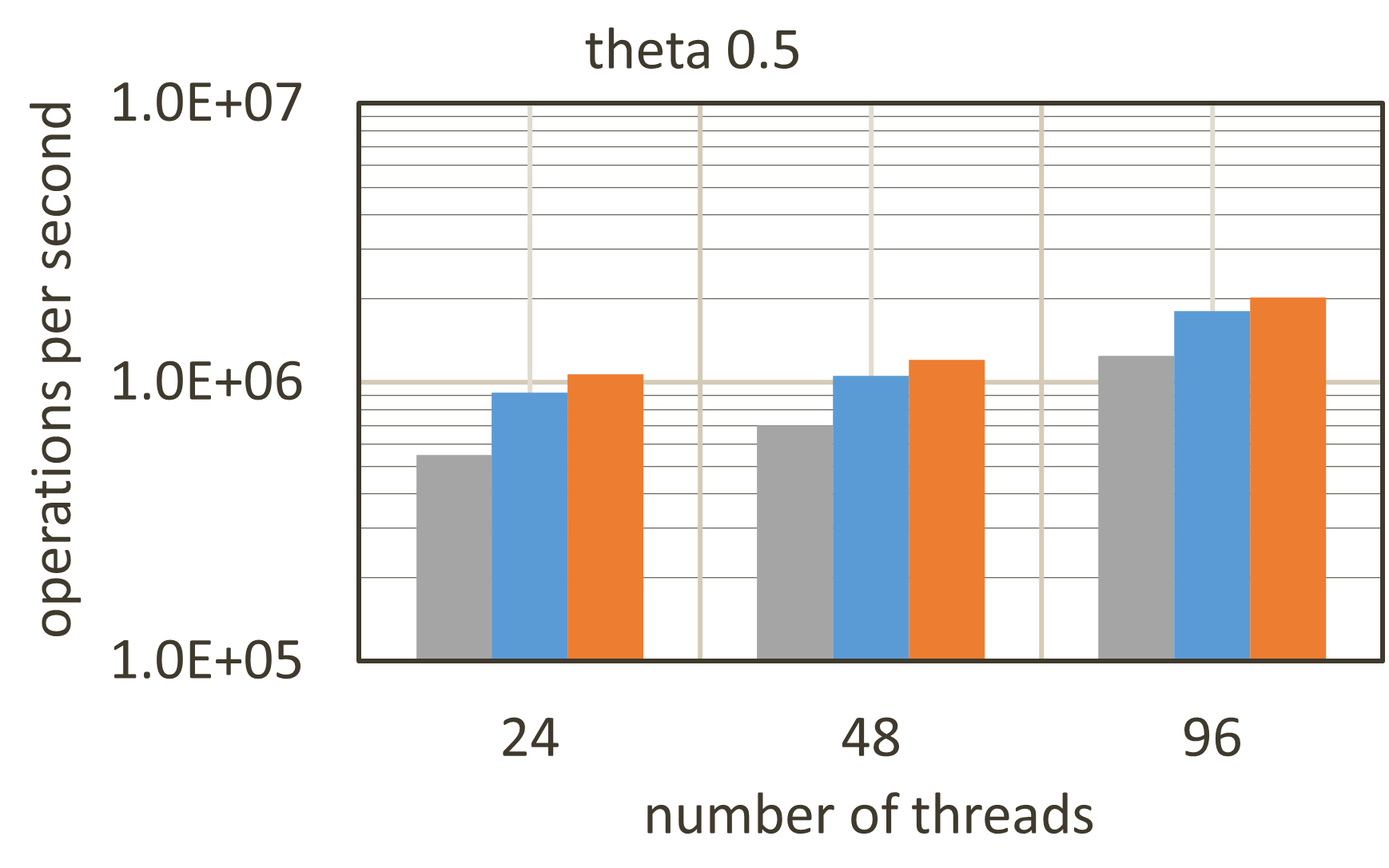}
\includegraphics[width=0.45\textwidth]{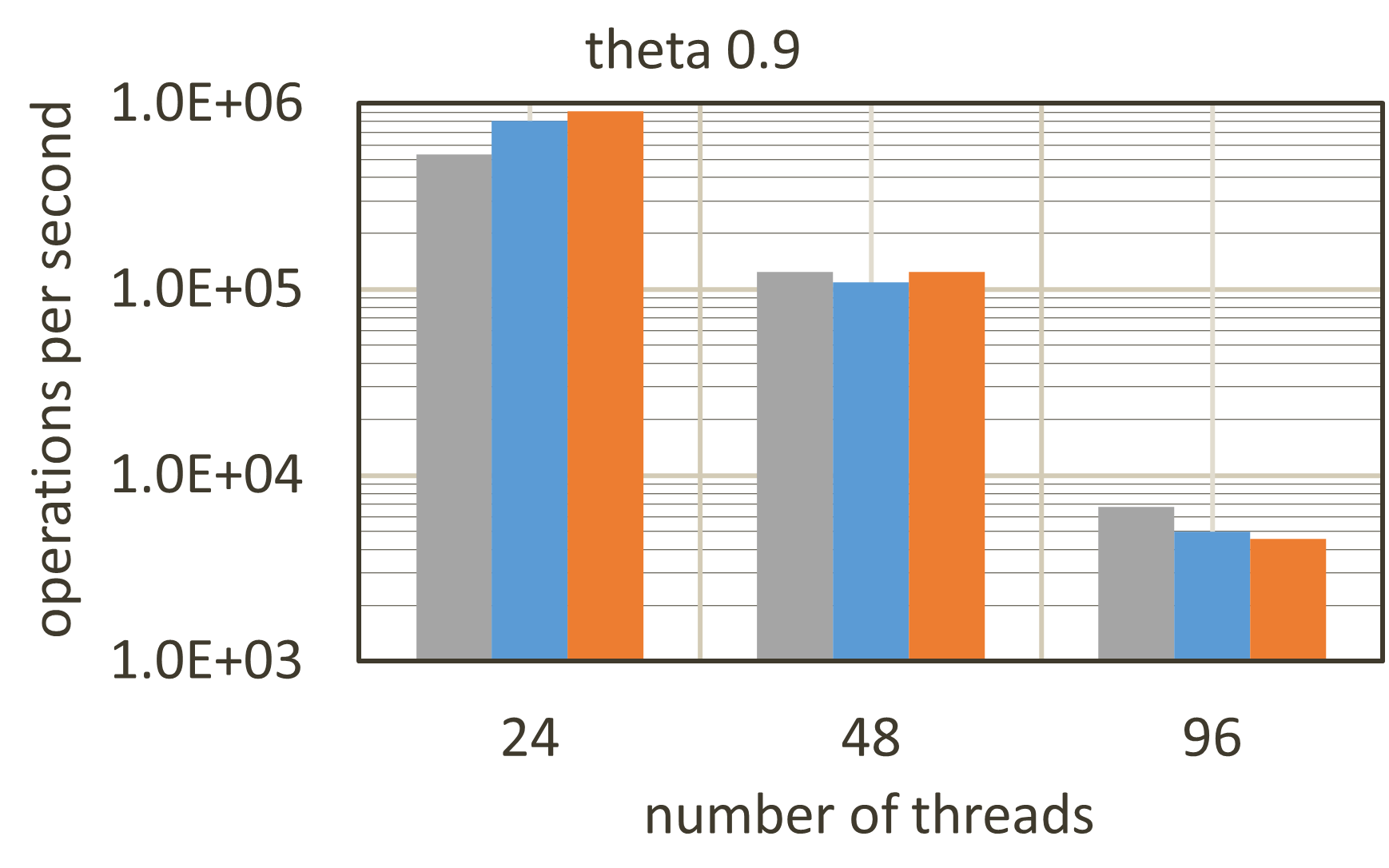}

\includegraphics[scale=0.5]{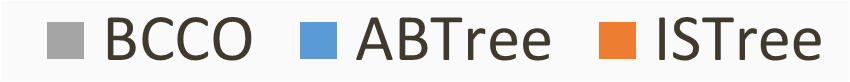}
\caption{YCSB database performance with different index data structures, and a skewed \textit{key access pattern}. (NM omitted because it is slower than BCCO.)}
\label{fig:appendix:experiments:ycsb}
\end{figure*}

Following the approaches in~\cite{dbx1000, ArbelRaviv2018GettingTT}, we run a subset of the well
known Yahoo! Cloud Serving Benchmark (YCSB) core with a single table containing 100 million
rows, and a single index. Each thread performs a fixed number of transactions (100,000 in our
runs), and the execution terminates when the first thread finishes performing its transactions.
Each transaction accesses 16 different rows in the table, which are determined by index lookups
on randomly generated keys. Each row is read with probability 0.9 and written with probability
0.1. The keys that a transaction will \textit{access} are generated according to a
\textbf{Zipfian} distribution 
following the approach in~\cite{Gray:1994}.

The results, which appear in Figure~\ref{fig:appendix:experiments:ycsb}, show how performance
degrades as the distribution of \textit{accesses} to keys becomes highly skewed. (Higher
$\theta$ values imply a more extreme skew. A $\theta$ value of 0.9 is \textit{extremely}
skewed.)

\paratitle{Trees containing Zipfian-distributed keys} Since the performance of the ISTree can
theoretically degrade when the tree contains a highly skewed set of keys, we construct a
synthetic benchmark to study such scenarios. In this benchmark, $n$ threads access a single
instance of the ISTree, and there is a \textit{prefilling} phase followed by a \textit{measured}
phase. In the prefilling phase, each thread repeatedly generates a key from a Zipfian
distribution ($\theta=0.5$) over the key range $[1, 10^8]$ (picking one of 100 million
\textit{possible} keys), and inserts this key into the data structure (if it is not already
present). This continues until the data structure contains 10 million keys (only 10\% of the key
range), at which point the prefilling phase ends. In the measured phase, all threads perform
$u$\% updates and ($100-u$)\% searches (for $u \in \{0, 1, 10\}$) on keys drawn from the same
Zipfian distribution, for 30 seconds. This entire process is repeated for multiple trials, and
for thread counts $n \in \{24, 48, 96, 144, 190\}$ (with at least one core left idle to run
system processes).
\begin{figure*}
\includegraphics[width=0.3\textwidth]{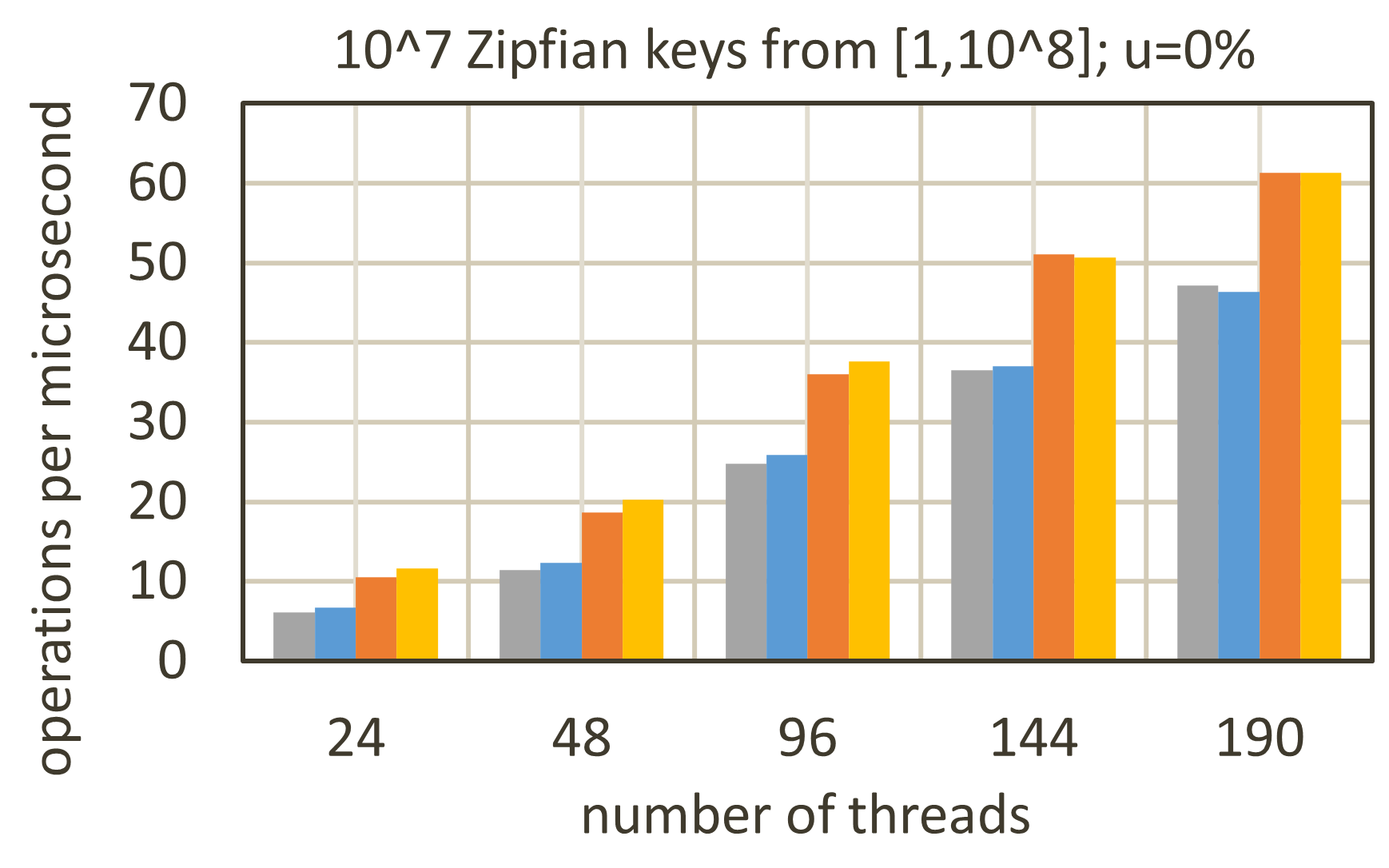}
\includegraphics[width=0.3\textwidth]{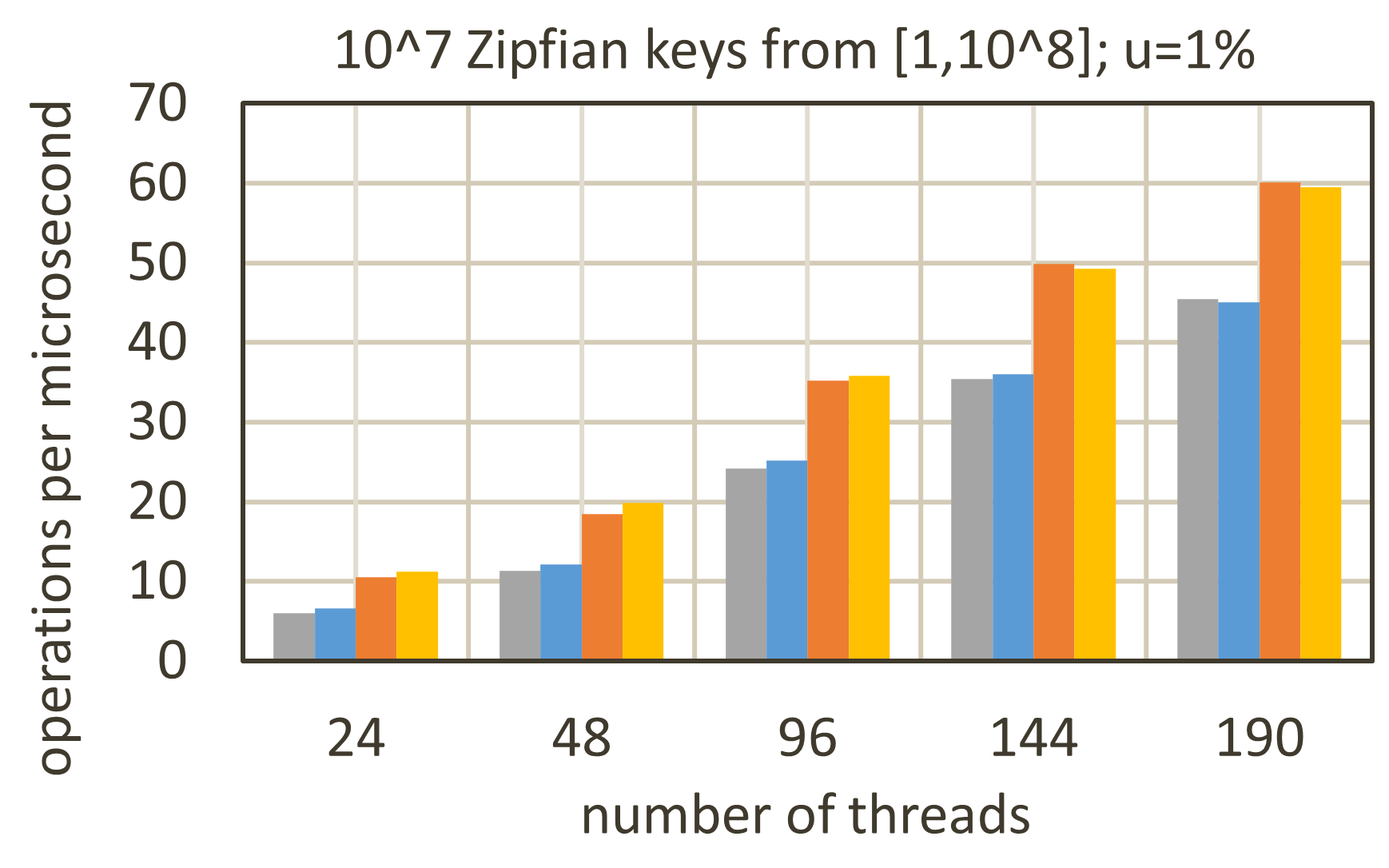}
\includegraphics[width=0.3\textwidth]{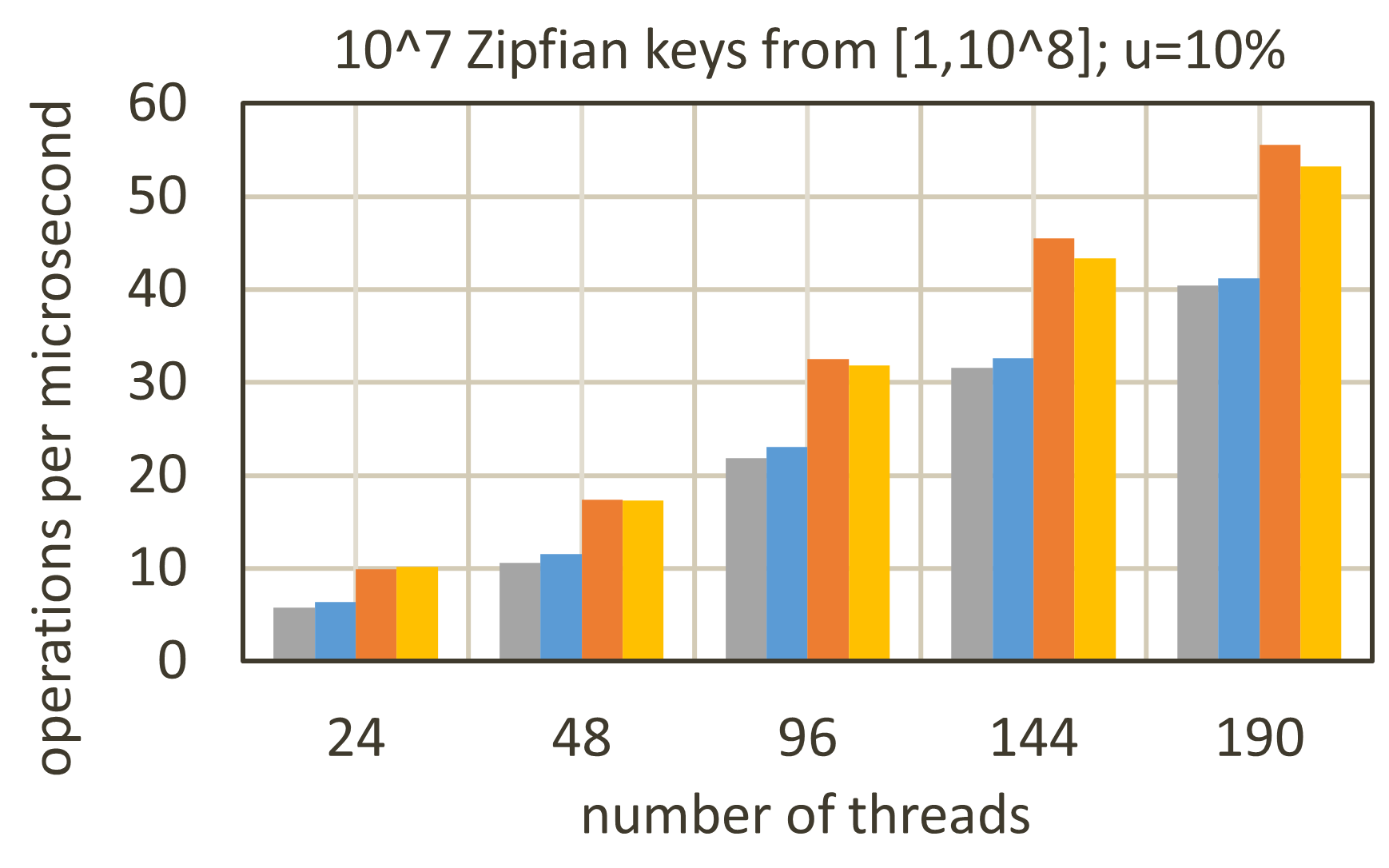}

\includegraphics[scale=0.5]{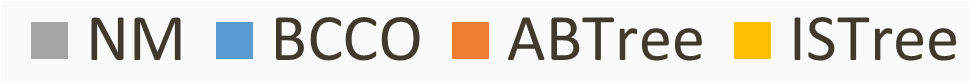}
\caption{Synthetic benchmark in which the \textit{set of keys stored} in a data structure is highly skewed.}
\label{fig:appendix:experiments:zipf}
\end{figure*}
The results, which appear in Figure~\ref{fig:appendix:experiments:zipf}, suggest that the ISTree
can be surprisingly robust in scenarios where it contains a highly skewed distribution.


\section{Related Work}
\label{sec:related}

Sequential interpolation search was first proposed by Peterson~\cite{Peterson},
and subsequently analyzed by~\cite{TwoYaos, Pearl, Gonnet}.
The \emph{dynamic} case, where insertions and deletions are possible,
was proposed by Frederickson~\cite{Fred}.
The sequential IST variant we build on is by Mehlhorn and Tsakalidis~\cite{IST}. 
This data structure supports amortized insertions and deletions in $O( \log n )$ time,
under arbitrary distributions, and amortized insertion, deletion,
and search, in $O(\log \log n)$ time under smoothness assumptions on the key distribution.
To improve scalability, we augmented C-IST
with parallel marking (to prevent updates during rebuilding),
and a parallel rebuilding phase.

For \emph{concurrent} search data structures ensuring predecessor queries,
the work that is closest to ours is the SkipTrie~\cite{SkipTrie},
which allows predecessor queries
in amortized expected $O( \log \log u  + \gamma )$ steps,
and insertions and deletions in $O( \gamma \log \log u )$ time,
where $u$ is the size of the key space, and $\gamma$ is an upper bound on contention. 
The C-IST provides inferior runtime bounds in the worst case
(e.g., $O( \log n )$ versus $O( \log \log u )$ amortized);
however, the guarantees provided under distributional assumptions
are asymptotically the same.
We believe the C-IST should provide superior
practical performance due to better cache behavior.
We have attempted to provide a comparison of the C-IST
with an open-source implementation of the SkipTrie~\cite{skip-trie-github};
we found that this implementation had significant stability
and performance issues, which render a fair comparison impossible.

There is considerable work on designing
efficient concurrent search tree data structures with predecessor queries,
e.g.~\cite{Natarajan:2014, BrownPhD, Drachsler, Braginsky2012, ArbelRaviv2018GettingTT, DBLP:conf/swat/Brown14, DBLP:conf/opodis/BrownA12, DBLP:conf/opodis/BrownA12}.
The average-case complexity of these operations is usually logarithmic
in the number of keys.
For large key counts (our target application) this search term dominates,
giving the C-IST a significant performance advantage.
This effect is apparent in our experimental section.

Other work on concurrent search tree data structures
includes early work by Kung~\cite{Kung:1980:CMB:320613.320619},
Bronson's lock-based concurrent AVL trees~\cite{bronsonavl},
Pugh's concurrent skip list~\cite{pugh90conc},
and later improvements by Herlihy et al.~\cite{Herlihy_aprovably}
(which the JDK implementation is based on),
non-blocking BSTs due to Ellen et al.~\cite{Ellen:2010:NBS:1835698.1835736},
and the KiWi data structure due to Basin et al.~\cite{Basin:2017:KKM:3018743.3018761}.

The §DCSS§ and §DCSS\_READ§ primitives that we rely on
were originally proposed by Harris~\cite{Harris:2002:PMC:645959.676137}.
The §DCSS§ primitive needs to allocate a descriptor object to synchronize
multiple memory locations.
Our C++ implementation of §DCSS§,
due to Arbel-Raviv and Brown~\cite{DBLP:conf/wdag/Arbel-Raviv017},
is able to recycle the descriptors.
There are alternative primitives to §DCSS§ with similar expressive power,
such as the §GCAS§ instruction~\cite{Prokopec:2012:CTE:2145816.2145836},
used to achieve snapshots in the Ctrie data structure.

Many concurrent data structures use the technique of snapshotting
the entire data structure or some part thereof,
with the goal of implementing a specific operation.
The SnapQueue data structure~\cite{Prokopec:2015:SLQ:2774975.2774976}
uses a \emph{freezing} technique
in which the writable locations are overwritten with special values
such that the subsequent §CAS§ operations fail.
Ctries~\cite{EPFL-REPORT-166908,Prokopec2011,Prokopec:2012:CTE:2145816.2145836,Prokopec:200977}
use the afore-mentioned §GCAS§ operation to prevent further updates to the data structure.
Work-stealing iterators~\cite{iterators-7092728,Prokopec:196627},
used in work-stealing schedulers~\cite{Prokopec2014-wstree,10.1145/2491956.2462193}
for data-parallel collections~\cite{prokopec11-pc},
use similar techniques to capture a snapshot of the iterator state.

The core motivation behind C-IST is to decrease
the number of pointer hops during the key search.
The underlying reason for this is that cache misses, which are incurred during the key search,
are the dominating factor in the operation's running time.
The motivation behind the recently proposed Cache-Trie data structure,
a non-blocking Ctrie variant, is similar -- Cache-Tries use
an auxiliary, quiescently-consistent table to speed up
the key searches~\cite{Prokopec:2018:CCL:3178487.3178498,Prokopec2018,techreport-17-prokopec}.

Our implementation of the C-IST data structure
uses a scalable concurrent counter in the root node to track the number of updates
since the last rebuild of the root node.
In the past, a large body of research focused on scalable concurrent counters,
both deterministic and probabilistic variant
thereof~\cite{10.1145/210223.210225,10.1145/1639950.1639954,10.1145/185675.185815,10.1145/2851141.2851147,10.1145/2486159.2486182,ABKLN}.
Scalable counters are useful in a number of other non-blocking data structures,
which use counters to track their size or various statistics about the data structure.
These include non-blocking queues~\cite{10.1145/248052.248106},
FlowPools~\cite{prokopec12flowpools,EPFL-REPORT-181098,Schlatter:198208},
concurrent hash maps in the JDK~\cite{dougleahome},
certain concurrent skip list implementations~\cite{10.1145/2688500.2688501},
and graphs with reachability queries~\cite{10.1145/3288599.3288617}.

Our C-IST implementation is done in C++,
and it uses a custom concurrent memory management scheme
due to Brown~\cite{DBLP:conf/podc/Brown15,DBLP:journals/corr/abs-1712-01044}.
In addition, our implementation uses techniques that decrease memory-allocator pressure
by reusing the descriptors that are typically used
in lock-free algorithms~\cite{DBLP:conf/wdag/Arbel-Raviv017,DBLP:journals/corr/abs-1708-01797,DBLP:conf/ppopp/Arbel-RavivB17}.


\section{Conclusion}
\label{sec:conclusion}

We proved the correctness, linearizability and non-blocking behavior
of the C-IST data structure.
We furthermore analyzed the doubly-logarithmic complexity
of the basic C-IST operations.
An extended experimental evaluation suggest that
C-IST significantly improves upon the performance
of competitive state-of-the-art search data structures,
in some cases by up to $\approx3.5\times$,
and the current best-performing alternative by up to $50\%$.
In highly skewed workloads, C-IST loses some of its performance advantages
but nevertheless exhibits a similar performance
as the next best concurrent tree data structure.

Given the importance of Big Data workloads and very large databases,
we believe that the techniques used in the C-IST data structure
have the potential to trigger an intriguing line of future work.

\bibliography{main}

\end{document}